\documentclass[manuscript,screen]{acmart} %,review
\usepackage{mathtools}
\usepackage{amsmath,amsfonts}
\usepackage{graphicx}
\usepackage{textcomp}
\usepackage{xcolor}
\usepackage{soul}
\usepackage{subfigure}
\usepackage{tikz}
\usepackage{tabu}
\usepackage{threeparttable}
\usepackage{balance}
\usepackage{colortbl}
\usepackage{balance}
\usepackage{colortbl}
\usepackage{amsmath}
\usepackage{tcolorbox}
\usepackage{multirow}
\usepackage{adjustbox}
\usepackage{xcolor}
\usepackage{amsmath}
\usepackage{hyperref}  % Load hyperref first
\usepackage{hyperxmp}  % Then load hyperxmp
\usepackage{listings}
\newcommand\myeq{\stackrel{\mathclap{\normalfont\mbox{def}}}{=}}

\definecolor{light-gray-fork}{gray}{0.925}
\definecolor{light-gray-mainline}{gray}{0.8}
\definecolor{javared}{rgb}{0.6,0,0} % for strings
\definecolor{javagreen}{rgb}{0.25,0.5,0.35} % comments
\definecolor{javapurple}{rgb}{0.5,0,0.35} % keywords
\definecolor{javadocblue}{rgb}{0.25,0.35,0.75} % javadoc

\usepackage{listings}
\usepackage{multirow}
\usepackage{microtype}
\usepackage[figuresleft]{rotating}
\usepackage{cleveref}

\def\baseCommit END{\emph{Base}}
\def\leftCommit END{\emph{Left}}
\def\rightCommit END{\emph{Right}}
\def\mergeCommit END{\emph{Merge}}

\usepackage{xcolor}
\definecolor{excerptgreen}{HTML}{AAEEB7}
\definecolor{excerptred}{HTML}{E4A598}

\AtBeginDocument{%
  \providecommand\BibTeX{{%
    \normalfont B\kern-0.5em{\scshape i\kern-0.25em b}\kern-0.8em\TeX}}}

\setcopyright{acmcopyright}
\copyrightyear{2018}
\acmYear{2018}
\acmDOI{XXXXXXX.XXXXXXX}

\DeclareUnicodeCharacter{2212}{-}
\begin{document}

\title{Continuously Learning Bug Locations}


\author{Paulina Stevia Nouwou Mindom}
\email{paulina-stevia.nouwou-mindom@polymtl.ca}
\affiliation{%
  \institution{Polytechnique Montréal}
  \country{Canada}}
  
\author{Leuson Da Silva}
\email{leuson-mario-pedro.da-silva@polymtl.ca}
\affiliation{%
  \institution{Polytechnique Montréal}
  \country{Canada}}
%\orcid{0000-0002-9086-9038}

\author{Amin Nikanjam}
\email{amin.nikanjam@polymtl.ca}
\affiliation{%
  \institution{Polytechnique Montréal}
  \country{Canada}}
  
  \author{Foutse Khomh}
\email{foutse.khomh@polymtl.ca}
\affiliation{%
  \institution{Polytechnique Montréal}
  \country{Canada}}

\newcommand{\Paulina}[1]{\textcolor{blue}{{\it [Paulina: #1]}}}
\newcommand{\lm}[1]{\textcolor{purple}{{\it [Leuson: #1]}}}
\newcommand{\Foutse}[1]{\textcolor{red}{{\it [Foutse: #1]}}}
\newcommand{\Amin}[1]{\textcolor{orange}{{\it [Amin: #1]}}}
%% the author to define a more concise list Aiming to address this challenge, in this paper, we evaluate the potential of using Continual Learning (CL) techniques in a multitask \Foutse{multi-task? what are the different tasks? it seems we are doing only bug localization, or this is broken down into multiple sub-tasks?} non-stationary setting for bug localization, comparing it against a bug localization with BERT model, a deep reinforcement learning-based model that leverages the A2C algorithm and a DL-based function-level interaction model for semantic bug localization.
\renewcommand{\shortauthors}{Mindom et al.}

\begin{abstract}
%\lm{The evaluation is done based on file and changeset level. Maybe adopt a more generic terminology and then specify the level of granularity you're considering in the study?} \Paulina{Changesets is actually the generic terms. It is for changeset-files hunks and commits}
Automatically locating buggy changesets associated with bug reports is crucial in the software development process. Deep Learning (DL)-based techniques show promising results by leveraging structural information from the code and learning links between changesets and bug reports. However, since source code associated with changesets evolves, the performance of such models tends to degrade over time due to concept drift. Aiming to address this challenge, in this paper, we evaluate the potential of using Continual Learning (CL) techniques in multiple sub-tasks
%\Foutse{multi-task? what are the different tasks? it seems we are doing only bug localization, or this is broken down into multiple sub-tasks?}  
setting for bug localization (each of which operates on either stationary or non-stationary data), comparing it against a bug localization technique that leverages the BERT model, a deep reinforcement learning-based technique that leverages the A2C algorithm, and a DL-based function-level interaction model for semantic bug localization.  
Additionally, we enhanced the CL techniques by using logistic regression to identify and integrate the most significant bug-inducing factors. Our empirical evaluation across seven widely used software projects shows that CL techniques perform better than  DL-based techniques by up to 61\% in terms of Mean Reciprocal Rank (MRR),  44\% in terms of Mean Average Precision (MAP), 83\% in terms of top@1, 56\% in terms of top@5, and 66\% in terms of top@10 metrics in non-stationary setting. Further, we show that the CL techniques we studied are effective at localizing changesets relevant to a bug report while being able to mitigate catastrophic forgetting across the studied tasks and require up to 5x less computational effort during training. Our findings demonstrate the potential of adopting CL for bug localization in non-stationary settings, and we hope it helps to improve bug localization activities in Software Engineering using CL techniques.
\end{abstract}

\begin{CCSXML}
<ccs2012>
   <concept>
       <concept_id>10011007.10011006.10011071</concept_id>
       <concept_desc>Software and its engineering~Software configuration management and version control systems</concept_desc>
       <concept_significance>500</concept_significance>
       </concept>
 </ccs2012>
\end{CCSXML}

\ccsdesc[500]{Software and its engineering~Software configuration management and version control systems}

\keywords{Reinforcement Learning, Continual Learning, Bug Localization, Software Maintenance}


\maketitle

\section{Introduction}
\label{sec:introduction}
\noindent
\looseness=-1
\noindent
%Bug localization techniques can reduce the time spent by developers to find bugs and reduce the cost of software maintenance. 
%\Amin{the intro is tool long!I indicated th parts that can be summarized} \Paulina{I shrinked all the part you mentioned. But not at the exact number of sentences you suggested. I have now 1 to 3 more sentences. In my opinion shrinking that much  could disrupt the flow of reading. Moreover the paper changed a bit, as now I considered the hunk level. There is a couple more of sentences to justify the choice of hunk level. Also I think choosing CL agents lacked some motivations compared to the previous version. In anycase please have a run at it, if it still does not make sense to you, I will shrink to the exact amount. Thanks}\Amin{ok}
%\Amin{this paragraph is good}
Researchers and practitioners are using software systems for various purposes across different domains, such as education, financial services, and healthcare. Such systems are getting larger and more complex, which makes it challenging for developers to find bugs when they occur and eventually fix them \cite{chakraborty2023rlocator}. Developers spend approximately one-third of their time trying to locate and fix bugs \cite{zhang2015survey}; when not done in time, a bug can pose serious concerns regarding the usability and security of software systems \cite{zhang2015survey}. For example, extended debugging periods can delay product releases, or a bug that remains unfixed can degrade the user experience \cite{cotroneo2016bugs}. In the literature, different techniques have been adopted to assist developers in trying to locate and fix bugs. While debugging tools can assist in locating bugs \cite{petrillo2016towards}, regression testing ensures that any changes in the software system artifacts do not adversely affect existing features \cite{wong1997study}. However, both approaches demand significant human effort by analyzing code through debugging or creating and maintaining regression tests, respectively.
Various bug localization techniques have been proposed, including Machine Learning (ML) approaches, which locate bugs by measuring the cosine similarity between bug reports and source code artifacts~\cite{zhou2012should,liang2022modeling,chakraborty2023rlocator,moser2008comparative,chidamber1994metrics,antoniol2005feature}.
%Various techniques have been proposed for bug localization, which includes leveraging Machine Learning (ML) techniques ~\cite{zhou2012should,liang2022modeling,chakraborty2023rlocator,moser2008comparative,chidamber1994metrics,antoniol2005feature}. Among ML techniques, models like Convolutional Neural Networks (CNNs) and Recurrent Neural Networks (RNNs) have been employed to locate bugs at the changeset-files (i.e., a collection of source code files that were modified as part of a particular commit) level by measuring cosine similarity between bug reports and changeset-files~\cite{xiao2018machine,liang2019deep}. 
%Liang et al. \cite{liang2022modeling} explore the semantic relationships between source code and bug reports using a fine-tuned language model to derive semantic features for a learn-to-rank model. Additionally, Chakraborty et al. ~\cite{chakraborty2023rlocator} utilized Deep Reinforcement Learning (DRL) to train an agent to pick buggy changeset-files based on their relevance to bug reports and the distance between the already picked changeset-files.
 % 
%However, recent studies show that bug localization at the changeset-files demands significant effort from developers, particularly when dealing with large files. This has prompted a shift towards exploring more practical, fine-grained bug localization approaches targeting 
Over the years, researchers have investigated the process of locating bugs at both the changeset-file level (i.e., collections of source code files modified in a specific commit)~\cite{corley2018changeset,kim2013should,nguyen2011topic,chakraborty2023rlocator,liang2022modeling}, and the code segment level ~\cite{murali2021industry,zou2018practitioners}. %level requires developers considerable effort, especially with large files, leading to the exploration of more practical fine-grained code segments like the hunks \cite{murali2021industry,zou2018practitioners}.
 While these studies have reported promising results, they primarily focus on fault localization in stationary settings~\cite{chakraborty2023rlocator,liang2022modeling}. 
This presents a notable limitation, as software projects are inherently non-stationary. Between the time a bug is reported and resolved, multiple versions of changeset files or hunks may be created. %exist, leading to evolving and dynamic data. 
Consequently, the data associated with bug reports and fixes is subject to non-stationary distribution shifts, a phenomenon increasingly recognized in the literature~\cite{kurle2019continual}. 
When ML-based techniques are evaluated on non-stationary data, their performance can degrade due to concept drift~\cite{lu2018learning}. However, existing ML approaches for bug localization often fail to address this drift, as they typically consider only a single version of changeset-files or hunks~\cite{chakraborty2023rlocator,liang2022modeling}. This limitation can reduce their effectiveness in ranking relevant buggy changeset-files across multiple evolving versions of a software program.
Multiple strategies have been proposed in the literature to address concept drift. These strategies include retraining machine learning models from scratch on new changeset files (or hunks) or incrementally updating the models using all available changeset files (or hunks) up to the point of bug localization.  However, these approaches are very resource intensive, as the rapid evolution of modern software generates expansive search spaces and repositories containing numerous commits~\cite{olewicki2023towards,rosen2015commit,savor2016continuous}.
%Retraining models from scratch is costly as software projects grow \cite{olewicki2023towards}, and trained the model on all changesets-files (or hunks)  can becomes impractical due to the vast search space in rapidly evolving software
To mitigate these challenges, researchers have investigated Continual Learning (CL) approaches, which aim to address evolving data distributions and the associated concept drift, over time~\cite{rolnick2019experience}. 
%CL techniques can be applied for supervised learning by replay or generating samples for more updates ~\cite{lopez2017gradient,javed2019meta}, and for DRL, to train CL agents on a variety of tasks in non-stationary settings ~\cite{powers2022cora,rolnick2019experience,kirkpatrick2017overcoming}. 
Researchers such as Rolnick et al. \cite{rolnick2019experience} and Kirkpatrick et al. \cite{kirkpatrick2017overcoming} have implemented DRL-based Continual Learning (CL) agents (e.g., CLEAR \cite{rolnick2019experience} and EWC \cite{kirkpatrick2017overcoming}), by cyclically training them across diverse and dynamic environments with shifting data distributions. Evaluation results reveal that these CL agents can achieve performance comparable to that of agents trained separately in each environment; effectively handling distribution shifts while reducing training costs \cite{powers2022cora,rolnick2019experience,kirkpatrick2017overcoming}. 
Inspired by these findings, we hypothesize that \textit{DRL-based continual learning (CL) agents have the potential to improve bug localization}. However, their application in this context remains unexplored, leaving the software engineering community with little understanding of how these agents can be adapted or how effective they might be in addressing the concept drift challenge of software engineering data.
%Inspired by these findings, we postulate that \textit{DRL-based CL agents can enhance the bug localization process}. However, previous research has yet to explore their application in this context, leaving the software engineering community with limited insights into how these agents can be adapted and their effectiveness in this domain.

To address this gap, this paper empirically evaluates the effectiveness of DRL-based CL techniques for bug localization, comparing them against three baseline machine learning techniques. The evaluation focuses on performance and computational efficiency during training across different data granularities (i.e., changeset-files and hunks). We adopt a CL framework that incorporates two well-established approaches to continual learning, tailored for DRL in multi-task, non-stationary environments: CLEAR \cite{rolnick2019experience} and EWC \cite{kirkpatrick2017overcoming}. CLEAR employs an actor-critic model that combines new and replayed experiences, updating the model using the V-Trace off-policy algorithm \cite{espeholt2018impala}, which corrects for off-policy distribution shifts via importance weights. EWC, on the other hand, is a weight consolidation technique that maintains the model's adaptability (i.e., its ability to acquire new knowledge) by restricting significant parameters to stay close to their original values.

We further enhance these CL techniques, by training a logistic regression model using known bug-inducing factors (e.g., source and executable lines of code, cyclomatic complexity, number of lines modified, and pre-release bugs), selecting only non-collinear, statistically significant factors. These factors are effective indicators of potential bugs in software programs \cite{taba2013predicting}. The scalar output of the logistic regression model was then incorporated into the reward function of the DRL agents, boosting their effectiveness in identifying and localizing bugs. 

To evaluate the proposed technique, we conducted extensive empirical studies on seven large-scale, open-source software projects that contain non-stationary data at changeset-files and hunks levels, namely AspectJ, SWT, Birt, Tomcat, PDE, Zxing, and Eclipse (i.e., Eclipse Platform UI). Results show that the CL techniques outperform the baseline studies when evaluated on the non-stationary changeset-files data. Furthermore, incorporating prior knowledge about bug-inducing factors through logistic regression enhances performance by up to 51\% across the studied software projects at the changeset-files level. At the hunk level, CL agents outperform the baselines by up to 67\% in ranking relevant hunks for a bug report, using top@5 and top@10 metrics. Our findings highlight that combining CL techniques with prior knowledge of bug-inducing factors significantly enhances performance, offering practical insights for advancing debugging practices. This approach not only underscores the potential of CL techniques to reduce debugging time but also contributes to improving overall software reliability; paving the way for more efficient and adaptive debugging solutions.

%Furthermore, prior knowledge about bug-inducing factors \Foutse{injected?}through logistic regression boosts their performance by up to $51\%$, across the studied software projects at the changeset-files level. At the hunk level, CL agents outperform the baseline studies by up to 67\% in ranking relevant hunks to bug reports with the top@5 and top@10 metrics. Our research empirically demonstrates that incorporating CL and prior knowledge about bug-inducing factors significantly boosts performance, providing valuable insights for improving debugging practices. This also drives the development of advanced techniques like CL, which can reduce debugging time and enhance software quality.
%\lm{I missed more discussion about the results. Maybe make the previous paragraph shorter by presenting the state-of-the-art tools in a concise way.} 
%\lm{what is the impact for research, practitioners etc.}
To summarize, in this paper, we make the following contributions:
\begin{itemize}
    \item We examine the effectiveness of DRL-based CL techniques for bug localization.
    \item We adopt both regularization and rehearsal approaches for our CL agents to locate buggy changeset-files and hunks in non-stationary settings.
    \item We enhance the performance of CL agents by incorporating prior knowledge of bug-inducing factors through a logistic regression model integrated into the agents' reward function.
    \item We train and evaluate the CL agents on non-stationary data collected from seven Apache software projects. %(AspectJ, Birt, Eclipse Platform UI, SWT, Tomcat)     through extensive mining. 
    \item We make our data and models publicly available in our replication package \cite{replication-package}.
\end{itemize}
%\lm{the contributions should be concise and direct, like one single sentence. Btw, for the last contribution, I'd focus only on the new data generated by you.}
The rest of this paper is organized as follows. In Section \ref{Background}, we review the necessary background knowledge on DRL, CL, and the bug localization problem. Section \ref{sec:Motivating example} presents a concrete illustration of the problem we are addressing and our approach to solving it. The methodology followed in our study is described in Section \ref{studydesign}. We discuss the obtained results in Section \ref{sec:Experimental results}. In Section \ref{sec:Experimental results}, we also conduct ablation studies to validate the choice of bug-inducing factors incorporated into the CL agents. In Section \ref{sec:Discussions}, we interpret our findings, discuss their implications, and highlight some avenues for future works. We review related work in Section \ref{Related work}. Threats to the validity of our study are discussed in Section \ref{threats}. Finally, we conclude the paper in Section \ref{Conclusion}.

\section{Background}\label{Background}
\noindent
%\subsection{Preliminaries}
%\subsection{Deep Reinforcement Learning}
%\Paulina{Not sure about the following, we should check it out}
%Deep Learning (DL) is a supervised learning technique based on DNNs that learn from big, static, labeled datasets,~\cite{goodfellow2016deep} commonly applied to different problems, including in software engineering~\cite{tong2018software,wang2018deep,zhou2019deep}. 
%These techniques mainly focus on minimizing the loss function and quantifying the error between predictions and reported results.
%However, DL struggles when dealing with dynamic environments due to its static training nature. %for example, in bug localization, DL cannot effectively treat concept drift~\cite{munappy2019data}.
%Deep Reinforcement Learning (DRL) arises as an approach to address the previous limitations by handling sequential decision-making tasks and learning from interactions with an environment~\cite{li2017deep}.
%In these applications, an agent takes actions in an environment, providing feedback to the model based on rewards or penalties.
%Regarding the training, DRL explores the environment, gets experiences, and uses them as input to update the deep network.
%Previous studies have investigated the adoption of DRL in SE tasks, revealing that their approaches achieved better results than state-of-the-art tools~\cite{romdhana2022deep}.
%However, none of them have investigated adopting DRL for bug localization yet.

\subsection{Bug Localization and Non-stationary Data}
Bug localization systems are designed to rank software artifacts based on their likelihood of being associated with a given bug~\cite{wang2023systematic}.
Initially, these systems relied heavily on testing information, analyzing test statuses (passing or failing) and the files repeatedly involved in these tests~\cite{wong2010family}.
Over time, ML techniques have been adopted, improving these systems by incorporating information from several sources, including bug reports, logs, and the changeset-files themselves~\cite{wong2016survey}.
However, even after narrowing down the target changeset-files, identifying the specific code snippets associated with a bug remains a challenging task. This has led researchers to explore finer-grained information, such as hunks, for more precise bug localization~\cite{wen2016locus}.
Recent studies have highlighted the potential of elements from version control systems, such as changesets (i.e., changeset-files and hunks), to improve bug localization~\cite{ye2014learning, wu2018changelocator,ciborowska2022fast}.
Changesets (i.e., changeset-files and hunks) %represent a powerful source with promising findings, as they 
provide code level (lines added, removed, surrounding context) and metadata information, such as the name of the author responsible for some applied changes, the date at which the changes occurred, the commit message, etc. 

However, the inconsistent use of terminology to describe similar information in bug reports often challenges ML techniques, hindering their ability to effectively learn from changesets~\cite{ye2014learning}. Deep learning (DL) techniques alleviate this issue by leveraging word embeddings, contextualized representations, and data augmentation; achieving notable success in tasks such as classification, recommendation, and regression~\cite{lam2017bug,huo2019deep,zhang2019cnn,xiao2019improving}. However, despite their improved performance, DL models struggle to handle the evolutionary nature of changesets~\cite{chen2014big} and the associated concept drift~\cite{munappy2019data}, as they are typically trained on stationary data. To address this limitation, CL techniques have emerged as a potential solution. CL techniques are designed to adapt to evolving data, making them better suited for non-stationary changesets information. % and bug reports, and enabling more effective bug localization in dynamic software environments.
%So, in bug localization, DL may fail to effectively handle concept drift~\cite{munappy2019data} as they are trained on static data related to changesets and bug reports. A possible solution to the evolutionary nature of changesets is CL techniques that can adapt to these changes. 
%\Foutse{explain why....}.
%\Foutse{too much usage of "this way" and it is unclear what it means!}This way, \Foutse{why not simply say that "a possible solution to the evolutionary nature of source code artifacts is CL techniques..etc}DRL techniques should be further explored to accomplish the challenges associated with the evolution of source code (non-stationary data). 

\subsection{Continual Learning in the context of DRL}
RL aims to create autonomous agents that learn optimal policies through trial and error interactions within an environment, modeled as a Markov Decision Process (MDP) \cite{sutton1998reinforcement,li2017deep}. During an episode, an RL agent perceives the environment's state, selects actions, transits to the new state, and obtains reward as feedback which measures the quality of its actions. Policies map states to actions, guiding the agent's decisions, and can be derived from human expertise or learned from experience.

DRL algorithms employ DNNs to approximate the value function or model (state transition function and reward function), consequently providing a more manageable solution space in extensive and complex environments. DRL algorithms can be classified based on the following properties.
\textbf{Value-based, policy-based, and actor-critic learning:} Value-based methods estimate Q-values for each state and select actions with the highest Q-values. Policy-based methods use a parameterized policy, updated through optimization during training. Actor-critic methods combine both approaches, with the actor selecting actions and the critic estimating their Q-values. In this work, our CL agents employ an actor-critic architecture for training. \textbf{On-policy vs Off-policy:} On-policy methods use the same policy for generating trajectories, evaluating, and improving the target policy. Off-policy methods use a different policy (behavior policy) for generating trajectories than the target policy. %I this work, the CL agents employ an off-policy methods (V\_TRACE \cite{espeholt2018impala}) to learn the optimal behavior to find bugs.
In ML, CL represents a paradigm that explores learning based on the model's ability to acquire, update, accumulate, and exploit knowledge continually.
Different from conventional ML models, CL is mainly known for learning from dynamic data, focusing on learning without forgetting previous knowledge (catastrophic forgetting)~\cite{mccloskey1989catastrophic} and transferring knowledge from previous tasks for new ones~\cite{silver2013lifelong}. 
Continual Learning for DRL extends CL by incorporating dynamic environments in the decision-making process through an agent that directly interacts with these dynamic environments~\cite {abel2024definition}.
Continual Learning in Deep Reinforcement Learning (DRL) faces similar challenges as traditional CL. It requires agents to continuously adapt to changes in the environment while retaining the ability to apply and build upon strategies learned from previous tasks. In this work, the CL agent environment's state describes the bug reports and source code files associated with a particular commit SHA, its action corresponds to picking a file from the list of source code files and ranking it; %into a ranked list; 
similar to the agents in the baseline study by Chakraborty et al. \cite{chakraborty2023rlocator}.

\section{Motivating example}
\label{sec:Motivating example}
\begin{figure}
    \centering
    \includegraphics[width=0.5\textwidth]{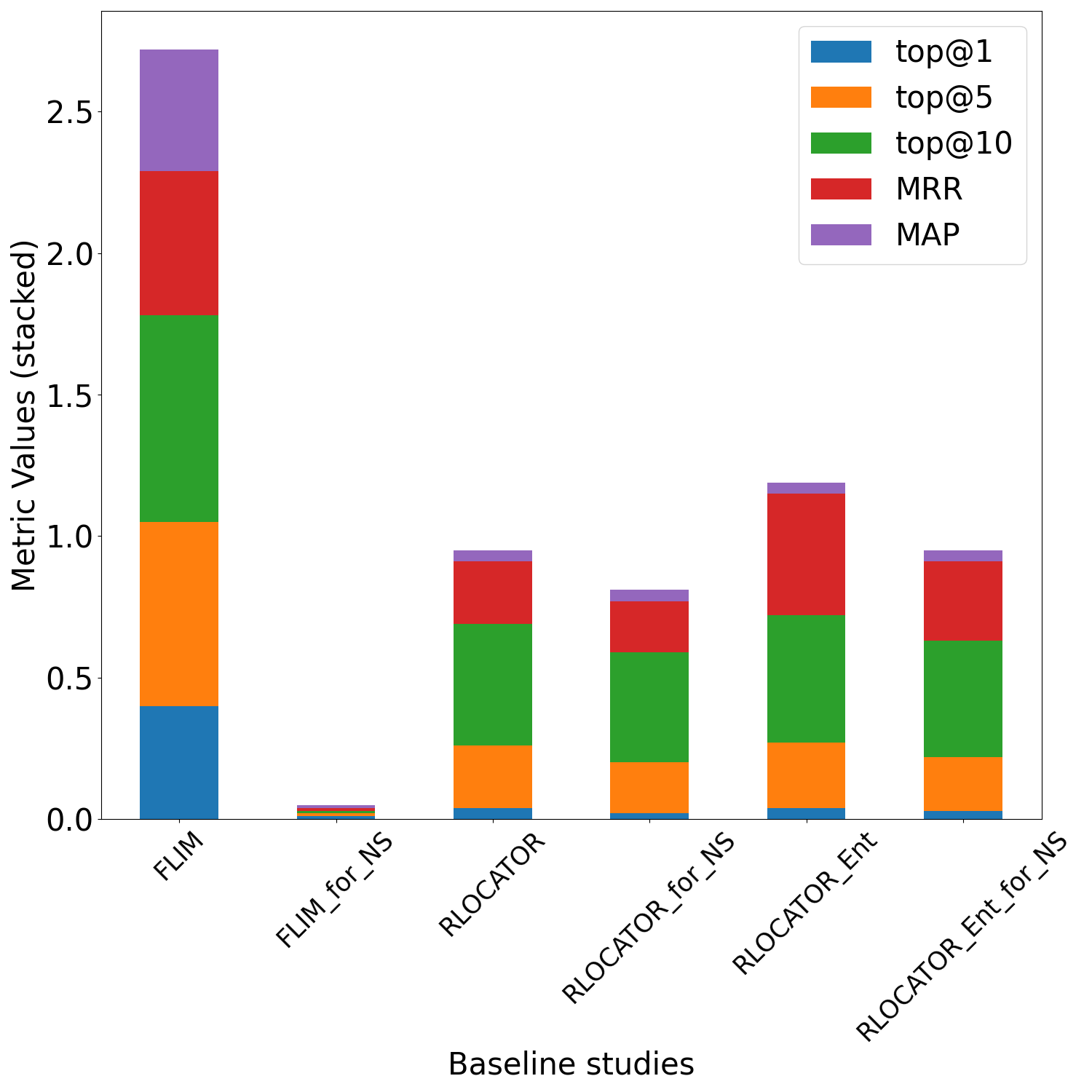}
    \caption{Performance of state-of-art baseline studies on Eclipse project.}
    \label{fig:BASELINES}
    \vspace{-1em}
\end{figure}

To motivate the problem under investigation, let's consider a scenario from the Eclipse Apache project \cite{Eclipse-Apache-project}. This project, due to its complexity, requires developers to collaborate closely, each handling specific tasks. However, bugs are occasionally reported, and developers must then spend valuable time analyzing and fixing these issues. In this context, developer \textbf{A}, responsible for maintaining the project, decides to implement a bug localization technique to streamline the debugging process and improve productivity. The goal is to reduce the manual effort involved in identifying the bug’s location in the code.

To achieve this, developer \textbf{A} adopts a bug localization technique, such as FLIM \cite{liang2022modeling} or RLOCATOR \cite{chakraborty2023rlocator}, both of which rank changeset-files (stationary data) according to their relevance to a given bug report, with and without using entropy.

For instance, when bug \#420210 is reported \cite{Link-for-Bug-420210}, developer \textbf{A} uses the previously implemented bug localization technique to assist in the bug-fixing process. The technique works by analyzing the list of changeset-files associated with the latest commit following the bug report. It then ranks these changeset-files based on their likelihood of containing the bug. Figure \ref{fig:BASELINES} illustrates the performance of FLIM, RLOCATOR, and RLOCATOR with entropy (RLOCATOR\_Ent) in ranking changeset-files on the Apache Eclipse project for bug reports similar to bug \#384108. Based on the ranking, developer \textbf{A} can prioritize the files most likely to be related to the bug, thereby accelerating the bug-fixing process.

However, when another bug is reported (bug \#384108 \cite{Link-for-Bug-384108}), the scenario becomes more complex. Developer \textbf{B}, who is working on a different task, modifies certain changeset-files that developer \textbf{A} had previously identified as important for bug \#384108. These modifications (e.g., code updates, code additions, or removals) introduce concept drift into the data used to rank bug-prone files. As a result, the bug localization technique's performance may degrade.

Due to the concept drift in the data, the performance of DL-based models used by developer \textbf{A} tends to decrease. Figure \ref{fig:BASELINES} shows a noticeable decline in the performance of FLIM, RLOCATOR (with and without entropy), on non-stationary data (i.e., FLIM\_for\_NS, RLOCATOR\_for\_NS, and RLOCATOR\_Ent\_for\_NS). Specifically, there is a drop of 9\% to 194\% in top@1, top@5, and mean reciprocal rank (MRR) metrics.

This highlights the challenge of maintaining the effectiveness of bug localization techniques in the face of dynamic code changes and concept drift, underscoring the need for methods that can adapt to such changes over time.

To address or minimize the concept drift associated with the data, we propose a solution involving CL agents capable of adapting to both stationary and non-stationary data. On non-stationary data, the performance of the CL agents decreased by 4 - 62\%. They also outperformed FLIM and RLOCATOR by 14-167\% in terms of top@1, top@5, and MRR metrics. Our CL agents can adapt more effectively to non-stationary data by reducing catastrophic forgetting while demanding less computational resources.

\section{Study design}\label{studydesign}
In this section, we describe the methodology of our study which aims to leverage CL agents for bug localization in software projects. 
\subsection{Methodology} \label{sec:Method}
\begin{figure*}
    \centering
    \includegraphics[width=1\textwidth]{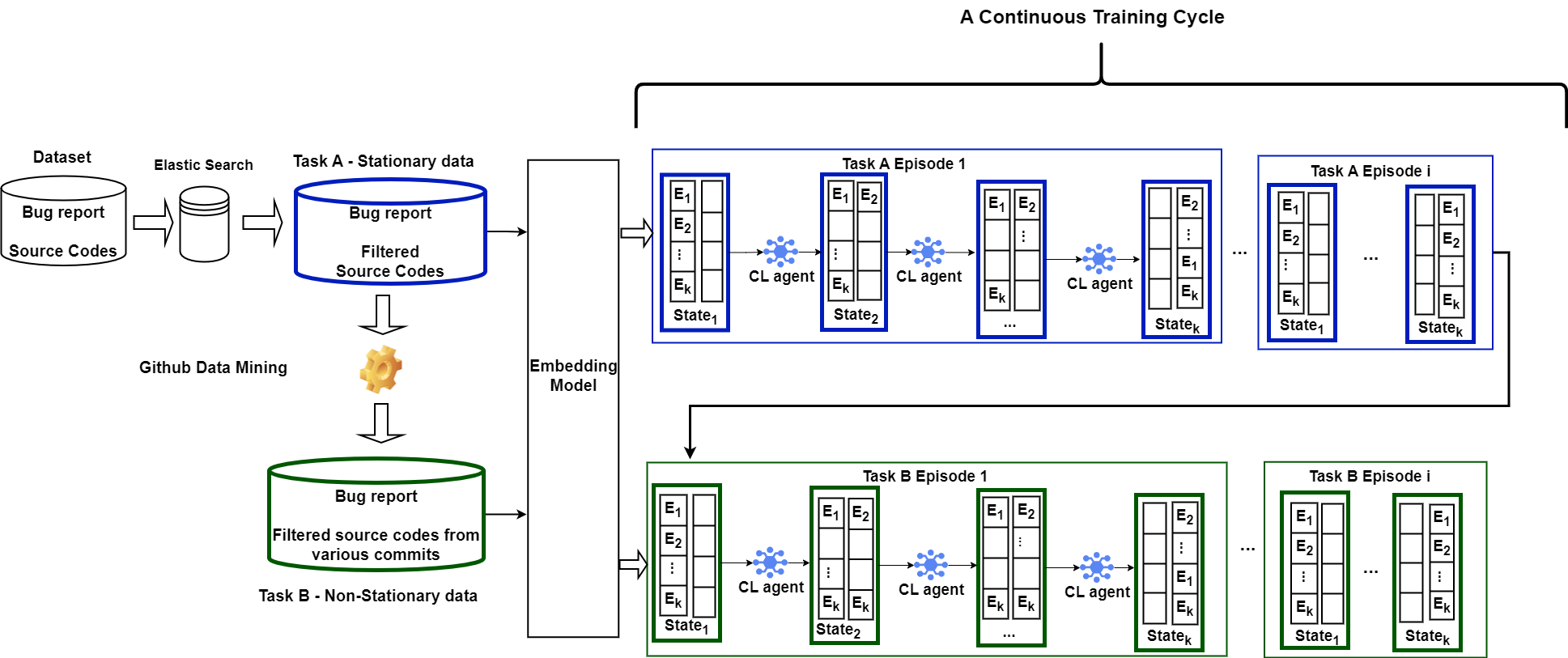}
    \caption{Illustration of the CL framework.}
    \label{fig:pipeline}
\end{figure*}

To leverage CL for bug localization we powered our proposed technique with two CL agents to rank changeset-files and hunks related to bug reports. Our approach draws inspiration from state-of-the-art CL techniques, such as CL with experience and replay \cite{rolnick2019experience} and elastic weight consolidation \cite{kirkpatrick2017overcoming}. We employed these techniques to train DRL agents in a cyclical, sequential manner, using both stationary and non-stationary data. Figure \ref{fig:pipeline} provides an illustration of our proposed CL framework. In the following sections, we describe in detail the components and setup used in our CL framework. %that we used to train DRL agents cyclically in sequence on stationary and non-stationary data. Figure \ref{fig:pipeline} illustrates our proposed CL setup. In the rest of this section, we detail the CL setup and used components.
\subsubsection{Continual learning with experience and replay (CLEAR)} CLEAR is a CL technique used to mitigate catastrophic forgetting by training an actor-critic deep neural network on a mixture of new and replayed experiences. Inspired by IMPALA~\cite{espeholt2018impala} (Importance Weighted Actor-Learner Architecture), CLEAR uses the actor-network to generate trajectories of the environments (e.g., ranking tasks) which are sent to the critic-network to learn an off-policy algorithm V\_TRACE. V\_TRACE mitigates catastrophic forgetting by correcting the distribution shift of the experiences generated by the actor. As stated in Section \ref{Background}, off-policy algorithms like V\_TRACE use the behavior policy $\mu$ to generate trajectories in the form of:

\[ (x_{t},a_{t},r_{t})_{t=s}^{t=s+n} \]
then, it uses those trajectories to learn the value function $V^{\pi}$ of another policy called target policy. V\_TRACE target policy is formulated as:

\begin{equation}
    v_s ~ \myeq ~ V(x_s) + \sum _{t=s}^{s+n-1} \gamma ^{t-s} (\prod _{i=s}^{t-1} c_i ) \delta _t V
\end{equation}
where $\gamma$ is the discount factor and a temporal difference for $V$ is defined as \[\delta _t V ~ \myeq ~ \rho_{t}(r_t + \gamma V(x_{t+1}) - V(x_{t}))\] 
where $\rho_{t}$ and $c_i$ are truncated Importance Sampling (IS) weights defined as:
\[\rho_{t} ~ \myeq ~ min(\bar{\rho}, \frac{\pi(a_{t} \mid x_{t})}{\mu(a_{t} \mid x_{t})}),\] \[c_i ~ \myeq ~ min(\bar{c}, \frac{\pi(a_{i} \mid x_{i})}{\mu(a_{i} \mid x_{i})})\] 

At training time $s$, the value parameters $\theta$ are updated by the gradient descent on $l2$ loss to the target function $v_s$ in the direction of \[(v_s - V_{\theta}(x_{s}))\Delta_{\theta}V_{\theta}(x_{s}),\] and the policy parameters $\omega$ are updated in the direction of the policy gradient: 
\[\rho_s \Delta_{\omega} log\pi_{\omega}(a_{s} \mid x_{s})(r_s + \gamma v_{s+1} - V_{\theta}(x_{s}))\] 

In bug localization setting, the actor generates trajectories that are a mixture of stationary and non-stationary data. With two tasks to be learned (i.e., ranking source code files from stationary and non-stationary data), we leverage CLEAR to train a CL agent that learns cyclically in sequence the optimal ranking strategy on both tasks. In this paper, we adopt the CLEAR implementation with the default parameters.
\subsubsection{Elastic weight consolidation (EWC)} EWC is a CL algorithm that enables plasticity on deep neural networks by constraining important parameters to stay close to their old values. More specifically, EWC works by over-parameterizing the neural networks such that there is a solution to a task B that is close to the solution of a previous task A. When learning task B, the performance of task A is protected by constraining the parameters of the neural network to stay in a region of low error for task A, i.e., centered around the parameters obtained after learning task A. To define which parameters are more important for a task B given the learned parameters $\theta_{A}^*$ of a task B, the EWC loss function is defined as follows:
\begin{equation}
\mathcal{L}(\theta)=\mathcal{L}_B(\theta)+\sum_i \frac{\lambda}{2} F_i\left(\theta_i-\theta_{A, i}^*\right)^2
\end{equation}
where $\mathcal{L}_B(\theta)$ is the loss function for task B, $F_i$ is the Fisher information matrix,  $\lambda$ sets how important the old task is compared to the new one, and $i$ labels each parameter. 

In the context of DRL, the DQN algorithm is augmented with EWC to achieve CL across the Atari 2600 task set \cite{bellemare2013arcade}. 

To the best of our knowledge, EWC implementation is not publicly available, therefore we utilized it in this paper from a popular implementation made by Powers et al.~\cite{powers2022cora}. In their implementation, they use the IMPALA architecture to achieve comparable results. In this paper, we adopt the EWC implementation with the default parameters. Similar to CLEAR, the CL agent is exposed to experiences from stationary and non-stationary data. To apply the EWC algorithm, the Fisher information matrix is computed before switching to each bug localization task.

\subsubsection{Bug-inducing factors for bug localization}\label{sec:Bug-inducing factors for bug localization}
Bug-inducing factors associated with changesets have shown to be predictors of bugs across software projects \cite{taba2013predicting, zimmermann2007predicting,wen2016locus}. Taba et al. \cite{taba2013predicting} used antipattern information as metrics to improve traditional bug prediction models at the changeset-files level. Table \ref{tab:measured-metrics} presents the bug-inducing factors metrics used by Taba et al.~\cite{taba2013predicting}.   They combine metrics representing bug-inducing factors to construct a logistic regression model for predicting the presence of bugs. In this model, the independent variables correspond to the bug-inducing factor metrics, while the dependent variable is a binary variable indicating whether a file contains one or more bugs. Following a similar approach, we collected these metrics at both the changeset-file and hunk levels. 
%\lm{Could you compute the VG for hunks, considering they have just code snippets?} \Paulina{TBD}
A detailed description of the bug-inducing factors is provided below: 

\begin{table}[]
\caption{Measured Source Code and Bug Report Metrics}
\label{tab:measured-metrics}
\begin{tabular}{crl}
\hline \hline
& \multicolumn{1}{c}{Metrics} & Description \\ \hline \hline
\multicolumn{1}{c|}{\multirow{3}{*}{Source Code}} & LOC & Source Lines of Code \\
\multicolumn{1}{c|}{} & MLOC & Executable Lines of Code \\
\multicolumn{1}{c|}{} & VG & Cyclomatic complexity \\ \hline
\multicolumn{1}{c|}{\multirow{2}{*}{Bug Report}} & PRE & Number of pre-released bugs \\
\multicolumn{1}{c|}{} & Churn & \begin{tabular}[c]{@{}l@{}}Number of lines of code modified \\ (added, removed)\end{tabular} \\ \hline
\end{tabular}
\vspace{-1em}
\end{table}
\begin{itemize}
    \item \textbf{Lines of Code (LOC):} For a given buggy version, we do a checkout at the buggy \emph{commit} and compute the associated lines of code.
   \item \textbf{Executable LOC (MLOC):} This metric is generated based on the previous one (LOC), but this time considering only executable lines, excluding lines with comments and blank ones.
   \item \textbf{Cyclomatic Complexity (VG):} This metric computes the number of linearly independent paths of the given  source code~\cite{mccabe1996cyclomatic}. For computing this and the previous metrics, we consider the CCC tool \cite{scc-tool}. 
   \item \textbf{Pre-released Bugs (PRE):} Based on the set of files updated when fixing a given bug, we compute the number of times these files were changed in other bug fixes. For that, we get the changed files in each commit associated with a bug fix; then, we search for occurrences considering the files associated with the bug report under analysis.
    \item \textbf{Code Churn (Churn):} We compute the number of lines of code added, modified, or deleted. We get the \emph{diff} associated with the bug commit under analysis and get the required information. %\lm{In case we need space, we can replace the previous bullet points with a table.}
\end{itemize}
Since the metrics collected can be dependent and highly collinear, we follow the iterative process described below to retain only important and non-collinear metrics. %remove the ones that fit those characteristics \Foutse{which characteristics? be specific please!!! did you use VIF to remove multi-colinear features? explain/be specific!!!}. 
The following elaborates more on the process.
\begin{enumerate}
    \item Removing statistically insignificant variables: In this step, we first build the logistic regression model with all metrics mentioned above as variables of the model; then, in an iterative process, we remove the statistically insignificant metrics. We use the threshold $p$-value $< 0.05$ to determine whether a variable is statistically significant or not.
    \item Collinearity analysis: In a regression model, two or more variables are multicollinear when they are highly correlated. We removed the highly collinear variables and only kept %. With this collinear analysis, we only keep the 
    independent variables that produce an effect on the dependent variable \cite{taba2013predicting}. We used the Variance Inflation Factor (VIF) to measure the level of multicollinearity of the regression model. We set the maximum VIF value to be 2.5, as suggested in \cite{taba2013predicting}. Independent variables with VIF values that exceed 2.5 are considered highly collinear and, hence, are removed.
\end{enumerate}
After removing statistically insignificant and highly collinear variables, we narrow down our list of independent variables to the Churn and the PRE metrics. In this paper, we integrated the output of the logistic regression model as an additional component of the reward function of the CL agents. To validate this approach, we conduct ablation studies in Section \ref{sec:Ablation study on using bug-inducing factors}. These studies compare the retrieval performance of the CL agents under three conditions: (1) when each of the bug-inducing factors is individually added to the reward function of the CL agents, (2) when the output of a logistic regression model incorporating all bug-inducing factors as independent variables is added to the reward function of the CL agents and (3) when the output of a logistic regression model, using only statistically significant and non-collinear bug-inducing factors, is added to the reward function of the CL agents.

\subsubsection{Formulation of the bug localization task as a DRL problem}\label{sec:Creation of the DRL algorithms}
In this paper, %we consider two formulations given the type of data being learned by the CL agents
we explore two formulations tailored to the type of data processed by CL agents: The ranking of changesets (i.e., changeset-files or hunks) on stationary and non-stationary data. The bug localization process is considered a DRL problem where the agent picks buggy changesets during an episode and ranks them based on their relevance to a bug report \cite{chakraborty2023rlocator}. The CL agents' inputs are bug reports and changesets associated with a particular commit SHA. However, the nature of DRL does not allow a variable number of changesets to be passed to the CL model. Therefore, similar to the baseline \cite{chakraborty2023rlocator}, we filtered the number of changesets.

We utilize ElasticSearch (ES), an open-source search and analytics engine that analyzes and indexes data using the BM25 algorithm \cite{gormley2015elasticsearch}. We created an ES index from the changeset-files and queried it with the bug report description to retrieve the top $k$ files most textually similar to the description of the bug report. ES ranks files by their textual similarity to a query.

After choosing $k$ most relevant changesets from the output of ES, for each bug report, the ranking of changesets and their associated hunks on stationary and non-stationary data can be mapped into a DRL process by defining the states (i.e., observations), actions, reward function, and end of an episode. The goal is to navigate the DRL process until all files are ranked.\\
\textbf{Observation space: }An observation describes a bug report and the top $k$ relevant changesets retrieved from ES. We used CodeBERT \cite{feng2020codebert}, a transformer-based BERT model to convert changesets and bug report texts into embeddings. After obtaining the embeddings for the source code $(F_1, F_2,.., F_k)$ and the bug report $(R)$, we concatenate them to form part of the observation: $(E_1,E_2,..,E_k) = (F_1\|R,F_2 \| R,..,F_k\| R)$. The final observation comprises the list of embeddings $(E_1, E_2,.., E_k)$ and a ranked list of changesets based on their relevance to the bug report $(R)$. Initially, the ranked list is empty, at each transition during an episode the agent moves one embedding from the list of embeddings $(E_1, E_2,.., E_k)$ into the ranked list based on its relevance to the bug report $(R)$. \\
 \textbf{Action space: }An action corresponds to picking a changeset from the list of embeddings $(E_1, E_2,.., E_k)$ and moving it to the ranked list. Specifically, selecting actions based on the observation space involves using CodeBERT to generate embeddings for a bug report and the top $k$ relevant changesets. These embeddings constitute the observation of the CL agent. At each step, the CL agent observes the embeddings of the bug report and the remaining changesets (i.e., those not yet moved to the ranked list). As changesets are moved from the candidate list to the ranked list, the observation changes, altering the set of embeddings the CL agent observes. This allows the CL agent to sequentially select the most relevant changeset at each step until all changesets are ranked. For a fair comparison with the baseline study, the action space corresponds to a set of $k=31$ changesets. Any values of $k$ can be chosen depending on the capacity of the hardware used for training \cite{chakraborty2023rlocator}. To avoid out-of-memory errors during training  $k$ is set to 31, the same value used by the baseline study of Chakraborty et al. \cite{chakraborty2023rlocator}\\
\textbf{Reward function: } The agent is rewarded based on the rank of the relevant changesets and the distance between them in the ranked list \cite{chakraborty2023rlocator}. The reward function is formulated as below:
\begin{equation}
   \mathcal{R}(o,a)= \frac{3 \times changeset~relevance}{\log_2(t+1) \times distance(o)} 
\end{equation}
if $a$ is an action that has not been selected before, otherwise:
\begin{equation}
   \mathcal{R}(o,a)= -\log_2(t+1)
\end{equation}
$t$ is a step during an episode, $o$ is an observation and $a$
is an action. $M$ is a value that scales the reward given to the generalist agent when it picks a relevant changeset. Through experimentation~\cite{chakraborty2023rlocator}, different values ($1, 3, 6,$ and $9$) were tested to see how they impacted the reward and, consequently, the performance of the model. The value $3$ resulted in the highest reward for the DRL model. The distance \(\text{distance}(o)\) is calculated as the average of \(\delta(i)\), where \(\delta(i)\) represents the distance between consecutive relevant changesets of observation \(o\). Given the positions of relevant changesets \(\{c_1, c_2, \ldots, c_n\}\), \(\delta(i)\) is defined as
\[
\delta(i) = c_{i+1} - c_i, \quad \text{for } i = 1, 2, \ldots, n-1.
\]
Finally, \(\text{distance}(o)\) is calculated as
\[
\text{distance}(o) = \frac{1}{n-1} \sum_{i=1}^{n-1} \delta(i).
\]

When considering bug-inducing factors as prior knowledge for the CL agents, we use a logistic model to learn the probability that a changeset is buggy (see Section \ref{sec:Bug-inducing factors for bug localization}). The reward function for this case is as follows:
\begin{equation}
   \mathcal{R}(o,a)= \mathcal{R}(o,a) + bug~ probability~ indicator
\end{equation}
where $bug~ probability~ indicator$ is the output of the logistic regression model. The bug probability indicator is a scalar. At each step during an episode, we add the bug probability indicator to the reward earned by the CL agent.
\subsection{Datasets} \label{sec:Dataset}

\begin{table}[t]
\caption{Benchmark statistics}
%\lm{Before, you had the information of changed files for JDT. Why did you remove them? I know that you're not evaluating it because you're not able to have non-stationary data.} \Paulina{It was a mistake, I copy paste one of your table in comments and forgot to remove JDT the first time around. Then fix it. If you remember it was the same number as bug reported}} FIXED.
    \centering
\begin{tabular}{|c|c|c|}
\hline 
Project & \#Bugs reported & \#Changeset-files and hunks \\ \hline \hline
AspectJ & 593             & 27,127                      \\ \hline
Birt    & 4,178           & 1,939                       \\ \hline
Eclipse & 6,495           & 6,645                       \\ \hline
SWT     & 4,151           & 598                         \\ \hline
Tomcat  & 1,056           & 41,271                      \\ \hline
PDE     & 60              & 123,053                      \\ \hline
Zxing   & 20              & 19,084                       \\ \hline
\end{tabular}
        \label{tab:benchmark-dataset}
        
\end{table}
We evaluate our approach using bugs reported in widely used software projects, leveraging their associated changeset-files and hunks~\cite{ye2014learning,wen2016locus}.
These software projects are commonly used in previous and related studies, valued for their quality and the diversity of bug scenarios they present~\cite{chakraborty2023rlocator,wen2016locus,ciborowska2022fast,liang2022modeling}.
From the eight open-source Java repositories available from Apache (AspectJ, Birt, JDT, PDE, Eclipse, Zxing, Tomcat, and Birt), we selected seven, excluding JDT due to the absence of non-stationary data. Specifically, when analyzing changes between bug report dates and commit fix dates in the JDT project, we found no modifications in the files associated with the reported bug fixes, resulting in no intersection with the ground truth. 
Among the different information provided for a given bug, we highlight the bug commit and its associated changeset-files — key elements needed for computing our metrics on stationary data (see Section~\ref{sec:Bug-inducing factors for bug localization}).
From the commits associated with bugs, we checkout on the commits and access the full version of the buggy code.
%In the same way, with the reported files, we check the frequency of these files that were also referred to in other bug fixes. 

%For our non-stationary data, based on the changeset-files associated to each bug, \Foutse{what do you mean by 'verifying number of commits'? for what purpose exactly?}we first verify if there is any  commits changing these files. 
For the non-stationary data, we analyze all commits that modify the changeset-files associated with a bug report. The objective is to identify all changeset-files potentially impacted by a bug. To achieve this, we collect commits made between the dates of the bug report and the corresponding bug-fixing commit. Once we have the list of commits, we extract the different versions of the target changeset-files and use them as non-stationary data for our model. To collect the hunks associated with each changeset-files we collect the output of the \emph{git diff} command in which added lines of code are annotated with +, and removed lines with -. Table \ref{tab:benchmark-dataset} shows for each project (i.e., AspectJ, Birt,  PDE, Eclipse, Zxing, Birt, and Tomcat), the total number of changed buggy changeset-files and hunks relevant to the bug reports. To create our training and testing datasets, we sort the bug reports based on their report date, following an ascending order. Then, we divide the dataset into a 60:40 percent split (training and test) similar to our baselines~\cite{chakraborty2023rlocator}.

%\lm{DEADLINE: Friday, end of the day. https://arxiv.org/pdf/2305.05586 - Section 4.1 Table 2, except JDT project.}

\subsection{Baselines studies} 
\label{sec:Baselines studies}
As baselines for our study, we employ three methods detailed below, each representing state-of-the-art approaches utilizing DL techniques for bug localization.

\begin{itemize}
    \item Rlocator~\cite{chakraborty2023rlocator}: a DRL model based on an MDP to directly optimize the ranking of considered metrics. For evaluation, we consider the model provided with and without Entropy for A2C. A2C with entropy incorporates the entropy of the probability of the possible action to the loss of the
actor model. Moreover, we evaluated the CL agents against RLOCATOR on AspectJ, Birt, Eclipse, SWT, and Tomcat projects. 
%with a dataset division of a 60:40 percent split for training and testing (see Section \ref{sec:Dataset}). In this work, we adopted the same ratio to train and test the CL agents.
    \item FLIM~\cite{liang2022modeling}: a framework that extracts semantic features from code at the function level and calculates the relevance between natural and programming language. Then, it uses a DL model to fuse the function-level semantic features with IR features to calculate the final relevance. The DL model employed by FLIM uses 19 independent variables. These variables capture the similarity between a source code file and a bug report, the API specifications of the source code file, as well as the recency and frequency of bug fixes associated with the source code files. In this work, we evaluated the CL agents against FLIM on AspectJ, Birt, Eclipse, SWT, and Tomcat projects.
    %For evaluation purposes, we use the original implementation \cite{FLIM-implementation}. 
    \item FBL-BERT~\cite{ciborowska2022fast}: A supervised learning approach that uses BERT model for bug localization at changeset-files, hunks, and commits (i.e., set of hunks) levels. The paper explores three strategies (i.e., QD, QARC, and QARCL) for encoding code modifications and data granularities to optimize BERT's performance: 1) QD considers a changeset as a single document, with a special token pre-appended to signal the start of the code sequence for the model 2) QARC encodes a changeset by dividing it into lines grouped by their type: added (+), removed (-), or context (empty space). Each group is prefixed with a special token, and the sequences are then concatenated to form the model's input 3) In QARCL, similar to the QARC strategy, a changeset is divided into lines, but the original line order is preserved, with special tokens pre-appended at each change in the type of modification. We compare the performance of the CL agents on the AspectJ, Zxing, PDE, SWT, and Tomcat projects against all three strategies at the changeset-files and hunks levels. We do not evaluate the CL agents against FBL-BERT at the commits level as the CL agent action consists of picking one changeset-file or hunk at a time during an episode, as presented in Section \ref{sec:Creation of the DRL algorithms}.  
\end{itemize}

\subsection{Evaluation metrics} 
\label{sec:Evaluation metrics}
To evaluate our proposed model and compare it with the baseline studies, we rely on the ground truth provided by the benchmark considered for this study (see Section ~\ref{sec:Dataset}). 
For each bug reported, the dataset provides the changed files to fix the bug. This way, we can evaluate the performance of our model based on five criteria, which are widely recognized and used by related studies when conducting bug localization studies \cite{chakraborty2023rlocator,liang2022modeling,ye2015mapping,mindom2022comparison,powers2022cora}. \textbf{Mean Reciprocal Rank (MRR)}, \textbf{Mean Average Precision (MAP)}, and \textbf{Top K} are the metrics we used to evaluate the effectiveness of the CL agents when ranking the source code files. Following previous studies, we consider three values for N: 1, 5, and 10 \cite{chakraborty2023rlocator}. Finally, the \textbf{training time} measures the time consumed by the CL agents to get trained. We collected the training time for 7,500 episodes per task, same as the baseline agent.
\begin{table*}[t]
\label{tab:Algorithm evaluation on Tomcat (left) and Birt (right) projects}
\begin{minipage}{0.8\textwidth}
\resizebox{\columnwidth}{!}{
\centering
\begin{tabular}{|cc|cc|cc|cc|cc|cc|}
\hline
\multicolumn{2}{|c|}{} & \multicolumn{2}{c|}{top@1} & \multicolumn{2}{c|}{top@5} & \multicolumn{2}{c|}{top@10} & \multicolumn{2}{c|}{MRR} & \multicolumn{2}{c|}{MAP} \\ \cline{3-12} 
\multicolumn{2}{|c|}{\multirow{-2}{*}{}} & \multicolumn{1}{c|}{ST} & NS & \multicolumn{1}{c|}{ST} & NS & \multicolumn{1}{c|}{ST} & NS & \multicolumn{1}{c|}{ST} & NS & \multicolumn{1}{c|}{ST} & NS \\ \hline
\multicolumn{1}{|c|}{FLIM} & -files & \multicolumn{1}{c|}{\cellcolor[HTML]{656565}0.47} & 0.00 & \multicolumn{1}{c|}{\cellcolor[HTML]{656565}0.70} & 0.00 & \multicolumn{1}{c|}{\cellcolor[HTML]{656565}0.77} & 0.00 & \multicolumn{1}{c|}{\cellcolor[HTML]{656565}0.57} & 0.01 & \multicolumn{1}{c|}{\cellcolor[HTML]{656565}0.51} & 0.01 \\ \hline
\multicolumn{1}{|c|}{RLO.} & -files & \multicolumn{1}{c|}{0.04} & 0.04 & \multicolumn{1}{c|}{\cellcolor[HTML]{C0C0C0}0.24} & 0.21 & \multicolumn{1}{c|}{\cellcolor[HTML]{C0C0C0}0.48} & 0.42 & \multicolumn{1}{c|}{0.24} & 0.18 & \multicolumn{1}{c|}{0.04} & 0.04 \\ \hline
\multicolumn{1}{|c|}{RLO. + Reg.} & -files & \multicolumn{1}{c|}{0.04} & 0.04 & \multicolumn{1}{c|}{0.23} & 0.22 & \multicolumn{1}{c|}{0.46} & 0.43 & \multicolumn{1}{c|}{0.25} & 0.19 & \multicolumn{1}{c|}{0.04} & 0.04 \\ \hline
\multicolumn{1}{|c|}{RLO. + Ent.} & -files & \multicolumn{1}{c|}{0.04} & 0.04 & \multicolumn{1}{c|}{0.21} & 0.22 & \multicolumn{1}{c|}{0.45} & \cellcolor[HTML]{C0C0C0}0.45 & \multicolumn{1}{c|}{0.34} & 0.28 & \multicolumn{1}{c|}{0.04} & 0.04 \\ \hline
\multicolumn{1}{|c|}{RLO. + Ent. + Reg.} & -files & \multicolumn{1}{c|}{0.04} & 0.04 & \multicolumn{1}{c|}{0.21} & 0.21 & \multicolumn{1}{c|}{0.41} & 0.44 & \multicolumn{1}{c|}{0.27} & 0.32 & \multicolumn{1}{c|}{0.04} & 0.04 \\ \hline
\multicolumn{1}{|c|}{} & -files & \multicolumn{1}{c|}{\cellcolor[HTML]{FFFFFF}0.05} & \cellcolor[HTML]{FFFFFF}0.05 & \multicolumn{1}{c|}{\cellcolor[HTML]{C0C0C0}0.24} & 0.22 & \multicolumn{1}{c|}{0.44} & \cellcolor[HTML]{656565}0.48 & \multicolumn{1}{c|}{\cellcolor[HTML]{C0C0C0}0.45} & 0.25 & \multicolumn{1}{c|}{\cellcolor[HTML]{C0C0C0}0.05} & \cellcolor[HTML]{C0C0C0}0.05 \\ \cline{2-12} 
\multicolumn{1}{|c|}{\multirow{-2}{*}{CLEAR}} & hunks & \multicolumn{1}{c|}{0.03} & 0.03 & \multicolumn{1}{c|}{0.21} & 0.20 & \multicolumn{1}{c|}{0.42} & 0.40 & \multicolumn{1}{c|}{0.39} & 0.29 & \multicolumn{1}{c|}{0.04} & 0.04 \\ \hline
\multicolumn{1}{|c|}{} & -files & \multicolumn{1}{c|}{0.04} & \cellcolor[HTML]{C0C0C0}0.06 & \multicolumn{1}{c|}{0.21} & \cellcolor[HTML]{656565}0.29 & \multicolumn{1}{c|}{0.32} & 0.43 & \multicolumn{1}{c|}{0.31} & \cellcolor[HTML]{C0C0C0}0.27 & \multicolumn{1}{c|}{0.04} & \cellcolor[HTML]{656565}0.06 \\ \cline{2-12} 
\multicolumn{1}{|c|}{\multirow{-2}{*}{CLEAR + Reg.}} & hunks & \multicolumn{1}{c|}{\cellcolor[HTML]{FFFFFF}0.05} & \cellcolor[HTML]{656565}0.07 & \multicolumn{1}{c|}{0.22} & 0.28 & \multicolumn{1}{c|}{0.35} & 0.42 & \multicolumn{1}{c|}{0.26} & 0.17 & \multicolumn{1}{c|}{0.04} & 0.06 \\ \hline
\multicolumn{1}{|c|}{} & -files & \multicolumn{1}{c|}{0.03} & 0.03 & \multicolumn{1}{c|}{0.21} & 0.20 & \multicolumn{1}{c|}{0.43} & 0.42 & \multicolumn{1}{c|}{0.30} & \cellcolor[HTML]{656565}0.34 & \multicolumn{1}{c|}{0.04} & 0.04 \\ \cline{2-12} 
\multicolumn{1}{|c|}{\multirow{-2}{*}{EWC}} & hunks & \multicolumn{1}{c|}{\cellcolor[HTML]{C0C0C0}0.06} & 0.04 & \multicolumn{1}{c|}{0.23} & 0.22 & \multicolumn{1}{c|}{\cellcolor[HTML]{FFFFFF}0.46} & 0.44 & \multicolumn{1}{c|}{0.28} & 0.22 & \multicolumn{1}{c|}{\cellcolor[HTML]{C0C0C0}0.05} & 0.04 \\ \hline
\multicolumn{1}{|c|}{} & -files & \multicolumn{1}{c|}{0.04} & 0.04 & \multicolumn{1}{c|}{0.20} & \cellcolor[HTML]{C0C0C0}0.24 & \multicolumn{1}{c|}{0.42} & 0.42 & \multicolumn{1}{c|}{0.18} & 0.26 & \multicolumn{1}{c|}{0.04} & \cellcolor[HTML]{C0C0C0}0.05 \\ \cline{2-12} 
\multicolumn{1}{|c|}{\multirow{-2}{*}{EWC+Reg.}} & hunks & \multicolumn{1}{c|}{0.05} & 0.05 & \multicolumn{1}{c|}{0.23} & \cellcolor[HTML]{C0C0C0}0.24 & \multicolumn{1}{c|}{\cellcolor[HTML]{C0C0C0}0.48} & \cellcolor[HTML]{656565}0.48 & \multicolumn{1}{c|}{0.20} & 0.32 & \multicolumn{1}{c|}{\cellcolor[HTML]{C0C0C0}0.05} & \cellcolor[HTML]{C0C0C0}0.05 \\ \hline
\end{tabular}
}
\caption{Tomcat}
\label{tab:Tomcat}
\end{minipage}%
    \hfill
    \begin{minipage}{0.8\textwidth}
        \resizebox{\columnwidth}{!}{
\centering
\begin{tabular}{|cc|cc|cc|cc|cc|cc|}
\hline
\multicolumn{2}{|c|}{} & \multicolumn{2}{c|}{top@1} & \multicolumn{2}{c|}{top@5} & \multicolumn{2}{c|}{top@10} & \multicolumn{2}{c|}{MRR} & \multicolumn{2}{c|}{MAP} \\ \cline{3-12} 
\multicolumn{2}{|c|}{\multirow{-2}{*}{}} & \multicolumn{1}{c|}{ST} & NS & \multicolumn{1}{c|}{ST} & NS & \multicolumn{1}{c|}{ST} & NS & \multicolumn{1}{c|}{ST} & NS & \multicolumn{1}{c|}{ST} & NS \\ \hline
\multicolumn{1}{|c|}{FLIM} & -files & \multicolumn{1}{c|}{\cellcolor[HTML]{656565}0.13} & \cellcolor[HTML]{656565}0.14 & \multicolumn{1}{c|}{\cellcolor[HTML]{656565}0.30} & 0.14 & \multicolumn{1}{c|}{0.38} & \cellcolor[HTML]{656565}0.57 & \multicolumn{1}{c|}{0.21} & 0.21 & \multicolumn{1}{c|}{\cellcolor[HTML]{656565}0.16} & \cellcolor[HTML]{656565}0.24 \\ \hline
\multicolumn{1}{|c|}{RLO.} & -files & \multicolumn{1}{c|}{0.05} & 0.02 & \multicolumn{1}{c|}{0.24} & 0.16 & \multicolumn{1}{c|}{0.51} & 0.33 & \multicolumn{1}{c|}{0.27} & 0.16 & \multicolumn{1}{c|}{\cellcolor[HTML]{C0C0C0}0.05} & 0.03 \\ \hline
\multicolumn{1}{|c|}{RLO. + Reg.} & -files & \multicolumn{1}{c|}{0.05} & 0.03 & \multicolumn{1}{c|}{0.25} & 0.19 & \multicolumn{1}{c|}{0.51} & 0.33 & \multicolumn{1}{c|}{\cellcolor[HTML]{FFFFFF}0.44} & 0.16 & \multicolumn{1}{c|}{\cellcolor[HTML]{C0C0C0}0.05} & 0.03 \\ \hline
\multicolumn{1}{|c|}{RLO. + Ent.} & -files & \multicolumn{1}{c|}{0.05} & 0.02 & \multicolumn{1}{c|}{0.25} & 0.14 & \multicolumn{1}{c|}{\cellcolor[HTML]{C0C0C0}0.52} & 0.28 & \multicolumn{1}{c|}{0.34} & 0.30 & \multicolumn{1}{c|}{\cellcolor[HTML]{C0C0C0}0.05} & 0.03 \\ \hline
\multicolumn{1}{|c|}{RLO. + Ent. + Reg.} & -files & \multicolumn{1}{c|}{0.05} & 0.03 & \multicolumn{1}{c|}{0.25} & 0.18 & \multicolumn{1}{c|}{\cellcolor[HTML]{C0C0C0}0.52} & 0.37 & \multicolumn{1}{c|}{0.38} & 0.42 & \multicolumn{1}{c|}{\cellcolor[HTML]{C0C0C0}0.05} & 0.03 \\ \hline
\multicolumn{1}{|c|}{} & -files & \multicolumn{1}{c|}{0.05} & 0.05 & \multicolumn{1}{c|}{0.26} & \cellcolor[HTML]{656565}0.25 & \multicolumn{1}{c|}{\cellcolor[HTML]{C0C0C0}0.52} & \cellcolor[HTML]{C0C0C0}0.49 & \multicolumn{1}{c|}{0.31} & \cellcolor[HTML]{656565}0.46 & \multicolumn{1}{c|}{\cellcolor[HTML]{C0C0C0}0.05} & \cellcolor[HTML]{C0C0C0}0.05 \\ \cline{2-12} 
\multicolumn{1}{|c|}{\multirow{-2}{*}{CLEAR}} & hunks & \multicolumn{1}{c|}{0.05} & \cellcolor[HTML]{C0C0C0}0.07 & \multicolumn{1}{c|}{0.23} & 0.25 & \multicolumn{1}{c|}{0.50} & 0.47 & \multicolumn{1}{c|}{0.29} & 0.24 & \multicolumn{1}{c|}{\cellcolor[HTML]{C0C0C0}0.05} & \cellcolor[HTML]{C0C0C0}0.05 \\ \hline
\multicolumn{1}{|c|}{} & -files & \multicolumn{1}{c|}{0.05} & \cellcolor[HTML]{FFFFFF}0.06 & \multicolumn{1}{c|}{0.25} & \cellcolor[HTML]{656565}0.25 & \multicolumn{1}{c|}{0.49} & 0.46 & \multicolumn{1}{c|}{\cellcolor[HTML]{C0C0C0}0.36} & \cellcolor[HTML]{FFFFFF}0.33 & \multicolumn{1}{c|}{\cellcolor[HTML]{C0C0C0}0.05} & \cellcolor[HTML]{C0C0C0}0.05 \\ \cline{2-12} 
\multicolumn{1}{|c|}{\multirow{-2}{*}{CLEAR + Reg.}} & hunks & \multicolumn{1}{c|}{0.05} & 0.02 & \multicolumn{1}{c|}{\cellcolor[HTML]{C0C0C0}0.28} & 0.23 & \multicolumn{1}{c|}{0.54} & 0.42 & \multicolumn{1}{c|}{0.44} & \cellcolor[HTML]{C0C0C0}0.39 & \multicolumn{1}{c|}{\cellcolor[HTML]{C0C0C0}0.05} & 0.04 \\ \hline
\multicolumn{1}{|c|}{} & -files & \multicolumn{1}{c|}{0.05} & 0.03 & \multicolumn{1}{c|}{\cellcolor[HTML]{FFFFFF}0.27} & 0.17 & \multicolumn{1}{c|}{\cellcolor[HTML]{656565}0.53} & 0.39 & \multicolumn{1}{c|}{\cellcolor[HTML]{C0C0C0}0.45} & \cellcolor[HTML]{FFFFFF}0.33 & \multicolumn{1}{c|}{\cellcolor[HTML]{C0C0C0}0.05} & 0.04 \\ \cline{2-12} 
\multicolumn{1}{|c|}{\multirow{-2}{*}{EWC}} & hunks & \multicolumn{1}{c|}{0.05} & 0.02 & \multicolumn{1}{c|}{0.25} & 0.19 & \multicolumn{1}{c|}{0.51} & 0.41 & \multicolumn{1}{c|}{0.36} & 0.17 & \multicolumn{1}{c|}{\cellcolor[HTML]{C0C0C0}0.05} & 0.04 \\ \hline
\multicolumn{1}{|c|}{} & -files & \multicolumn{1}{c|}{\cellcolor[HTML]{C0C0C0}0.06} & \cellcolor[HTML]{FFFFFF}0.06 & \multicolumn{1}{c|}{\cellcolor[HTML]{FFFFFF}0.27} & \cellcolor[HTML]{C0C0C0}0.20 & \multicolumn{1}{c|}{0.52} & 0.41 & \multicolumn{1}{c|}{\cellcolor[HTML]{C0C0C0}0.45} & 0.26 & \multicolumn{1}{c|}{\cellcolor[HTML]{C0C0C0}0.05} & 0.04 \\ \cline{2-12} 
\multicolumn{1}{|c|}{\multirow{-2}{*}{EWC+Reg.}} & hunks & \multicolumn{1}{c|}{0.05} & 0.03 & \multicolumn{1}{c|}{\cellcolor[HTML]{C0C0C0}0.28} & 0.17 & \multicolumn{1}{c|}{0.51} & 0.40 & \multicolumn{1}{c|}{\cellcolor[HTML]{C0C0C0}0.50} & 0.17 & \multicolumn{1}{c|}{\cellcolor[HTML]{C0C0C0}0.05} & 0.04 \\ \hline
\end{tabular}
}
\caption{Birt}
\label{tab:Birt}
    \end{minipage}
\begin{tablenotes}
     \item Algorithms evaluation across software projects. "RLO." refers to the baseline RLOCATOR. "+ Reg" indicates the inclusion of logistic regression, while "+ Ent" denotes the addition of entropy. "-files"=changeset-files. The best and second best average performances are highlighted in dark grey and light grey respectively. For each metric, the ST column contains the performance of algorithms on stationary data, and the NS column contains the performance on non-stationary data. %\lm{Can you combine the tables into subtables? Just to avoid replicating the same caption.}
 \end{tablenotes}
\end{table*}
\begin{table*}[t]
\label{tab:Algorithm evaluation on Tomcat (left) and Birt (right) projects}
\begin{minipage}{0.8\textwidth}
\resizebox{\columnwidth}{!}{
\centering
\begin{tabular}{|cc|cc|cc|cc|cc|cc|}
\hline
\multicolumn{2}{|c|}{} & \multicolumn{2}{c|}{top@1} & \multicolumn{2}{c|}{top@5} & \multicolumn{2}{c|}{top@10} & \multicolumn{2}{c|}{MRR} & \multicolumn{2}{c|}{MAP} \\ \cline{3-12} 
\multicolumn{2}{|c|}{\multirow{-2}{*}{}} & \multicolumn{1}{c|}{ST} & NS & \multicolumn{1}{c|}{ST} & NS & \multicolumn{1}{c|}{ST} & NS & \multicolumn{1}{c|}{ST} & NS & \multicolumn{1}{c|}{ST} & NS \\ \hline
\multicolumn{1}{|c|}{FLIM} & -files & \multicolumn{1}{c|}{\cellcolor[HTML]{656565}0.40} & 0.00 & \multicolumn{1}{c|}{\cellcolor[HTML]{656565}0.65} & 0.00 & \multicolumn{1}{c|}{\cellcolor[HTML]{656565}0.73} & 0.00 & \multicolumn{1}{c|}{\cellcolor[HTML]{C0C0C0}0.51} & 0.00 & \multicolumn{1}{c|}{\cellcolor[HTML]{656565}0.43} & 0.00 \\ \hline
\multicolumn{1}{|c|}{RLO.} & -files & \multicolumn{1}{c|}{0.04} & 0.02 & \multicolumn{1}{c|}{0.22} & 0.18 & \multicolumn{1}{c|}{0.43} & \cellcolor[HTML]{FFFFFF}0.39 & \multicolumn{1}{c|}{0.22} & 0.18 & \multicolumn{1}{c|}{\cellcolor[HTML]{FFFFFF}0.04} & \cellcolor[HTML]{656565}0.04 \\ \hline
\multicolumn{1}{|c|}{RLO. + Reg.} & -files & \multicolumn{1}{c|}{0.04} & \cellcolor[HTML]{656565}0.05 & \multicolumn{1}{c|}{0.22} & \cellcolor[HTML]{C0C0C0}0.19 & \multicolumn{1}{c|}{0.43} & 0.36 & \multicolumn{1}{c|}{0.28} & \cellcolor[HTML]{656565}0.48 & \multicolumn{1}{c|}{\cellcolor[HTML]{FFFFFF}0.04} & \cellcolor[HTML]{656565}0.04 \\ \hline
\multicolumn{1}{|c|}{RLO. + Ent.} & -files & \multicolumn{1}{c|}{0.04} & 0.03 & \multicolumn{1}{c|}{\cellcolor[HTML]{C0C0C0}0.23} & \cellcolor[HTML]{C0C0C0}0.19 & \multicolumn{1}{c|}{\cellcolor[HTML]{C0C0C0}0.45} & \cellcolor[HTML]{C0C0C0}0.41 & \multicolumn{1}{c|}{0.43} & 0.28 & \multicolumn{1}{c|}{\cellcolor[HTML]{FFFFFF}0.04} & \cellcolor[HTML]{656565}0.04 \\ \hline
\multicolumn{1}{|c|}{RLO. + Ent. + Reg.} & -files & \multicolumn{1}{c|}{0.04} & \cellcolor[HTML]{C0C0C0}0.04 & \multicolumn{1}{c|}{0.22} & \cellcolor[HTML]{C0C0C0}0.19 & \multicolumn{1}{c|}{0.43} & 0.38 & \multicolumn{1}{c|}{\cellcolor[HTML]{656565}0.56} & \cellcolor[HTML]{C0C0C0}0.47 & \multicolumn{1}{c|}{\cellcolor[HTML]{FFFFFF}0.04} & \cellcolor[HTML]{656565}0.04 \\ \hline
\multicolumn{1}{|c|}{} & -files & \multicolumn{1}{c|}{\cellcolor[HTML]{C0C0C0}0.05} & 0.03 & \multicolumn{1}{c|}{0.22} & \cellcolor[HTML]{C0C0C0}0.19 & \multicolumn{1}{c|}{0.44} & 0.38 & \multicolumn{1}{c|}{0.34} & 0.27 & \multicolumn{1}{c|}{\cellcolor[HTML]{FFFFFF}0.04} & \cellcolor[HTML]{656565}0.04 \\ \cline{2-12} 
\multicolumn{1}{|c|}{\multirow{-2}{*}{CLEAR}} & hunks & \multicolumn{1}{c|}{\cellcolor[HTML]{C0C0C0}0.05} & 0.03 & \multicolumn{1}{c|}{0.22} & 0.16 & \multicolumn{1}{c|}{0.44} & 0.37 & \multicolumn{1}{c|}{0.35} & 0.34 & \multicolumn{1}{c|}{0.04} & \cellcolor[HTML]{656565}0.04 \\ \hline
\multicolumn{1}{|c|}{} & -files & \multicolumn{1}{c|}{0.04} & \cellcolor[HTML]{C0C0C0}0.04 & \multicolumn{1}{c|}{0.21} & \cellcolor[HTML]{656565}0.20 & \multicolumn{1}{c|}{0.35} & 0.31 & \multicolumn{1}{c|}{0.34} & 0.24 & \multicolumn{1}{c|}{\cellcolor[HTML]{FFFFFF}0.04} & \cellcolor[HTML]{656565}0.04 \\ \cline{2-12} 
\multicolumn{1}{|c|}{\multirow{-2}{*}{CLEAR + Reg.}} & hunks & \multicolumn{1}{c|}{0.04} & 0.04 & \multicolumn{1}{c|}{0.20} & 0.21 & \multicolumn{1}{c|}{0.34} & 0.36 & \multicolumn{1}{c|}{0.31} & 0.23 & \multicolumn{1}{c|}{0.04} & \cellcolor[HTML]{656565}0.04 \\ \hline
\multicolumn{1}{|c|}{} & -files & \multicolumn{1}{c|}{0.04} & \cellcolor[HTML]{656565}0.05 & \multicolumn{1}{c|}{0.21} & \cellcolor[HTML]{C0C0C0}0.19 & \multicolumn{1}{c|}{\cellcolor[HTML]{C0C0C0}0.45} & \cellcolor[HTML]{FFFFFF}0.39 & \multicolumn{1}{c|}{0.48} & 0.25 & \multicolumn{1}{c|}{\cellcolor[HTML]{FFFFFF}0.04} & \cellcolor[HTML]{656565}0.04 \\ \cline{2-12} 
\multicolumn{1}{|c|}{\multirow{-2}{*}{EWC}} & hunks & \multicolumn{1}{c|}{0.04} & \cellcolor[HTML]{656565}0.05 & \multicolumn{1}{c|}{0.21} & 0.20 & \multicolumn{1}{c|}{\cellcolor[HTML]{C0C0C0}0.45} & 0.40 & \multicolumn{1}{c|}{0.26} & 0.38 & \multicolumn{1}{c|}{0.04} & \cellcolor[HTML]{656565}0.04 \\ \hline
\multicolumn{1}{|c|}{} & -files & \multicolumn{1}{c|}{0.04} & \cellcolor[HTML]{C0C0C0}0.04 & \multicolumn{1}{c|}{0.21} & \cellcolor[HTML]{C0C0C0}0.19 & \multicolumn{1}{c|}{0.44} & 0.38 & \multicolumn{1}{c|}{0.26} & 0.28 & \multicolumn{1}{c|}{\cellcolor[HTML]{FFFFFF}0.04} & \cellcolor[HTML]{656565}0.04 \\ \cline{2-12} 
\multicolumn{1}{|c|}{\multirow{-2}{*}{EWC+Reg.}} & hunks & \multicolumn{1}{c|}{\cellcolor[HTML]{C0C0C0}0.05} & \cellcolor[HTML]{656565}0.05 & \multicolumn{1}{c|}{\cellcolor[HTML]{C0C0C0}0.23} & 0.22 & \multicolumn{1}{c|}{0.44} & \cellcolor[HTML]{656565}0.42 & \multicolumn{1}{c|}{0.31} & 0.30 & \multicolumn{1}{c|}{\cellcolor[HTML]{C0C0C0}0.05} & \cellcolor[HTML]{656565}0.04 \\ \hline
\end{tabular}
}
\caption{Eclipse}
\label{tab:Eclipse}
\end{minipage}
\hfill
\begin{minipage}{0.8\textwidth}
\resizebox{\columnwidth}{!}{
\centering
\begin{tabular}{|cc|cc|cc|cc|cc|cc|}
\hline
\multicolumn{2}{|c|}{} & \multicolumn{2}{c|}{top@1} & \multicolumn{2}{c|}{top@5} & \multicolumn{2}{c|}{top@10} & \multicolumn{2}{c|}{MRR} & \multicolumn{2}{c|}{MAP} \\ \cline{3-12} 
\multicolumn{2}{|c|}{\multirow{-2}{*}{}} & \multicolumn{1}{c|}{ST} & NS & \multicolumn{1}{c|}{ST} & NS & \multicolumn{1}{c|}{ST} & NS & \multicolumn{1}{c|}{ST} & NS & \multicolumn{1}{c|}{ST} & NS \\ \hline
\multicolumn{1}{|c|}{FLIM} & -files & \multicolumn{1}{c|}{\cellcolor[HTML]{656565}0.14} & 0.00 & \multicolumn{1}{c|}{\cellcolor[HTML]{656565}0.44} & \cellcolor[HTML]{656565}0.37 & \multicolumn{1}{c|}{\cellcolor[HTML]{656565}0.63} & \cellcolor[HTML]{656565}1.00 & \multicolumn{1}{c|}{0.29} & 0.17 & \multicolumn{1}{c|}{\cellcolor[HTML]{656565}0.24} & \cellcolor[HTML]{656565}0.05 \\ \hline
\multicolumn{1}{|c|}{RLO.} & -files & \multicolumn{1}{c|}{0.04} & 0.03 & \multicolumn{1}{c|}{0.20} & 0.17 & \multicolumn{1}{c|}{0.33} & 0.33 & \multicolumn{1}{c|}{0.36} & 0.30 & \multicolumn{1}{c|}{0.04} & 0.03 \\ \hline
\multicolumn{1}{|c|}{RLO. + Reg.} & -files & \multicolumn{1}{c|}{0.04} & 0.03 & \multicolumn{1}{c|}{0.21} & 0.17 & \multicolumn{1}{c|}{0.45} & 0.30 & \multicolumn{1}{c|}{0.32} & 0.30 & \multicolumn{1}{c|}{0.04} & \cellcolor[HTML]{C0C0C0}0.04 \\ \hline
\multicolumn{1}{|c|}{RLO. + Ent.} & -files & \multicolumn{1}{c|}{0.04} & 0.03 & \multicolumn{1}{c|}{0.22} & 0.15 & \multicolumn{1}{c|}{0.47} & 0.31 & \multicolumn{1}{c|}{0.36} & 0.24 & \multicolumn{1}{c|}{0.04} & 0.03 \\ \hline
\multicolumn{1}{|c|}{RLO. + Ent. + Reg.} & -files & \multicolumn{1}{c|}{0.04} & 0.02 & \multicolumn{1}{c|}{0.23} & 0.14 & \multicolumn{1}{c|}{0.46} & 0.29 & \multicolumn{1}{c|}{0.32} & 0.12 & \multicolumn{1}{c|}{0.04} & 0.03 \\ \hline
\multicolumn{1}{|c|}{} & -files & \multicolumn{1}{c|}{0.04} & 0.03 & \multicolumn{1}{c|}{0.22} & 0.22 & \multicolumn{1}{c|}{0.45} & 0.40 & \multicolumn{1}{c|}{0.37} & 0.14 & \multicolumn{1}{c|}{0.04} & \cellcolor[HTML]{C0C0C0}0.04 \\ \cline{2-12} 
\multicolumn{1}{|c|}{\multirow{-2}{*}{CLEAR}} & hunks & \multicolumn{1}{c|}{0.04} & \cellcolor[HTML]{656565}0.06 & \multicolumn{1}{c|}{0.25} & 0.21 & \multicolumn{1}{c|}{0.48} & 0.40 & \multicolumn{1}{c|}{0.32} & 0.29 & \multicolumn{1}{c|}{\cellcolor[HTML]{C0C0C0}0.05} & \cellcolor[HTML]{C0C0C0}0.04 \\ \hline
\multicolumn{1}{|c|}{} & -files & \multicolumn{1}{c|}{0.04} & 0.04 & \multicolumn{1}{c|}{0.24} & 0.17 & \multicolumn{1}{c|}{0.48} & 0.37 & \multicolumn{1}{c|}{0.39} & 0.26 & \multicolumn{1}{c|}{\cellcolor[HTML]{C0C0C0}0.05} & \cellcolor[HTML]{C0C0C0}0.04 \\ \cline{2-12} 
\multicolumn{1}{|c|}{\multirow{-2}{*}{CLEAR + Reg.}} & hunks & \multicolumn{1}{c|}{\cellcolor[HTML]{C0C0C0}0.06} & \cellcolor[HTML]{C0C0C0}0.05 & \multicolumn{1}{c|}{0.23} & \cellcolor[HTML]{C0C0C0}0.24 & \multicolumn{1}{c|}{0.44} & 0.46 & \multicolumn{1}{c|}{0.37} & 0.30 & \multicolumn{1}{c|}{\cellcolor[HTML]{C0C0C0}0.05} & \cellcolor[HTML]{656565}0.05 \\ \hline
\multicolumn{1}{|c|}{} & -files & \multicolumn{1}{c|}{0.04} & 0.04 & \multicolumn{1}{c|}{0.25} & 0.20 & \multicolumn{1}{c|}{0.48} & 0.41 & \multicolumn{1}{c|}{0.34} & \cellcolor[HTML]{C0C0C0}0.37 & \multicolumn{1}{c|}{\cellcolor[HTML]{C0C0C0}0.05} & \cellcolor[HTML]{C0C0C0}0.04 \\ \cline{2-12} 
\multicolumn{1}{|c|}{\multirow{-2}{*}{EWC}} & hunks & \multicolumn{1}{c|}{0.05} & 0.03 & \multicolumn{1}{c|}{0.23} & 0.14 & \multicolumn{1}{c|}{0.46} & 0.33 & \multicolumn{1}{c|}{\cellcolor[HTML]{C0C0C0}0.41} & 0.30 & \multicolumn{1}{c|}{\cellcolor[HTML]{C0C0C0}0.05} & 0.03 \\ \hline
\multicolumn{1}{|c|}{} & -files & \multicolumn{1}{c|}{0.04} & \cellcolor[HTML]{C0C0C0}0.05 & \multicolumn{1}{c|}{\cellcolor[HTML]{C0C0C0}0.26} & 0.23 & \multicolumn{1}{c|}{\cellcolor[HTML]{C0C0C0}0.49} & 0.44 & \multicolumn{1}{c|}{\cellcolor[HTML]{656565}0.43} & 0.42 & \multicolumn{1}{c|}{\cellcolor[HTML]{C0C0C0}0.05} & \cellcolor[HTML]{C0C0C0}0.04 \\ \cline{2-12} 
\multicolumn{1}{|c|}{\multirow{-2}{*}{EWC+Reg.}} & hunks & \multicolumn{1}{c|}{0.05} & 0.04 & \multicolumn{1}{c|}{0.22} & 0.22 & \multicolumn{1}{c|}{0.43} & \cellcolor[HTML]{C0C0C0}0.45 & \multicolumn{1}{c|}{0.33} & \cellcolor[HTML]{656565}0.45 & \multicolumn{1}{c|}{0.04} & \cellcolor[HTML]{C0C0C0}0.04 \\ \hline
\end{tabular}
}
\caption{AspectJ}
\label{tab:AspectJ}
\end{minipage}
\begin{tablenotes}
     \item Algorithms evaluation across software projects. "RLO." refers to the baseline RLOCATOR. "+ Reg" indicates the inclusion of logistic regression, while "+ Ent" denotes the addition of entropy. "-files"=changeset-files. The best and second best average performances are highlighted in dark grey and light grey respectively. For each metric, the ST column contains the performance of algorithms on stationary data, and the NS column contains the performance on non-stationary data.%\lm{Can you combine the tables into subtables? Just to avoid replicating the same caption.}
 \end{tablenotes}
\end{table*}

We also evaluate the forgetting of the CL agents as part of our evaluation metrics. Specifically, at the training stage, we measure how much each CL agent forgets previously learned bug localization skills on non-stationary data while learning bug localization on stationary data.  The \textbf{forgetting  metric} \cite{powers2022cora} $\mathcal{F}$ is computed as follows:
\begin{equation}
    \mathcal{F}=\frac{r_{i,j-1,end}-r_{i,j,end}}{\mid r_{i,all,max} \mid}
\end{equation}
where $r_{i,j,end}$ is expected return achieved on task $i$ after training on task $j$ and $r_{i,all,max}$ is the maximum  expected return achieved on task $i$. $r_{i,all,max}$ ensures the normalization of the rewards across tasks for better comparison among them.
\begin{table}[t]
%\caption{Algorithms evaluation across software projects with different  data granularities.}
\resizebox{0.8\textwidth}{!}{
\centering
\begin{tabular}{|cc|cc|cc|cc|cc|cc|}
\hline
\multicolumn{2}{|c|}{} & \multicolumn{2}{c|}{top@1} & \multicolumn{2}{c|}{top@5} & \multicolumn{2}{c|}{top@10} & \multicolumn{2}{c|}{MRR} & \multicolumn{2}{c|}{MAP} \\ \cline{3-12} 
\multicolumn{2}{|c|}{\multirow{-2}{*}{}} & \multicolumn{1}{c|}{ST} & NS & \multicolumn{1}{c|}{ST} & NS & \multicolumn{1}{c|}{ST} & NS & \multicolumn{1}{c|}{ST} & NS & \multicolumn{1}{c|}{ST} & NS \\ \hline
\multicolumn{1}{|c|}{FLIM} & -files & \multicolumn{1}{c|}{\cellcolor[HTML]{656565}0.32} & 0.00 & \multicolumn{1}{c|}{\cellcolor[HTML]{656565}0.62} & 0.00 & \multicolumn{1}{c|}{\cellcolor[HTML]{656565}0.73} & 0.50 & \multicolumn{1}{c|}{\cellcolor[HTML]{C0C0C0}0.46} & 0.05 & \multicolumn{1}{c|}{\cellcolor[HTML]{656565}0.39} & \cellcolor[HTML]{C0C0C0}0.10 \\ \hline
\multicolumn{1}{|c|}{RLO.} & -files & \multicolumn{1}{c|}{0.04} & 0.07 & \multicolumn{1}{c|}{0.22} & 0.37 & \multicolumn{1}{c|}{0.44} & 0.88 & \multicolumn{1}{c|}{0.26} & 0.32 & \multicolumn{1}{c|}{0.04} & 0.08 \\ \hline
\multicolumn{1}{|c|}{RLO. + Reg.} & -files & \multicolumn{1}{c|}{0.04} & 0.10 & \multicolumn{1}{c|}{0.22} & 0.34 & \multicolumn{1}{c|}{0.44} & 0.70 & \multicolumn{1}{c|}{0.26} & 0.37 & \multicolumn{1}{c|}{\cellcolor[HTML]{C0C0C0}0.05} & 0.07 \\ \hline
\multicolumn{1}{|c|}{RLO. + Ent.} & -files & \multicolumn{1}{c|}{0.04} & 0.10 & \multicolumn{1}{c|}{0.23} & 0.35 & \multicolumn{1}{c|}{0.43} & 0.83 & \multicolumn{1}{c|}{0.19} & 0.58 & \multicolumn{1}{c|}{0.04} & \cellcolor[HTML]{C0C0C0}0.10 \\ \hline
\multicolumn{1}{|c|}{RLO. + Ent. + Reg.} & -files & \multicolumn{1}{c|}{0.04} & 0.10 & \multicolumn{1}{c|}{0.22} & 0.41 & \multicolumn{1}{c|}{0.44} & 0.80 & \multicolumn{1}{c|}{\cellcolor[HTML]{656565}0.64} & \cellcolor[HTML]{656565}0.68 & \multicolumn{1}{c|}{0.04} & \cellcolor[HTML]{C0C0C0}0.10 \\ \hline
\multicolumn{1}{|c|}{} & -files & \multicolumn{1}{c|}{\cellcolor[HTML]{C0C0C0}0.05} & \cellcolor[HTML]{656565}0.17 & \multicolumn{1}{c|}{0.22} & 0.51 & \multicolumn{1}{c|}{0.45} & 0.86 & \multicolumn{1}{c|}{\cellcolor[HTML]{C0C0C0}0.46} & \cellcolor[HTML]{C0C0C0}0.59 & \multicolumn{1}{c|}{0.04} & \cellcolor[HTML]{C0C0C0}0.10 \\ \cline{2-12} 
\multicolumn{1}{|c|}{\multirow{-2}{*}{CLEAR}} & hunks & \multicolumn{1}{c|}{0.03} & 0.08 & \multicolumn{1}{c|}{0.21} & 0.45 & \multicolumn{1}{c|}{0.44} & 0.91 & \multicolumn{1}{c|}{0.25} & 0.54 & \multicolumn{1}{c|}{0.04} & 0.09 \\ \hline
\multicolumn{1}{|c|}{} & -files & \multicolumn{1}{c|}{\cellcolor[HTML]{C0C0C0}0.05} & 0.13 & \multicolumn{1}{c|}{0.24} & 0.41 & \multicolumn{1}{c|}{0.43} & 0.91 & \multicolumn{1}{c|}{0.24} & 0.41 & \multicolumn{1}{c|}{\cellcolor[HTML]{C0C0C0}0.05} & 0.09 \\ \cline{2-12} 
\multicolumn{1}{|c|}{\multirow{-2}{*}{CLEAR + Reg.}} & hunks & \multicolumn{1}{c|}{0.04} & 0.09 & \multicolumn{1}{c|}{0.22} & 0.46 & \multicolumn{1}{c|}{0.45} & 0.87 & \multicolumn{1}{c|}{0.23} & 0.40 & \multicolumn{1}{c|}{0.04} & \cellcolor[HTML]{C0C0C0}0.10 \\ \hline
\multicolumn{1}{|c|}{} & -files & \multicolumn{1}{c|}{0.04} & \cellcolor[HTML]{C0C0C0}0.15 & \multicolumn{1}{c|}{0.21} & 0.35 & \multicolumn{1}{c|}{0.44} & 0.85 & \multicolumn{1}{c|}{0.33} & 0.40 & \multicolumn{1}{c|}{0.04} & 0.08 \\ \cline{2-12} 
\multicolumn{1}{|c|}{\multirow{-2}{*}{EWC}} & hunks & \multicolumn{1}{c|}{0.04} & 0.13 & \multicolumn{1}{c|}{0.24} & \cellcolor[HTML]{656565}0.56 & \multicolumn{1}{c|}{0.45} & \cellcolor[HTML]{656565}1.0 & \multicolumn{1}{c|}{0.31} & 0.56 & \multicolumn{1}{c|}{\cellcolor[HTML]{C0C0C0}0.05} & \cellcolor[HTML]{656565}0.11 \\ \hline
\multicolumn{1}{|c|}{} & -files & \multicolumn{1}{c|}{\cellcolor[HTML]{C0C0C0}0.05} & 0.12 & \multicolumn{1}{c|}{\cellcolor[HTML]{C0C0C0}0.25} & \cellcolor[HTML]{C0C0C0}0.47 & \multicolumn{1}{c|}{\cellcolor[HTML]{C0C0C0}0.49} & \cellcolor[HTML]{C0C0C0}0.94 & \multicolumn{1}{c|}{0.36} & 0.48 & \multicolumn{1}{c|}{\cellcolor[HTML]{C0C0C0}0.05} & \cellcolor[HTML]{C0C0C0}0.10 \\ \cline{2-12} 
\multicolumn{1}{|c|}{\multirow{-2}{*}{EWC+Reg.}} & hunks & \multicolumn{1}{c|}{0.04} & 0.12 & \multicolumn{1}{c|}{0.22} & 0.44 & \multicolumn{1}{c|}{0.43} & 0.91 & \multicolumn{1}{c|}{0.19} & 0.51 & \multicolumn{1}{c|}{0.04} & 0.09 \\ \hline
\end{tabular}
}
\caption{Algorithms evaluation  on SWT project}
\label{tab:SWT}
\begin{tablenotes}
     \item  "RLO." refers to the baseline RLOCATOR. "+ Reg" indicates the inclusion of logistic regression, while "+ Ent" denotes the addition of entropy. "-files"=changeset-files. The best and second best average performances are highlighted in dark grey and light grey respectively. For each metric, the ST column contains the performance of algorithms on stationary data, and the NS column contains the performance on non-stationary data. %\lm{Can you combine the tables into subtables? Just to avoid replicating the same caption.}
 \end{tablenotes}
\end{table}
\begin{table}[t]
%\caption{Algorithms evaluation across software projects with different  data granularities.}
\resizebox{0.8\textwidth}{!}{
\centering
\begin{tabular}{|ccc|ll|ll|ll|ll|ll|}
\hline
\multicolumn{3}{|c|}{}                                                                                       & \multicolumn{2}{c|}{top@1}                                                       & \multicolumn{2}{c|}{top@5}                                                       & \multicolumn{2}{c|}{top@10}                                                      & \multicolumn{2}{c|}{MRR}                                                         & \multicolumn{2}{c|}{MAP}                                                         \\ \cline{4-13} 
\multicolumn{3}{|c|}{\multirow{-2}{*}{}}                                                                     & \multicolumn{1}{c|}{ST}                           & \multicolumn{1}{c|}{NS}      & \multicolumn{1}{c|}{ST}                           & \multicolumn{1}{c|}{NS}      & \multicolumn{1}{c|}{ST}                           & \multicolumn{1}{c|}{NS}      & \multicolumn{1}{c|}{ST}                           & \multicolumn{1}{c|}{NS}      & \multicolumn{1}{c|}{ST}                           & \multicolumn{1}{c|}{NS}      \\ \hline
\multicolumn{1}{|c|}{}                        & \multicolumn{1}{c|}{}                               & -files & \multicolumn{1}{l|}{0.08}                         & 0.08                         & \multicolumn{1}{l|}{0.40}                         & 0.41                         & \multicolumn{1}{l|}{\cellcolor[HTML]{C0C0C0}0.87} & \cellcolor[HTML]{C0C0C0}0.87 & \multicolumn{1}{l|}{\cellcolor[HTML]{656565}0.51} & 0.28                         & \multicolumn{1}{l|}{\cellcolor[HTML]{C0C0C0}0.08} & \cellcolor[HTML]{656565}0.09 \\ \cline{3-13} 
\multicolumn{1}{|c|}{}                        & \multicolumn{1}{c|}{\multirow{-2}{*}{CLEAR}}        & hunks  & \multicolumn{1}{l|}{0.07}                         & 0.08                         & \multicolumn{1}{l|}{0.42}                         & \cellcolor[HTML]{C0C0C0}0.44 & \multicolumn{1}{l|}{0.81}                         & 0.83                         & \multicolumn{1}{l|}{0.38}                         & \cellcolor[HTML]{656565}0.45 & \multicolumn{1}{l|}{\cellcolor[HTML]{C0C0C0}0.08} & \cellcolor[HTML]{656565}0.09 \\ \cline{2-13} 
\multicolumn{1}{|c|}{}                        & \multicolumn{1}{c|}{}                               & -files & \multicolumn{1}{l|}{0.08}                         & 0.08                         & \multicolumn{1}{l|}{0.42}                         & 0.39                         & \multicolumn{1}{l|}{0.77}                         & 0.77                         & \multicolumn{1}{l|}{0.33}                         & \cellcolor[HTML]{C0C0C0}0.41 & \multicolumn{1}{l|}{\cellcolor[HTML]{C0C0C0}0.08} & \cellcolor[HTML]{C0C0C0}0.08 \\ \cline{3-13} 
\multicolumn{1}{|c|}{}                        & \multicolumn{1}{c|}{\multirow{-2}{*}{CLEAR + Reg.}} & hunks  & \multicolumn{1}{l|}{\cellcolor[HTML]{C0C0C0}0.09} & \cellcolor[HTML]{C0C0C0}0.09 & \multicolumn{1}{l|}{0.39}                         & 0.41                         & \multicolumn{1}{l|}{0.69}                         & 0.73                         & \multicolumn{1}{l|}{0.36}                         & 0.33                         & \multicolumn{1}{l|}{\cellcolor[HTML]{C0C0C0}0.08} & \cellcolor[HTML]{656565}0.09 \\ \cline{2-13} 
\multicolumn{1}{|c|}{}                        & \multicolumn{1}{c|}{}                               & -files & \multicolumn{1}{l|}{\cellcolor[HTML]{C0C0C0}0.09} & \cellcolor[HTML]{C0C0C0}0.09 & \multicolumn{1}{l|}{0.44}                         & 0.43                         & \multicolumn{1}{l|}{\cellcolor[HTML]{656565}0.89} & \cellcolor[HTML]{656565}0.88 & \multicolumn{1}{l|}{0.45}                         & 0.33                         & \multicolumn{1}{l|}{\cellcolor[HTML]{656565}0.09} & \cellcolor[HTML]{656565}0.09 \\ \cline{3-13} 
\multicolumn{1}{|c|}{}                        & \multicolumn{1}{c|}{\multirow{-2}{*}{EWC}}          & hunks  & \multicolumn{1}{l|}{0.07}                         & \cellcolor[HTML]{C0C0C0}0.09 & \multicolumn{1}{l|}{0.42}                         & \cellcolor[HTML]{656565}0.45 & \multicolumn{1}{l|}{0.85}                         & \cellcolor[HTML]{C0C0C0}0.87 & \multicolumn{1}{l|}{\cellcolor[HTML]{C0C0C0}0.47} & 0.29                         & \multicolumn{1}{l|}{\cellcolor[HTML]{C0C0C0}0.08} & \cellcolor[HTML]{656565}0.09 \\ \cline{2-13} 
\multicolumn{1}{|c|}{}                        & \multicolumn{1}{c|}{}                               & -files & \multicolumn{1}{l|}{\cellcolor[HTML]{656565}0.10} & \cellcolor[HTML]{C0C0C0}0.09 & \multicolumn{1}{l|}{\cellcolor[HTML]{C0C0C0}0.43} & \cellcolor[HTML]{C0C0C0}0.44 & \multicolumn{1}{l|}{0.85}                         & 0.86                         & \multicolumn{1}{l|}{0.32}                         & 0.39                         & \multicolumn{1}{l|}{\cellcolor[HTML]{656565}0.09} & \cellcolor[HTML]{656565}0.09 \\ \cline{3-13} 
\multicolumn{1}{|c|}{\multirow{-8}{*}{PDE}}   & \multicolumn{1}{c|}{\multirow{-2}{*}{EWC + Reg.}}   & hunks  & \multicolumn{1}{l|}{0.09}                         & \cellcolor[HTML]{656565}0.10 & \multicolumn{1}{l|}{\cellcolor[HTML]{656565}0.44} & \cellcolor[HTML]{656565}0.45 & \multicolumn{1}{l|}{0.85}                         & \cellcolor[HTML]{656565}0.88 & \multicolumn{1}{l|}{0.34}                         & 0.33                         & \multicolumn{1}{l|}{\cellcolor[HTML]{656565}0.09} & \cellcolor[HTML]{656565}0.09 \\ \hline
\multicolumn{1}{|c|}{}                        & \multicolumn{1}{c|}{}                               & -files & \multicolumn{1}{l|}{0.12}                         & 0.11                         & \multicolumn{1}{l|}{\cellcolor[HTML]{656565}0.65} & 0.56                         & \multicolumn{1}{l|}{\cellcolor[HTML]{656565}1.0}  & \cellcolor[HTML]{656565}1.0  & \multicolumn{1}{l|}{0.40}                         & \cellcolor[HTML]{C0C0C0}0.60 & \multicolumn{1}{l|}{\cellcolor[HTML]{C0C0C0}0.12} & \cellcolor[HTML]{C0C0C0}0.11 \\ \cline{3-13} 
\multicolumn{1}{|c|}{}                        & \multicolumn{1}{c|}{\multirow{-2}{*}{CLEAR}}        & hunks  & \multicolumn{1}{l|}{0.11}                         & 0.11                         & \multicolumn{1}{l|}{0.58}                         & 0.58                         & \multicolumn{1}{l|}{\cellcolor[HTML]{656565}1.0}  & \cellcolor[HTML]{656565}1.0  & \multicolumn{1}{l|}{0.54}                         & 0.55                         & \multicolumn{1}{l|}{\cellcolor[HTML]{C0C0C0}0.12} & \cellcolor[HTML]{C0C0C0}0.11 \\ \cline{2-13} 
\multicolumn{1}{|c|}{}                        & \multicolumn{1}{c|}{}                               & -files & \multicolumn{1}{l|}{0.12}                         & \cellcolor[HTML]{C0C0C0}0.12 & \multicolumn{1}{l|}{0.63}                         & \cellcolor[HTML]{656565}0.61 & \multicolumn{1}{l|}{\cellcolor[HTML]{656565}1.0}  & \cellcolor[HTML]{656565}1.0  & \multicolumn{1}{l|}{0.42}                         & 0.49                         & \multicolumn{1}{l|}{\cellcolor[HTML]{C0C0C0}0.12} & \cellcolor[HTML]{656565}0.12 \\ \cline{3-13} 
\multicolumn{1}{|c|}{}                        & \multicolumn{1}{c|}{\multirow{-2}{*}{CLEAR + Reg.}} & hunks  & \multicolumn{1}{l|}{\cellcolor[HTML]{656565}0.14} & \cellcolor[HTML]{C0C0C0}0.12 & \multicolumn{1}{l|}{0.65}                         & \cellcolor[HTML]{C0C0C0}0.59 & \multicolumn{1}{l|}{\cellcolor[HTML]{656565}1.0}  & \cellcolor[HTML]{656565}1.0  & \multicolumn{1}{l|}{0.51}                         & 0.51                         & \multicolumn{1}{l|}{\cellcolor[HTML]{656565}0.13} & \cellcolor[HTML]{656565}0.12 \\ \cline{2-13} 
\multicolumn{1}{|c|}{}                        & \multicolumn{1}{c|}{}                               & -files & \multicolumn{1}{l|}{0.12}                         & \cellcolor[HTML]{656565}0.14 & \multicolumn{1}{l|}{0.62}                         & 0.56                         & \multicolumn{1}{l|}{\cellcolor[HTML]{656565}1.0}  & \cellcolor[HTML]{656565}1.0  & \multicolumn{1}{l|}{\cellcolor[HTML]{656565}0.58} & 0.47                         & \multicolumn{1}{l|}{\cellcolor[HTML]{C0C0C0}0.12} & \cellcolor[HTML]{656565}0.12 \\ \cline{3-13} 
\multicolumn{1}{|c|}{}                        & \multicolumn{1}{c|}{\multirow{-2}{*}{EWC}}          & hunks  & \multicolumn{1}{l|}{0.12}                         & \cellcolor[HTML]{C0C0C0}0.12 & \multicolumn{1}{l|}{0.62}                         & 0.57                         & \multicolumn{1}{l|}{\cellcolor[HTML]{656565}1.0}  & \cellcolor[HTML]{656565}1.0  & \multicolumn{1}{l|}{\cellcolor[HTML]{C0C0C0}0.55} & 0.54                         & \multicolumn{1}{l|}{\cellcolor[HTML]{C0C0C0}0.12} & \cellcolor[HTML]{656565}0.12 \\ \cline{2-13} 
\multicolumn{1}{|c|}{}                        & \multicolumn{1}{c|}{}                               & -files & \multicolumn{1}{l|}{\cellcolor[HTML]{C0C0C0}0.13} & \cellcolor[HTML]{C0C0C0}0.12 & \multicolumn{1}{l|}{\cellcolor[HTML]{C0C0C0}0.63} & \cellcolor[HTML]{C0C0C0}0.59 & \multicolumn{1}{l|}{\cellcolor[HTML]{656565}1.0}  & \cellcolor[HTML]{656565}1.0  & \multicolumn{1}{l|}{\cellcolor[HTML]{656565}0.58} & \cellcolor[HTML]{656565}0.64 & \multicolumn{1}{l|}{\cellcolor[HTML]{656565}0.13} & \cellcolor[HTML]{656565}0.12 \\ \cline{3-13} 
\multicolumn{1}{|c|}{\multirow{-8}{*}{Zxing}} & \multicolumn{1}{c|}{\multirow{-2}{*}{EWC + Reg.}}   & hunks  & \multicolumn{1}{l|}{0.10}                         & \cellcolor[HTML]{C0C0C0}0.12 & \multicolumn{1}{l|}{0.61}                         & 0.56                         & \multicolumn{1}{l|}{\cellcolor[HTML]{656565}1.0}  & \cellcolor[HTML]{656565}1.0  & \multicolumn{1}{l|}{\cellcolor[HTML]{C0C0C0}0.55} & 0.57                         & \multicolumn{1}{l|}{\cellcolor[HTML]{C0C0C0}0.12} & \cellcolor[HTML]{C0C0C0}0.11 \\ \hline
\end{tabular}
}
\caption{Algorithms evaluation  on PDE and Zxing project}
\label{tab:PDEANDZXING}
\begin{tablenotes}
     \item  -files=changeset-files. The best and second best average performances are highlighted in dark grey and light grey respectively per project. For each metric, the ST column contains the performance of algorithms on stationary data, and the NS column contains the performance on non-stationary data. %\lm{Can you combine the tables into subtables? Just to avoid replicating the same caption.}
 \end{tablenotes}
\end{table}

To counter the randomness effect when collecting our evaluation metrics, we repeat each run 5 times to average the results. The experiments are conducted on the Cedar cluster servers provided by the Digital Research Alliance of Canada (the Alliance) \cite{Alliance}. Each server has 32 cores at 2.2GHz with 125GB of main memory and 32GB of GPU memory.  We employ Welch’s ANOVA and Games-Howell post-hoc test \cite{welch1947generalization}, \cite{games1976pairwise} to indicate the best bug localization technique. While Welch’s ANOVA test checks for significant differences between the CL agents across the datasets, the Games-Howell post-hoc test compares each pair of configurations. A difference with p-value $\leq 0.05$ is considered significant in our assessments.

\begin{table}[t]
\caption{Algorithms evaluation across software projects with different  data granularities. In bold are the best average performances per algorithm per metric for each project and data granularity. "-files"=changeset-files. "P@"=top@. }
\label{tab:Algorithms evaluation across software projects with different  data granularities}
\resizebox{0.8\textwidth}{!}{
\centering
\begin{tabular}{|c|c|c|c|c|c|c|c|c|}
\hline
     & \textbf{-files}        & SWT           & AspectJ       & Tomcat        & \textbf{hunks}         & SWT           & AspectJ       & Tomcat        \\ \hline
MRR  & \multirow{5}{*}{CLEAR} & 0.53          & \textbf{0.26} & 0.35          & \multirow{5}{*}{CLEAR} & 0.40          & \textbf{0.30}          & 0.34          \\ \cline{1-1} \cline{3-5} \cline{7-9} 
MAP  &                        & 0.07          & 0.04          & 0.05          &                        & 0.06          & 0.04          & 0.04          \\ \cline{1-1} \cline{3-5} \cline{7-9} 
P@1  &                        & 0.08          & 0.04          & 0.05          &                        & 0.04          & 0.04          & 0.03          \\ \cline{1-1} \cline{3-5} \cline{7-9} 
P@5  &                        & \textbf{0.36} & \textbf{0.22} & \textbf{0.23} &                        & 0.33          & \textbf{0.23} & 0.21          \\ \cline{1-1} \cline{3-5} \cline{7-9} 
P@10 &                        & \textbf{0.65} & 0.42          & \textbf{0.46} &                        & 0.68          & \textbf{0.44} & 0.41          \\ \hline
MRR  & \multirow{5}{*}{EWC}   & 0.36          & 0.35          & 0.32          & \multirow{5}{*}{EWC}   & 0.43          & 0.35          & 0.25          \\ \cline{1-1} \cline{3-5} \cline{7-9} 
MAP  &                        & 0.06          & 0.04          & 0.04          &                        & 0.08          & 0.04          & 0.05          \\ \cline{1-1} \cline{3-5} \cline{7-9} 
P@1  &                        & 0.05          & 0.04          & 0.03          &                        & 0.07          & 0.04          & 0.05          \\ \cline{1-1} \cline{3-5} \cline{7-9} 
P@5  &                        & 0.28          & \textbf{0.22} & 0.21          &                        &\textbf{ 0.40}          & 0.19          & \textbf{0.22} \\ \cline{1-1} \cline{3-5} \cline{7-9} 
P@10 &                        & \textbf{0.65} & \textbf{0.44} & 0.42          &                        & \textbf{0.75}          & 0.40          & \textbf{0.45} \\ \hline
MRR  & \multirow{5}{*}{QARC}  & \textbf{0.55} & 0.17          & \textbf{0.44} & \multirow{5}{*}{QARC}  & \textbf{0.51} & 0.17          & 0.38          \\ \cline{1-1} \cline{3-5} \cline{7-9} 
MAP  &                        & \textbf{0.13} & \textbf{0.08} & \textbf{0.11} &                        & \textbf{0.13} & 0.08          & \textbf{0.13} \\ \cline{1-1} \cline{3-5} \cline{7-9} 
P@1  &                        & \textbf{0.53} & 0.15          & 0.36          &                        & \textbf{0.47}          & \textbf{0.15} & 0.29          \\ \cline{1-1} \cline{3-5} \cline{7-9} 
P@5  &                        & 0.17          & 0.09          & 0.18          &                        & 0.18          & 0.08          & 0.18          \\ \cline{1-1} \cline{3-5} \cline{7-9} 
P@10 &                        & 0.15          & 0.10          & 0.18          &                        & 0.16          & 0.11          & 0.20          \\ \hline
MRR  & \multirow{5}{*}{QARCL} & 0.48          & 0.17          & 0.44          & \multirow{5}{*}{QARCL} & \textbf{0.51} & 0.17          & \textbf{0.48} \\ \cline{1-1} \cline{3-5} \cline{7-9} 
MAP  &                        & 0.12          & \textbf{0.08} & \textbf{0.11} &                        & \textbf{0.13} & 0.08          & \textbf{0.13} \\ \cline{1-1} \cline{3-5} \cline{7-9} 
P@1  &                        & 0.44          & 0.15          & \textbf{0.37} &                        & \textbf{0.47}          & \textbf{0.15} & \textbf{0.37} \\ \cline{1-1} \cline{3-5} \cline{7-9} 
P@5  &                        & 0.15          & 0.10          & 0.15          &                        & 0.18          & 0.08          & 0.20          \\ \cline{1-1} \cline{3-5} \cline{7-9} 
P@10 &                        & 0.14          & 0.11          & 0.15          &                        & 0.16          & 0.11          & 0.20          \\ \hline
MRR  & \multirow{5}{*}{QD}    & 0.54          & 0.18          & 0.40          & \multirow{5}{*}{QD}    & 0.49          & 0.17          & 0.40          \\ \cline{1-1} \cline{3-5} \cline{7-9} 
MAP  &                        & \textbf{0.13} & \textbf{0.08} & 0.09          &                        & \textbf{0.13} & 0.08          & 0.12          \\ \cline{1-1} \cline{3-5} \cline{7-9} 
P@1  &                        & \textbf{0.53} & \textbf{0.16} & 0.29          &                        & 0.42          & \textbf{0.15} & 0.28          \\ \cline{1-1} \cline{3-5} \cline{7-9} 
P@5  &                        & 0.16          & 0.10          & 0.15          &                        & 0.19          & 0.10          & 0.18          \\ \cline{1-1} \cline{3-5} \cline{7-9} 
P@10 &                        & 0.14          & 0.11          & 0.17          &                        & 0.17          & 0.10          & 0.20          \\ \hline
\end{tabular}
}
%\begin{tablenotes}
%     \item In bold are the best average performances per algorithm per metric for each project and data granularity. "-files"=changeset-files. "P@"=top@. 
 %\end{tablenotes}
 %\vspace{-1em}
\end{table}
\begin{table}[t]
\caption{Algorithms evaluation across PDE and Zxing projects with different  data granularities. In bold are the best average performances per algorithm per metric for each project and data granularity. "-files"=changeset-files. "P@"=top@. }
\label{tab:Algorithms evaluation across PDE and Zxing projects with different  data granularities}
\resizebox{0.6\textwidth}{!}{
\centering
\begin{tabular}{|c|c|c|c|c|c|c|}
\hline
     & \textbf{-files}        & PDE           & Zxing         & \textbf{hunks}         & PDE           & Zxing         \\ \hline
MRR  & \multirow{5}{*}{CLEAR} & \textbf{0.39} & 0.50          & \multirow{5}{*}{CLEAR} & \textbf{0.42} & \textbf{0.55} \\ \cline{1-1} \cline{3-4} \cline{6-7} 
MAP  &                        & 0.08          & 0.12          &                        & 0.08          & 0.11          \\ \cline{1-1} \cline{3-4} \cline{6-7} 
P@1  &                        & \textbf{0.08} & 0.11          &                        & 0.08          & 0.11          \\ \cline{1-1} \cline{3-4} \cline{6-7} 
P@5  &                        & 0.41          & 0.61          &                        & \textbf{0.43} & 0.58          \\ \cline{1-1} \cline{3-4} \cline{6-7} 
P@10 &                        & 0.87          & \textbf{1.00} &                        & 0.82          & \textbf{1.00} \\ \hline
MRR  & \multirow{5}{*}{EWC}   & \textbf{0.39} & \textbf{0.53} & \multirow{5}{*}{EWC}   & 0.38          & \textbf{0.55} \\ \cline{1-1} \cline{3-4} \cline{6-7} 
MAP  &                        & \textbf{0.09} & 0.12          &                        & 0.09          & 0.12          \\ \cline{1-1} \cline{3-4} \cline{6-7} 
P@1  &                        & 0.09          & 0.13          &                        & 0.08          & 0.12          \\ \cline{1-1} \cline{3-4} \cline{6-7} 
P@5  &                        & \textbf{0.44} & 0.59          &                        & \textbf{0.43} & \textbf{0.59} \\ \cline{1-1} \cline{3-4} \cline{6-7} 
P@10 &                        & \textbf{0.88} & \textbf{1.00} &                        & \textbf{0.86} & \textbf{1.00} \\ \hline
MRR  & \multirow{5}{*}{QARC}  & 0.17          & 0.24          & \multirow{5}{*}{QARC}  & 0.21          & 0.44          \\ \cline{1-1} \cline{3-4} \cline{6-7} 
MAP  &                        & 0.05          & 0.18          &                        & 0.08          & \textbf{0.29} \\ \cline{1-1} \cline{3-4} \cline{6-7} 
P@1  &                        & 0.07          & 0.20          &                        & 0.10          & \textbf{0.40} \\ \cline{1-1} \cline{3-4} \cline{6-7} 
P@5  &                        & 0.09          & 0.14          &                        & 0.17          & 0.31          \\ \cline{1-1} \cline{3-4} \cline{6-7} 
P@10 &                        & 0.12          & 0.34          &                        & 0.19          & 0.32          \\ \hline
MRR  & \multirow{5}{*}{QARCL} & 0.26          & 0.35          & \multirow{5}{*}{QARCL} & 0.26          & 0.47          \\ \cline{1-1} \cline{3-4} \cline{6-7} 
MAP  &                        & 0.08          & \textbf{0.21} &                        & 0.08          & \textbf{0.29} \\ \cline{1-1} \cline{3-4} \cline{6-7} 
P@1  &                        & 0.17          & 0.20          &                        & 0.17          & \textbf{0.40} \\ \cline{1-1} \cline{3-4} \cline{6-7} 
P@5  &                        & 0.11          & \textbf{0.39} &                        & 0.11          & 0.39          \\ \cline{1-1} \cline{3-4} \cline{6-7} 
P@10 &                        & 0.18          & 0.38          &                        & 0.18          & 0.42          \\ \hline
MRR  & \multirow{5}{*}{QD}    & 0.21          & 0.29          & \multirow{5}{*}{QD}    & 0.26          & 0.32          \\ \cline{1-1} \cline{3-4} \cline{6-7} 
MAP  &                        & 0.07          & 0.15          &                        & \textbf{0.10} & 0.14          \\ \cline{1-1} \cline{3-4} \cline{6-7} 
P@1  &                        & 0.13          & 0.20          &                        & \textbf{0.20} & 0.20          \\ \cline{1-1} \cline{3-4} \cline{6-7} 
P@5  &                        & 0.15          & 0.27          &                        & 0.18          & 0.29          \\ \cline{1-1} \cline{3-4} \cline{6-7} 
P@10 &                        & 0.15          & 0.38          &                        & 0.17          & 0.30          \\ \hline
\end{tabular}
}
%\begin{tablenotes}
 %    \item In bold are the best average performances per algorithm per metric for each project and data granularity. "-files"=changeset-files. "P@"=top@. 
% \end{tablenotes}
 %\vspace{-1em}
\end{table}

\section{Experimental results} \label{sec:Experimental results}
To evaluate the CL agents from our proposed CL framework, leveraged for bug localization, we define the following Research Questions (RQs). 
\begin{itemize}
    \item \textbf{RQ1: }How do DL techniques for bug localization perform in non-stationary setting?
    \item \textbf{RQ2: }Can CL techniques improve bug localization in both stationary and non-stationary settings? 
    \item \textbf{RQ3: }Can prior knowledge about bug-inducing factors improve the performance of CL techniques for bug localization?
    
\end{itemize}
We report the results associated to \textbf{RQ1, RQ2}, and \textbf{RQ3} in Subsection \ref{sec:rq1}, \ref{sec:rq2}, and \ref{sec:rq3}, respectively. When presenting our results, we consider a comparison significant if the effect size has a $p-value < 0.05$. Moreover, the p-values calculated are corrected for multiple comparisons. We employ Games-Howell and Tukey-HSD procedures, which control the family-wise type 1 error rate across two or more pairwise comparisons \cite{games1976pairwise,sauder2019updated} Finally, the results of the effect size analysis across all studied software projects and for different data granularities can be found in our replication package \cite{replication-package}.

\subsection{\textbf{RQ1: How do DL techniques for bug localization perform in non-stationary setting?}} \label{sec:rq1}
\textbf{Motivation:} The purpose of this RQ is to empirically evaluate the performance of our baseline studies on non-stationary setting at changeset-files level. Between the time a bug is reported and when it is fixed, the affected changeset-files in a software project can undergo various changes, such as content modifications~\cite{da2024detecting} or function renaming~\cite{da2022build}, yet still be associated with the bug report. RLOCATOR AND FLIM baseline studies trained and evaluated their proposed approaches on changeset-files collected when bugs are fixed only. Nonetheless, a bug localization technique should remain effective regardless of the different versions of the buggy changeset-files.
%For this RQ, we only consider RLOCATOR and FLIM baseline studies as they do not take into account the changes that changeset-files can undergo between the time a bug is reported and when it is fixed.\\
\textbf{Method:} We calculate the relevant metrics described in Section \ref{sec:Evaluation metrics} for FLIM and RLOCATOR baseline studies. \\
\textbf{Results:}  Table \ref{tab:Tomcat}, \ref{tab:Birt}, \ref{tab:Eclipse}, \ref{tab:AspectJ}, and \ref{tab:SWT} show the performance of the baselines studies (i.e., FLIM \cite{chakraborty2023rlocator} and RLOCATOR \cite{liang2022modeling}) on stationary and non-stationary data at the changeset-files level across Tomcat, Birt, Eclipse, AspectJ and SWT projects respectively. 
FLIM's performance significantly decreases (by 17 - 194\%) on non-stationary data compared to stationary data, as evaluated across the Tomcat (see Table \ref{tab:Tomcat}), Eclipse (see Table \ref{tab:Eclipse}),  AspectJ (see Table \ref{tab:AspectJ}) and SWT (see Table \ref{tab:SWT}) projects using the top@1, top@5, MRR, and MAP metrics. RLOCATOR (with and without entropy) demonstrates a similar trend on AspectJ, Birt, and Eclipse projects with a decrease in performance by 9 - 85\% on non-stationary data compared to stationary data using the top@1, top@5, top@10, and MRR metrics. 

\begin{table}[t]
\caption{Algorithms evaluation across software projects with different data granularity when incorporating bug-inducing factors on CL agents. In bold are the best average performances per algorithm per metric for each project and data granularity. "-files"=changeset-files. "P@"=top@.}
\label{tab:Algorithms evaluation across software projects with different data granularity when incorporating bug-inducing factors on CL agents}
\resizebox{0.8\textwidth}{!}{
\centering
%put it here|c|ccccc|ccccc|ccccc|ccccc|ccccc|
\begin{tabular}{|c|c|c|c|c|c|c|c|c|}
\hline
{\color[HTML]{000000} }     & {\color[HTML]{000000} \textbf{-files}}              & {\color[HTML]{000000} SWT}           & {\color[HTML]{000000} AspectJ}       & {\color[HTML]{000000} Tomcat}        & {\color[HTML]{000000} \textbf{hunks}}              & {\color[HTML]{000000} SWT}           & {\color[HTML]{000000} AspectJ}       & {\color[HTML]{000000} Tomcat}        \\ \hline
{\color[HTML]{000000} MRR}  & {\color[HTML]{000000} }                             & {\color[HTML]{000000} 0.33}          & {\color[HTML]{000000} 0.32}          & {\color[HTML]{000000} 0.29}          & {\color[HTML]{000000} }                             & {\color[HTML]{000000} 0.32}          & {\color[HTML]{000000} 0.33}          & {\color[HTML]{000000} 0.22}          \\ \cline{1-1} \cline{3-5} \cline{7-9} 
{\color[HTML]{000000} MAP}  & {\color[HTML]{000000} }                             & {\color[HTML]{000000} 0.07}          & {\color[HTML]{000000} 0.04}          & {\color[HTML]{000000} 0.05}          & {\color[HTML]{000000} }                             & {\color[HTML]{000000} 0.07}          & {\color[HTML]{000000} 0.05}          & {\color[HTML]{000000} 0.05}          \\ \cline{1-1} \cline{3-5} \cline{7-9} 
{\color[HTML]{000000} P@1}  & {\color[HTML]{000000} }                             & {\color[HTML]{000000} 0.07}          & {\color[HTML]{000000} 0.04}          & {\color[HTML]{000000} 0.05}          & {\color[HTML]{000000} }                             & {\color[HTML]{000000} 0.07}          & {\color[HTML]{000000} 0.05}          & {\color[HTML]{000000} 0.06}          \\ \cline{1-1} \cline{3-5} \cline{7-9} 
{\color[HTML]{000000} P@5}  & {\color[HTML]{000000} }                             & {\color[HTML]{000000} 0.33}          & {\color[HTML]{000000} 0.21}          & {\color[HTML]{000000} \textbf{0.25}} & {\color[HTML]{000000} }                             & {\color[HTML]{000000} \textbf{0.34}} & {\color[HTML]{000000} \textbf{0.23}} & {\color[HTML]{000000} \textbf{0.25}} \\ \cline{1-1} \cline{3-5} \cline{7-9} 
{\color[HTML]{000000} P@10} & \multirow{-5}{*}{{\color[HTML]{000000} CLEAR+Reg.}} & {\color[HTML]{000000} \textbf{0.67}} & {\color[HTML]{000000} 0.43}          & {\color[HTML]{000000} 0.38}          & \multirow{-5}{*}{{\color[HTML]{000000} CLEAR+Reg.}} & {\color[HTML]{000000} 0.66}          & {\color[HTML]{000000} \textbf{0.45}} & {\color[HTML]{000000} 0.39}          \\ \hline
{\color[HTML]{000000} MRR}  & {\color[HTML]{000000} }                             & {\color[HTML]{000000} 0.42}          & {\color[HTML]{000000} \textbf{0.42}} & {\color[HTML]{000000} 0.22}          & {\color[HTML]{000000} }                             & {\color[HTML]{000000} 0.35}          & {\color[HTML]{000000} \textbf{0.39}} & {\color[HTML]{000000} 0.26}          \\ \cline{1-1} \cline{3-5} \cline{7-9} 
{\color[HTML]{000000} MAP}  & {\color[HTML]{000000} }                             & {\color[HTML]{000000} 0.07}          & {\color[HTML]{000000} 0.05}          & {\color[HTML]{000000} 0.04}          & {\color[HTML]{000000} }                             & {\color[HTML]{000000} 0.06}          & {\color[HTML]{000000} 0.04}          & {\color[HTML]{000000} 0.05}          \\ \cline{1-1} \cline{3-5} \cline{7-9} 
{\color[HTML]{000000} P@1}  & {\color[HTML]{000000} }                             & {\color[HTML]{000000} 0.08}          & {\color[HTML]{000000} 0.05}          & {\color[HTML]{000000} 0.04}          & {\color[HTML]{000000} }                             & {\color[HTML]{000000} 0.08}          & {\color[HTML]{000000} 0.04}          & {\color[HTML]{000000} 0.05}          \\ \cline{1-1} \cline{3-5} \cline{7-9} 
{\color[HTML]{000000} P@5}  & {\color[HTML]{000000} }                             & {\color[HTML]{000000} \textbf{0.36}} & {\color[HTML]{000000} \textbf{0.24}} & {\color[HTML]{000000} 0.22}          & {\color[HTML]{000000} }                             & {\color[HTML]{000000} 0.33}          & {\color[HTML]{000000} 0.22}          & {\color[HTML]{000000} 0.24}          \\ \cline{1-1} \cline{3-5} \cline{7-9} 
{\color[HTML]{000000} P@10} & \multirow{-5}{*}{{\color[HTML]{000000} EWC+Reg.}}   & {\color[HTML]{000000} \textbf{0.71}} & {\color[HTML]{000000} \textbf{0.47}} & {\color[HTML]{000000} \textbf{0.42}} & \multirow{-5}{*}{{\color[HTML]{000000} EWC+Reg.}}   & {\color[HTML]{000000} \textbf{0.67}} & {\color[HTML]{000000} 0.44}          & {\color[HTML]{000000} \textbf{0.48}} \\ \hline
{\color[HTML]{000000} MRR}  & {\color[HTML]{000000} }                             & {\color[HTML]{000000} \textbf{0.55}} & {\color[HTML]{000000} 0.17}          & {\color[HTML]{000000} \textbf{0.44}} & {\color[HTML]{000000} }                             & {\color[HTML]{000000} \textbf{0.51}} & {\color[HTML]{000000} 0.17}          & {\color[HTML]{000000} 0.38}          \\ \cline{1-1} \cline{3-5} \cline{7-9} 
{\color[HTML]{000000} MAP}  & {\color[HTML]{000000} }                             & {\color[HTML]{000000} \textbf{0.13}} & {\color[HTML]{000000} \textbf{0.08}} & {\color[HTML]{000000} \textbf{0.11}} & {\color[HTML]{000000} }                             & {\color[HTML]{000000} \textbf{0.13}} & {\color[HTML]{000000} 0.08}          & {\color[HTML]{000000} \textbf{0.13}} \\ \cline{1-1} \cline{3-5} \cline{7-9} 
{\color[HTML]{000000} P@1}  & {\color[HTML]{000000} }                             & {\color[HTML]{000000} \textbf{0.53}} & {\color[HTML]{000000} 0.15}          & {\color[HTML]{000000} 0.36}          & {\color[HTML]{000000} }                             & {\color[HTML]{000000} \textbf{0.47}} & {\color[HTML]{000000} \textbf{0.15}} & {\color[HTML]{000000} 0.29}          \\ \cline{1-1} \cline{3-5} \cline{7-9} 
{\color[HTML]{000000} P@5}  & {\color[HTML]{000000} }                             & {\color[HTML]{000000} 0.17}          & {\color[HTML]{000000} 0.09}          & {\color[HTML]{000000} 0.18}          & {\color[HTML]{000000} }                             & {\color[HTML]{000000} 0.18}          & {\color[HTML]{000000} 0.08}          & {\color[HTML]{000000} 0.18}          \\ \cline{1-1} \cline{3-5} \cline{7-9} 
{\color[HTML]{000000} P@10} & \multirow{-5}{*}{{\color[HTML]{000000} QARC}}       & {\color[HTML]{000000} 0.15}          & {\color[HTML]{000000} 0.10}          & {\color[HTML]{000000} 0.18}          & \multirow{-5}{*}{{\color[HTML]{000000} QARC}}       & {\color[HTML]{000000} 0.16}          & {\color[HTML]{000000} 0.11}          & {\color[HTML]{000000} 0.20}          \\ \hline
{\color[HTML]{000000} MRR}  & {\color[HTML]{000000} }                             & {\color[HTML]{000000} 0.48}          & {\color[HTML]{000000} 0.17}          & {\color[HTML]{000000} 0.44}          & {\color[HTML]{000000} }                             & {\color[HTML]{000000} \textbf{0.51}} & {\color[HTML]{000000} 0.17}          & {\color[HTML]{000000} \textbf{0.48}} \\ \cline{1-1} \cline{3-5} \cline{7-9} 
{\color[HTML]{000000} MAP}  & {\color[HTML]{000000} }                             & {\color[HTML]{000000} 0.12}          & {\color[HTML]{000000} \textbf{0.08}} & {\color[HTML]{000000} \textbf{0.11}} & {\color[HTML]{000000} }                             & {\color[HTML]{000000} \textbf{0.13}} & {\color[HTML]{000000} 0.08}          & {\color[HTML]{000000} \textbf{0.13}} \\ \cline{1-1} \cline{3-5} \cline{7-9} 
{\color[HTML]{000000} P@1}  & {\color[HTML]{000000} }                             & {\color[HTML]{000000} 0.44}          & {\color[HTML]{000000} 0.15}          & {\color[HTML]{000000} \textbf{0.37}} & {\color[HTML]{000000} }                             & {\color[HTML]{000000} \textbf{0.47}} & {\color[HTML]{000000} \textbf{0.15}} & {\color[HTML]{000000} \textbf{0.37}} \\ \cline{1-1} \cline{3-5} \cline{7-9} 
{\color[HTML]{000000} P@5}  & {\color[HTML]{000000} }                             & {\color[HTML]{000000} 0.15}          & {\color[HTML]{000000} 0.10}          & {\color[HTML]{000000} 0.15}          & {\color[HTML]{000000} }                             & {\color[HTML]{000000} 0.18}          & {\color[HTML]{000000} 0.08}          & {\color[HTML]{000000} 0.20}          \\ \cline{1-1} \cline{3-5} \cline{7-9} 
{\color[HTML]{000000} P@10} & \multirow{-5}{*}{{\color[HTML]{000000} QARCL}}      & {\color[HTML]{000000} 0.14}          & {\color[HTML]{000000} 0.11}          & {\color[HTML]{000000} 0.15}          & \multirow{-5}{*}{{\color[HTML]{000000} QARCL}}      & {\color[HTML]{000000} 0.16}          & {\color[HTML]{000000} 0.11}          & {\color[HTML]{000000} 0.20}          \\ \hline
{\color[HTML]{000000} MRR}  & {\color[HTML]{000000} }                             & {\color[HTML]{000000} 0.54}          & {\color[HTML]{000000} 0.18}          & {\color[HTML]{000000} 0.40}          & {\color[HTML]{000000} }                             & {\color[HTML]{000000} 0.49}          & {\color[HTML]{000000} 0.17}          & {\color[HTML]{000000} 0.40}          \\ \cline{1-1} \cline{3-5} \cline{7-9} 
{\color[HTML]{000000} MAP}  & {\color[HTML]{000000} }                             & {\color[HTML]{000000} \textbf{0.13}} & {\color[HTML]{000000} \textbf{0.08}} & {\color[HTML]{000000} 0.09}          & {\color[HTML]{000000} }                             & {\color[HTML]{000000} \textbf{0.13}} & {\color[HTML]{000000} 0.08}          & {\color[HTML]{000000} 0.12}          \\ \cline{1-1} \cline{3-5} \cline{7-9} 
{\color[HTML]{000000} P@1}  & {\color[HTML]{000000} }                             & {\color[HTML]{000000} \textbf{0.53}} & {\color[HTML]{000000} \textbf{0.16}} & {\color[HTML]{000000} 0.29}          & {\color[HTML]{000000} }                             & {\color[HTML]{000000} 0.42}          & {\color[HTML]{000000} \textbf{0.15}} & {\color[HTML]{000000} 0.28}          \\ \cline{1-1} \cline{3-5} \cline{7-9} 
{\color[HTML]{000000} P@5}  & {\color[HTML]{000000} }                             & {\color[HTML]{000000} 0.16}          & {\color[HTML]{000000} 0.10}          & {\color[HTML]{000000} 0.15}          & {\color[HTML]{000000} }                             & {\color[HTML]{000000} 0.19}          & {\color[HTML]{000000} 0.10}          & {\color[HTML]{000000} 0.18}          \\ \cline{1-1} \cline{3-5} \cline{7-9} 
{\color[HTML]{000000} P@10} & \multirow{-5}{*}{{\color[HTML]{000000} QD}}         & {\color[HTML]{000000} 0.14}          & {\color[HTML]{000000} 0.11}          & {\color[HTML]{000000} 0.17}          & \multirow{-5}{*}{{\color[HTML]{000000} QD}}         & {\color[HTML]{000000} 0.17}          & {\color[HTML]{000000} 0.10}          & {\color[HTML]{000000} 0.20}          \\ \hline
\end{tabular}
}
%\begin{tablenotes}
 %    \item In bold are the best average performances per algorithm per metric for each project and data granularity. "-files"=changeset-files. "P@"=top@. 
% \end{tablenotes}
 %\vspace{-1em}
\end{table}

Overall, both studied baseline techniques struggle to adapt to non-stationary data, except for RLOCATOR (with and without entropy) on SWT, and FLIM on Birt and AspectJ (top@10 metric). As detailed in Section \ref{sec:Baselines studies}, FLIM utilizes 19 features within its DL framework. Among these features is the Pre-released Bugs metric (refer to Section \ref{sec:Bug-inducing factors for bug localization}), which is scaled by the maximum number of Pre-released Bugs of the source code files. For the Birt and AspectJ projects, over 80\% of the source code files have Pre-released Bugs greater than 0. 
\begin{table}[t]
\caption{Algorithms evaluation across PDE and Zxing projects with different data granularity when incorporating bug-inducing factors on CL agents. In bold are the best average performances per algorithm per metric for each project and data granularity. "-files"=changeset-files. "P@"=top@.}
\label{tab:Algorithms evaluation across PDE and Zxing projects with different data granularity when incorporating bug-inducing factors on CL agents}
\resizebox{0.6\textwidth}{!}{
\centering
%put it here|c|ccccc|ccccc|ccccc|ccccc|ccccc|
\begin{tabular}{|c|c|c|c|c|c|c|}
\hline
     & \textbf{-files}            & PDE           & Zxing         & \textbf{hunks}             & PDE           & Zxing         \\ \hline
MRR  & \multirow{5}{*}{CLEAR+Reg} & \textbf{0.37} & 0.46          & \multirow{5}{*}{CLEAR+Reg} & \textbf{0.35} & 0.51          \\ \cline{1-1} \cline{3-4} \cline{6-7} 
MAP  &                            & 0.08          & 0.12          &                            & 0.09          & 0.12          \\ \cline{1-1} \cline{3-4} \cline{6-7} 
P@1  &                            & 0.08          & 0.12          &                            & 0.09          & 0.13          \\ \cline{1-1} \cline{3-4} \cline{6-7} 
P@5  &                            & 0.41          & \textbf{0.62} &                            & 0.40          & \textbf{0.62} \\ \cline{1-1} \cline{3-4} \cline{6-7} 
P@10 &                            & 0.77          & \textbf{1.00} &                            & 0.71          & \textbf{1.00} \\ \hline
MRR  & \multirow{5}{*}{EWC+Reg}   & 0.36          & \textbf{0.61} & \multirow{5}{*}{EWC+Reg}   & 0.33          & \textbf{0.56} \\ \cline{1-1} \cline{3-4} \cline{6-7} 
MAP  &                            & \textbf{0.09} & 0.12          &                            & 0.09          & 0.12          \\ \cline{1-1} \cline{3-4} \cline{6-7} 
P@1  &                            & 0.10          & 0.12          &                            & 0.09          & 0.11          \\ \cline{1-1} \cline{3-4} \cline{6-7} 
P@5  &                            & \textbf{0.43} & 0.61          &                            & \textbf{0.44} & 0.59          \\ \cline{1-1} \cline{3-4} \cline{6-7} 
P@10 &                            & \textbf{0.85} & 1.00          &                            & \textbf{0.87} & \textbf{1.00} \\ \hline
MRR  & \multirow{5}{*}{QARC}      & 0.17          & 0.24          & \multirow{5}{*}{QARC}      & 0.21          & 0.44          \\ \cline{1-1} \cline{3-4} \cline{6-7} 
MAP  &                            & 0.05          & 0.18          &                            & 0.08          & \textbf{0.29} \\ \cline{1-1} \cline{3-4} \cline{6-7} 
P@1  &                            & 0.07          & \textbf{0.20} &                            & 0.10          & \textbf{0.40} \\ \cline{1-1} \cline{3-4} \cline{6-7} 
P@5  &                            & 0.09          & 0.14          &                            & 0.17          & 0.31          \\ \cline{1-1} \cline{3-4} \cline{6-7} 
P@10 &                            & 0.12          & 0.34          &                            & 0.19          & 0.32          \\ \hline
MRR  & \multirow{5}{*}{QARCL}     & 0.26          & 0.35          & \multirow{5}{*}{QARCL}     & 0.26          & 0.47          \\ \cline{1-1} \cline{3-4} \cline{6-7} 
MAP  &                            & 0.08          & \textbf{0.21} &                            & 0.08          & \textbf{0.29} \\ \cline{1-1} \cline{3-4} \cline{6-7} 
P@1  &                            & \textbf{0.17} & \textbf{0.20} &                            & 0.17          & \textbf{0.40} \\ \cline{1-1} \cline{3-4} \cline{6-7} 
P@5  &                            & 0.11          & 0.39          &                            & 0.11          & 0.39          \\ \cline{1-1} \cline{3-4} \cline{6-7} 
P@10 &                            & 0.18          & 0.38          &                            & 0.18          & 0.42          \\ \hline
MRR  & \multirow{5}{*}{QD}        & 0.21          & 0.29          & \multirow{5}{*}{QD}        & 0.26          & 0.32          \\ \cline{1-1} \cline{3-4} \cline{6-7} 
MAP  &                            & 0.07          & 0.15          &                            & \textbf{0.10} & 0.14          \\ \cline{1-1} \cline{3-4} \cline{6-7} 
P@1  &                            & 0.13          & \textbf{0.20} &                            & \textbf{0.20} & 0.20          \\ \cline{1-1} \cline{3-4} \cline{6-7} 
P@5  &                            & 0.15          & 0.27          &                            & 0.18          & 0.29          \\ \cline{1-1} \cline{3-4} \cline{6-7} 
P@10 &                            & 0.15          & 0.38          &                            & 0.17          & 0.30          \\ \hline
\end{tabular}
}
%\begin{tablenotes}
 %    \item In bold are the best average performances per algorithm per metric for each project and data granularity. "-files"=changeset-files. "P@"=top@. 
 %\end{tablenotes}
 %\vspace{-1em}
\end{table}
This likely accounts for FLIM's accurate ranking of buggy changeset-files in non-stationary setting for these datasets. RLOCATOR (with and without entropy) can adapt to non-stationary data on SWT, possibly because SWT project has less amount of changeset-files (see Table \ref{tab:benchmark-dataset}), allowing the A2C algorithm employed to stabilize more easily despite the non-stationary nature of the data. Chen et al. \cite{chen2022adaptive} observed similar performances, where DRL models performed relatively well in non-stationary setting due to the limited data available for training in each episode. 
\begin{table*}[t]
\caption{Ablation study 1: different bug report metrics added to CLEAR. In bold are the best average performances per algorithm per metric for each project.}
\label{tab:Ablation study1: different bug report metrics added to CLEAR.}
\resizebox{0.75\textwidth}{!}{
\centering
%put it here|c|ccccc|ccccc|ccccc|ccccc|ccccc|
\begin{tabular}{|c|cccccccccc|}
\hline
\multicolumn{1}{|l|}{{\color[HTML]{000000} }} & \multicolumn{2}{c|}{{\color[HTML]{000000} Eclipse}} & \multicolumn{2}{c|}{{\color[HTML]{000000} SWT}} & \multicolumn{2}{c|}{{\color[HTML]{000000} Birt}} & \multicolumn{2}{c|}{{\color[HTML]{000000} AspectJ}} & \multicolumn{2}{c|}{{\color[HTML]{000000} Tomcat}} \\ \cline{2-11} 
\multicolumn{1}{|l|}{{\color[HTML]{000000} }} & \multicolumn{10}{l|}{{\color[HTML]{000000} top@1}} \\ \cline{2-11} 
\multicolumn{1}{|l|}{\multirow{-3}{*}{{\color[HTML]{000000} }}} & \multicolumn{1}{c|}{{\color[HTML]{000000} ST}} & \multicolumn{1}{c|}{{\color[HTML]{000000} NS}} & \multicolumn{1}{c|}{{\color[HTML]{000000} ST}} & \multicolumn{1}{c|}{{\color[HTML]{000000} NS}} & \multicolumn{1}{c|}{{\color[HTML]{000000} ST}} & \multicolumn{1}{c|}{{\color[HTML]{000000} NS}} & \multicolumn{1}{c|}{{\color[HTML]{000000} ST}} & \multicolumn{1}{c|}{{\color[HTML]{000000} NS}} & \multicolumn{1}{c|}{{\color[HTML]{000000} ST}} & {\color[HTML]{000000} NS} \\ \hline
{\color[HTML]{000000} CLEAR+Churn} & \multicolumn{1}{c|}{{\color[HTML]{000000} 0.04}} & \multicolumn{1}{c|}{{\color[HTML]{000000} 0.05}} & \multicolumn{1}{c|}{{\color[HTML]{000000} 0.05}} & \multicolumn{1}{c|}{{\color[HTML]{000000} 0.14}} & \multicolumn{1}{c|}{{\color[HTML]{000000} \textbf{0.07}}} & \multicolumn{1}{c|}{{\color[HTML]{000000} 0.06}} & \multicolumn{1}{c|}{{\color[HTML]{000000} 0.06}} & \multicolumn{1}{c|}{{\color[HTML]{000000} 0.03}} & \multicolumn{1}{c|}{{\color[HTML]{000000} 0.04}} & {\color[HTML]{000000} 0.04} \\ \hline
{\color[HTML]{000000} CLEAR+PRE} & \multicolumn{1}{c|}{{\color[HTML]{000000} \textbf{0.05}}} & \multicolumn{1}{c|}{{\color[HTML]{000000} \textbf{0.05}}} & \multicolumn{1}{c|}{{\color[HTML]{000000} 0.04}} & \multicolumn{1}{c|}{{\color[HTML]{000000} 0.11}} & \multicolumn{1}{c|}{{\color[HTML]{000000} 0.06}} & \multicolumn{1}{c|}{{\color[HTML]{000000} 0.06}} & \multicolumn{1}{c|}{{\color[HTML]{000000} \textbf{0.06}}} & \multicolumn{1}{c|}{{\color[HTML]{000000} \textbf{0.06}}} & \multicolumn{1}{c|}{{\color[HTML]{000000} 0.03}} & {\color[HTML]{000000} 0.04} \\ \hline
{\color[HTML]{000000} CLEAR+Reg.} & \multicolumn{1}{c|}{{\color[HTML]{000000} 0.04}} & \multicolumn{1}{c|}{{\color[HTML]{000000} 0.04}} & \multicolumn{1}{c|}{{\color[HTML]{000000} \textbf{0.05}}} & \multicolumn{1}{c|}{{\color[HTML]{000000} \textbf{0.13}}} & \multicolumn{1}{c|}{{\color[HTML]{000000} 0.05}} & \multicolumn{1}{c|}{{\color[HTML]{000000} \textbf{0.06}}} & \multicolumn{1}{c|}{{\color[HTML]{000000} 0.04}} & \multicolumn{1}{c|}{{\color[HTML]{000000} 0.04}} & \multicolumn{1}{c|}{{\color[HTML]{000000} \textbf{0.04}}} & {\color[HTML]{000000} \textbf{0.06}} \\ \hline
\multicolumn{1}{|l|}{{\color[HTML]{000000} }} & \multicolumn{10}{l|}{{\color[HTML]{000000} top@5}} \\ \cline{2-11} 
\multicolumn{1}{|l|}{\multirow{-2}{*}{{\color[HTML]{000000} }}} & \multicolumn{1}{c|}{{\color[HTML]{000000} ST}} & \multicolumn{1}{c|}{{\color[HTML]{000000} NS}} & \multicolumn{1}{c|}{{\color[HTML]{000000} ST}} & \multicolumn{1}{c|}{{\color[HTML]{000000} NS}} & \multicolumn{1}{c|}{{\color[HTML]{000000} ST}} & \multicolumn{1}{c|}{{\color[HTML]{000000} NS}} & \multicolumn{1}{c|}{{\color[HTML]{000000} ST}} & \multicolumn{1}{c|}{{\color[HTML]{000000} NS}} & \multicolumn{1}{c|}{{\color[HTML]{000000} ST}} & {\color[HTML]{000000} NS} \\ \hline
{\color[HTML]{000000} CLEAR+Churn} & \multicolumn{1}{c|}{{\color[HTML]{000000} \textbf{0.24}}} & \multicolumn{1}{c|}{{\color[HTML]{000000} 0.20}} & \multicolumn{1}{c|}{{\color[HTML]{000000} 0.23}} & \multicolumn{1}{c|}{{\color[HTML]{000000} \textbf{0.49}}} & \multicolumn{1}{c|}{{\color[HTML]{000000} 0.24}} & \multicolumn{1}{c|}{{\color[HTML]{000000} 0.23}} & \multicolumn{1}{c|}{{\color[HTML]{000000} 0.23}} & \multicolumn{1}{c|}{{\color[HTML]{000000} 0.20}} & \multicolumn{1}{c|}{{\color[HTML]{000000} 0.23}} & {\color[HTML]{000000} 0.22} \\ \hline
{\color[HTML]{000000} CLEAR+PRE} & \multicolumn{1}{c|}{{\color[HTML]{000000} 0.19}} & \multicolumn{1}{c|}{{\color[HTML]{000000} 0.18}} & \multicolumn{1}{c|}{{\color[HTML]{000000} 0.23}} & \multicolumn{1}{c|}{{\color[HTML]{000000} 0.47}} & \multicolumn{1}{c|}{{\color[HTML]{000000} 0.24}} & \multicolumn{1}{c|}{{\color[HTML]{000000} \textbf{0.29}}} & \multicolumn{1}{c|}{{\color[HTML]{000000} 0.23}} & \multicolumn{1}{c|}{{\color[HTML]{000000} \textbf{0.25}}} & \multicolumn{1}{c|}{{\color[HTML]{000000} 0.21}} & {\color[HTML]{000000} 0.27} \\ \hline
{\color[HTML]{000000} CLEAR+Reg.} & \multicolumn{1}{c|}{{\color[HTML]{000000} 0.21}} & \multicolumn{1}{c|}{{\color[HTML]{000000} \textbf{0.20}}} & \multicolumn{1}{c|}{{\color[HTML]{000000} \textbf{0.24}}} & \multicolumn{1}{c|}{{\color[HTML]{000000} 0.41}} & \multicolumn{1}{c|}{{\color[HTML]{000000} \textbf{0.25}}} & \multicolumn{1}{c|}{{\color[HTML]{000000} 0.25}} & \multicolumn{1}{c|}{{\color[HTML]{000000} \textbf{0.24}}} & \multicolumn{1}{c|}{{\color[HTML]{000000} 0.17}} & \multicolumn{1}{c|}{{\color[HTML]{000000} \textbf{0.21}}} & {\color[HTML]{000000} \textbf{0.29}} \\ \hline
\multicolumn{1}{|l|}{{\color[HTML]{000000} }} & \multicolumn{10}{l|}{{\color[HTML]{000000} top@10}} \\ \cline{2-11} 
\multicolumn{1}{|l|}{\multirow{-2}{*}{{\color[HTML]{000000} }}} & \multicolumn{1}{c|}{{\color[HTML]{000000} ST}} & \multicolumn{1}{c|}{{\color[HTML]{000000} NS}} & \multicolumn{1}{c|}{{\color[HTML]{000000} ST}} & \multicolumn{1}{c|}{{\color[HTML]{000000} NS}} & \multicolumn{1}{c|}{{\color[HTML]{000000} ST}} & \multicolumn{1}{c|}{{\color[HTML]{000000} NS}} & \multicolumn{1}{c|}{{\color[HTML]{000000} ST}} & \multicolumn{1}{c|}{{\color[HTML]{000000} NS}} & \multicolumn{1}{c|}{{\color[HTML]{000000} ST}} & {\color[HTML]{000000} NS} \\ \hline
{\color[HTML]{000000} CLEAR+Churn} & \multicolumn{1}{c|}{{\color[HTML]{000000} \textbf{0.44}}} & \multicolumn{1}{c|}{{\color[HTML]{000000} \textbf{0.42}}} & \multicolumn{1}{c|}{{\color[HTML]{000000} \textbf{0.44}}} & \multicolumn{1}{c|}{{\color[HTML]{000000} 0.82}} & \multicolumn{1}{c|}{{\color[HTML]{000000} \textbf{0.54}}} & \multicolumn{1}{c|}{{\color[HTML]{000000} 0.46}} & \multicolumn{1}{c|}{{\color[HTML]{000000} \textbf{0.48}}} & \multicolumn{1}{c|}{{\color[HTML]{000000} 0.40}} & \multicolumn{1}{c|}{{\color[HTML]{000000} \textbf{0.44}}} & {\color[HTML]{000000} 0.41} \\ \hline
{\color[HTML]{000000} CLEAR+PRE} & \multicolumn{1}{c|}{{\color[HTML]{000000} 0.35}} & \multicolumn{1}{c|}{{\color[HTML]{000000} 0.32}} & \multicolumn{1}{c|}{{\color[HTML]{000000} 0.43}} & \multicolumn{1}{c|}{{\color[HTML]{000000} 0.83}} & \multicolumn{1}{c|}{{\color[HTML]{000000} 0.47}} & \multicolumn{1}{c|}{{\color[HTML]{000000} \textbf{0.47}}} & \multicolumn{1}{c|}{{\color[HTML]{000000} 0.45}} & \multicolumn{1}{c|}{{\color[HTML]{000000} \textbf{0.49}}} & \multicolumn{1}{c|}{{\color[HTML]{000000} 0.39}} & {\color[HTML]{000000} \textbf{0.46}} \\ \hline
{\color[HTML]{000000} CLEAR+Reg.} & \multicolumn{1}{c|}{{\color[HTML]{000000} 0.35}} & \multicolumn{1}{c|}{{\color[HTML]{000000} 0.31}} & \multicolumn{1}{c|}{{\color[HTML]{000000} 0.43}} & \multicolumn{1}{c|}{{\color[HTML]{000000} \textbf{0.91}}} & \multicolumn{1}{c|}{{\color[HTML]{000000} 0.49}} & \multicolumn{1}{c|}{{\color[HTML]{000000} \textbf{0.47}}} & \multicolumn{1}{c|}{{\color[HTML]{000000} \textbf{0.48}}} & \multicolumn{1}{c|}{{\color[HTML]{000000} 0.37}} & \multicolumn{1}{c|}{{\color[HTML]{000000} 0.32}} & {\color[HTML]{000000} 0.43} \\ \hline
\multicolumn{1}{|l|}{{\color[HTML]{000000} }} & \multicolumn{10}{l|}{{\color[HTML]{000000} MAP}} \\ \cline{2-11} 
\multicolumn{1}{|l|}{\multirow{-2}{*}{{\color[HTML]{000000} }}} & \multicolumn{1}{c|}{{\color[HTML]{000000} ST}} & \multicolumn{1}{c|}{{\color[HTML]{000000} NS}} & \multicolumn{1}{c|}{{\color[HTML]{000000} ST}} & \multicolumn{1}{c|}{{\color[HTML]{000000} NS}} & \multicolumn{1}{c|}{{\color[HTML]{000000} ST}} & \multicolumn{1}{c|}{{\color[HTML]{000000} NS}} & \multicolumn{1}{c|}{{\color[HTML]{000000} ST}} & \multicolumn{1}{c|}{{\color[HTML]{000000} NS}} & \multicolumn{1}{c|}{{\color[HTML]{000000} ST}} & {\color[HTML]{000000} NS} \\ \hline
{\color[HTML]{000000} CLEAR+Churn} & \multicolumn{1}{c|}{{\color[HTML]{000000} 0.04}} & \multicolumn{1}{c|}{{\color[HTML]{000000} 0.04}} & \multicolumn{1}{c|}{{\color[HTML]{000000} 0.05}} & \multicolumn{1}{c|}{{\color[HTML]{000000} 0.09}} & \multicolumn{1}{c|}{{\color[HTML]{000000} 0.05}} & \multicolumn{1}{c|}{{\color[HTML]{000000} 0.04}} & \multicolumn{1}{c|}{{\color[HTML]{000000} 0.05}} & \multicolumn{1}{c|}{{\color[HTML]{000000} 0.04}} & \multicolumn{1}{c|}{{\color[HTML]{000000} 0.04}} & {\color[HTML]{000000} 0.04} \\ \hline
{\color[HTML]{000000} CLEAR+PRE} & \multicolumn{1}{c|}{{\color[HTML]{000000} 0.04}} & \multicolumn{1}{c|}{{\color[HTML]{000000} 0.04}} & \multicolumn{1}{c|}{{\color[HTML]{000000} 0.04}} & \multicolumn{1}{c|}{{\color[HTML]{000000} 0.08}} & \multicolumn{1}{c|}{{\color[HTML]{000000} 0.05}} & \multicolumn{1}{c|}{{\color[HTML]{000000} \textbf{0.06}}} & \multicolumn{1}{c|}{{\color[HTML]{000000} 0.05}} & \multicolumn{1}{c|}{{\color[HTML]{000000} \textbf{0.05}}} & \multicolumn{1}{c|}{{\color[HTML]{000000} 0.04}} & {\color[HTML]{000000} 0.05} \\ \hline
{\color[HTML]{000000} CLEAR+Reg.} & \multicolumn{1}{c|}{{\color[HTML]{000000} \textbf{0.04}}} & \multicolumn{1}{c|}{{\color[HTML]{000000} \textbf{0.04}}} & \multicolumn{1}{c|}{{\color[HTML]{000000} \textbf{0.05}}} & \multicolumn{1}{c|}{{\color[HTML]{000000} \textbf{0.09}}} & \multicolumn{1}{c|}{{\color[HTML]{000000} \textbf{0.05}}} & \multicolumn{1}{c|}{{\color[HTML]{000000} 0.05}} & \multicolumn{1}{c|}{{\color[HTML]{000000} \textbf{0.05}}} & \multicolumn{1}{c|}{{\color[HTML]{000000} 0.04}} & \multicolumn{1}{c|}{{\color[HTML]{000000} \textbf{0.04}}} & {\color[HTML]{000000} \textbf{0.06}} \\ \hline
\multicolumn{1}{|l|}{{\color[HTML]{000000} }} & \multicolumn{10}{l|}{{\color[HTML]{000000} MRR}} \\ \cline{2-11} 
\multicolumn{1}{|l|}{\multirow{-2}{*}{{\color[HTML]{000000} }}} & \multicolumn{1}{c|}{{\color[HTML]{000000} ST}} & \multicolumn{1}{c|}{{\color[HTML]{000000} NS}} & \multicolumn{1}{c|}{{\color[HTML]{000000} ST}} & \multicolumn{1}{c|}{{\color[HTML]{000000} NS}} & \multicolumn{1}{c|}{{\color[HTML]{000000} ST}} & \multicolumn{1}{c|}{{\color[HTML]{000000} NS}} & \multicolumn{1}{c|}{{\color[HTML]{000000} ST}} & \multicolumn{1}{c|}{{\color[HTML]{000000} NS}} & \multicolumn{1}{c|}{{\color[HTML]{000000} ST}} & {\color[HTML]{000000} NS} \\ \hline
{\color[HTML]{000000} CLEAR+Churn} & \multicolumn{1}{c|}{{\color[HTML]{000000} 0.26}} & \multicolumn{1}{c|}{{\color[HTML]{000000} 0.24}} & \multicolumn{1}{c|}{{\color[HTML]{000000} 0.23}} & \multicolumn{1}{c|}{{\color[HTML]{000000} 0.40}} & \multicolumn{1}{c|}{{\color[HTML]{000000} \textbf{0.44}}} & \multicolumn{1}{c|}{{\color[HTML]{000000} \textbf{0.42}}} & \multicolumn{1}{c|}{{\color[HTML]{000000} 0.33}} & \multicolumn{1}{c|}{{\color[HTML]{000000} 0.18}} & \multicolumn{1}{c|}{{\color[HTML]{000000} \textbf{0.42}}} & {\color[HTML]{000000} 0.23} \\ \hline
{\color[HTML]{000000} CLEAR+PRE} & \multicolumn{1}{c|}{{\color[HTML]{000000} 0.26}} & \multicolumn{1}{c|}{{\color[HTML]{000000} 0.22}} & \multicolumn{1}{c|}{{\color[HTML]{000000} \textbf{0.39}}} & \multicolumn{1}{c|}{{\color[HTML]{000000} 0.30}} & \multicolumn{1}{c|}{{\color[HTML]{000000} 0.36}} & \multicolumn{1}{c|}{{\color[HTML]{000000} 0.34}} & \multicolumn{1}{c|}{{\color[HTML]{000000} 0.36}} & \multicolumn{1}{c|}{{\color[HTML]{000000} \textbf{0.33}}} & \multicolumn{1}{c|}{{\color[HTML]{000000} 0.39}} & {\color[HTML]{000000} \textbf{0.28}} \\ \hline
{\color[HTML]{000000} CLEAR+Reg.} & \multicolumn{1}{c|}{{\color[HTML]{000000} \textbf{0.34}}} & \multicolumn{1}{c|}{{\color[HTML]{000000} \textbf{0.24}}} & \multicolumn{1}{c|}{{\color[HTML]{000000} 0.24}} & \multicolumn{1}{c|}{{\color[HTML]{000000} \textbf{0.41}}} & \multicolumn{1}{c|}{{\color[HTML]{000000} 0.36}} & \multicolumn{1}{c|}{{\color[HTML]{000000} 0.33}} & \multicolumn{1}{c|}{{\color[HTML]{000000} \textbf{0.39}}} & \multicolumn{1}{c|}{{\color[HTML]{000000} 0.26}} & \multicolumn{1}{c|}{{\color[HTML]{000000} 0.31}} & {\color[HTML]{000000} \textbf{0.28}} \\ \hline
\end{tabular}
}
%\begin{tablenotes}
%     \item In bold are the best average performances per algorithm per metric for each project. 
% \end{tablenotes}
 %\vspace{-1em}
\end{table*}
\begin{tcolorbox}
\textbf{Finding 1.}  At the changeset-files level, FLIM and RLOCATOR (with and without entropy) struggle to adapt to non-stationary data in most projects (three out of five) and metrics (four out of five). 
\end{tcolorbox}

\subsection{\textbf{RQ2: Can CL techniques improve bug localization?}} \label{sec:rq2}
\textbf{Motivation:} 
In \textbf{RQ1}, the results show that FLIM and RLOCATOR (i.e., our baseline studies) techniques have difficulties adapting to the concept drift associated with data. One way to solve it is to employ CL agents that address the concept drift associated with data by being sequentially trained on both stationary and non-stationary data at the changeset-files and hunks level. Therefore, the purpose of this RQ is to evaluate the performance of CL agents on both stationary and non-stationary data across all datasets.
\begin{table*}[t]
\caption{Ablation study 2: different bug report metrics added to EWC. In bold are the best average performances per algorithm per metric for each project.}
\label{tab:Ablation study2: different bug report metrics added to EWC}
\resizebox{0.75\textwidth}{!}{
\centering
%put it here|c|ccccc|ccccc|ccccc|ccccc|ccccc|
\begin{tabular}{|c|cccccccccc|}
\hline
{\color[HTML]{000000} } & \multicolumn{2}{c|}{{\color[HTML]{000000} Eclipse}} & \multicolumn{2}{c|}{{\color[HTML]{000000} SWT}} & \multicolumn{2}{c|}{{\color[HTML]{000000} Birt}} & \multicolumn{2}{c|}{{\color[HTML]{000000} AspectJ}} & \multicolumn{2}{c|}{{\color[HTML]{000000} Tomcat}} \\ \cline{2-11} 
{\color[HTML]{000000} } & \multicolumn{10}{l|}{{\color[HTML]{000000} top@1}} \\ \cline{2-11} 
\multirow{-3}{*}{{\color[HTML]{000000} }} & \multicolumn{1}{c|}{{\color[HTML]{000000} ST}} & \multicolumn{1}{c|}{{\color[HTML]{000000} NS}} & \multicolumn{1}{c|}{{\color[HTML]{000000} ST}} & \multicolumn{1}{c|}{{\color[HTML]{000000} NS}} & \multicolumn{1}{c|}{{\color[HTML]{000000} ST}} & \multicolumn{1}{c|}{{\color[HTML]{000000} NS}} & \multicolumn{1}{c|}{{\color[HTML]{000000} ST}} & \multicolumn{1}{c|}{{\color[HTML]{000000} NS}} & \multicolumn{1}{c|}{{\color[HTML]{000000} ST}} & {\color[HTML]{000000} NS} \\ \hline
{\color[HTML]{000000} EWC+Churn} & \multicolumn{1}{c|}{{\color[HTML]{000000} 0.03}} & \multicolumn{1}{c|}{{\color[HTML]{000000} 0.02}} & \multicolumn{1}{c|}{{\color[HTML]{000000} 0.04}} & \multicolumn{1}{c|}{{\color[HTML]{000000} \textbf{0.13}}} & \multicolumn{1}{c|}{{\color[HTML]{000000} 0.05}} & \multicolumn{1}{c|}{{\color[HTML]{000000} 0.05}} & \multicolumn{1}{c|}{{\color[HTML]{000000} \textbf{0.04}}} & \multicolumn{1}{c|}{{\color[HTML]{000000} 0.04}} & \multicolumn{1}{c|}{{\color[HTML]{000000} 0.04}} & {\color[HTML]{000000} 0.04} \\ \hline
{\color[HTML]{000000} EWC+PRE} & \multicolumn{1}{c|}{{\color[HTML]{000000} \textbf{0.04}}} & \multicolumn{1}{c|}{{\color[HTML]{000000} \textbf{0.04}}} & \multicolumn{1}{c|}{{\color[HTML]{000000} \textbf{0.05}}} & \multicolumn{1}{c|}{{\color[HTML]{000000} 0.06}} & \multicolumn{1}{c|}{{\color[HTML]{000000} 0.05}} & \multicolumn{1}{c|}{{\color[HTML]{000000} 0.03}} & \multicolumn{1}{c|}{{\color[HTML]{000000} \textbf{0.04}}} & \multicolumn{1}{c|}{{\color[HTML]{000000} 0.04}} & \multicolumn{1}{c|}{{\color[HTML]{000000} \textbf{0.05}}} & {\color[HTML]{000000} \textbf{0.05}} \\ \hline
{\color[HTML]{000000} EWC+Reg.} & \multicolumn{1}{c|}{{\color[HTML]{000000} \textbf{0.04}}} & \multicolumn{1}{c|}{{\color[HTML]{000000} \textbf{0.04}}} & \multicolumn{1}{c|}{{\color[HTML]{000000} \textbf{0.05}}} & \multicolumn{1}{c|}{{\color[HTML]{000000} \textbf{0.13}}} & \multicolumn{1}{c|}{{\color[HTML]{000000} \textbf{0.06}}} & \multicolumn{1}{c|}{{\color[HTML]{000000} \textbf{0.06}}} & \multicolumn{1}{c|}{{\color[HTML]{000000} \textbf{0.04}}} & \multicolumn{1}{c|}{{\color[HTML]{000000} \textbf{0.05}}} & \multicolumn{1}{c|}{{\color[HTML]{000000} 0.04}} & {\color[HTML]{000000} 0.04} \\ \hline
\multicolumn{1}{|l|}{{\color[HTML]{000000} }} & \multicolumn{10}{l|}{{\color[HTML]{000000} top@5}} \\ \cline{2-11} 
\multicolumn{1}{|l|}{\multirow{-2}{*}{{\color[HTML]{000000} }}} & \multicolumn{1}{c|}{{\color[HTML]{000000} ST}} & \multicolumn{1}{c|}{{\color[HTML]{000000} NS}} & \multicolumn{1}{c|}{{\color[HTML]{000000} ST}} & \multicolumn{1}{c|}{{\color[HTML]{000000} NS}} & \multicolumn{1}{c|}{{\color[HTML]{000000} ST}} & \multicolumn{1}{c|}{{\color[HTML]{000000} NS}} & \multicolumn{1}{c|}{{\color[HTML]{000000} ST}} & \multicolumn{1}{c|}{{\color[HTML]{000000} NS}} & \multicolumn{1}{c|}{{\color[HTML]{000000} ST}} & {\color[HTML]{000000} NS} \\ \hline
{\color[HTML]{000000} EWC+Churn} & \multicolumn{1}{c|}{{\color[HTML]{000000} \textbf{0.22}}} & \multicolumn{1}{c|}{{\color[HTML]{000000} \textbf{0.19}}} & \multicolumn{1}{c|}{{\color[HTML]{000000} 0.23}} & \multicolumn{1}{c|}{{\color[HTML]{000000} 0.41}} & \multicolumn{1}{c|}{{\color[HTML]{000000} 0.26}} & \multicolumn{1}{c|}{{\color[HTML]{000000} 0.19}} & \multicolumn{1}{c|}{{\color[HTML]{000000} 0.21}} & \multicolumn{1}{c|}{{\color[HTML]{000000} 0.20}} & \multicolumn{1}{c|}{{\color[HTML]{000000} 0.22}} & {\color[HTML]{000000} 0.25} \\ \hline
{\color[HTML]{000000} EWC+PRE} & \multicolumn{1}{c|}{{\color[HTML]{000000} 0.21}} & \multicolumn{1}{c|}{{\color[HTML]{000000} \textbf{0.19}}} & \multicolumn{1}{c|}{{\color[HTML]{000000} 0.22}} & \multicolumn{1}{c|}{{\color[HTML]{000000} 0.39}} & \multicolumn{1}{c|}{{\color[HTML]{000000} \textbf{0.27}}} & \multicolumn{1}{c|}{{\color[HTML]{000000} 0.19}} & \multicolumn{1}{c|}{{\color[HTML]{000000} 0.24}} & \multicolumn{1}{c|}{{\color[HTML]{000000} 0.20}} & \multicolumn{1}{c|}{{\color[HTML]{000000} \textbf{0.25}}} & {\color[HTML]{000000} \textbf{0.26}} \\ \hline
{\color[HTML]{000000} EWC+Reg.} & \multicolumn{1}{c|}{{\color[HTML]{000000} \textbf{0.22}}} & \multicolumn{1}{c|}{{\color[HTML]{000000} \textbf{0.19}}} & \multicolumn{1}{c|}{{\color[HTML]{000000} \textbf{0.25}}} & \multicolumn{1}{c|}{{\color[HTML]{000000} \textbf{0.47}}} & \multicolumn{1}{c|}{{\color[HTML]{000000} \textbf{0.27}}} & \multicolumn{1}{c|}{{\color[HTML]{000000} \textbf{0.20}}} & \multicolumn{1}{c|}{{\color[HTML]{000000} \textbf{0.26}}} & \multicolumn{1}{c|}{{\color[HTML]{000000} \textbf{0.23}}} & \multicolumn{1}{c|}{{\color[HTML]{000000} 0.20}} & {\color[HTML]{000000} 0.24} \\ \hline
\multicolumn{1}{|l|}{{\color[HTML]{000000} }} & \multicolumn{10}{l|}{{\color[HTML]{000000} top@10}} \\ \cline{2-11} 
\multicolumn{1}{|l|}{\multirow{-2}{*}{{\color[HTML]{000000} }}} & \multicolumn{1}{c|}{{\color[HTML]{000000} ST}} & \multicolumn{1}{c|}{{\color[HTML]{000000} NS}} & \multicolumn{1}{c|}{{\color[HTML]{000000} ST}} & \multicolumn{1}{c|}{{\color[HTML]{000000} NS}} & \multicolumn{1}{c|}{{\color[HTML]{000000} ST}} & \multicolumn{1}{c|}{{\color[HTML]{000000} NS}} & \multicolumn{1}{c|}{{\color[HTML]{000000} ST}} & \multicolumn{1}{c|}{{\color[HTML]{000000} NS}} & \multicolumn{1}{c|}{{\color[HTML]{000000} ST}} & {\color[HTML]{000000} NS} \\ \hline
{\color[HTML]{000000} EWC+Churn} & \multicolumn{1}{c|}{{\color[HTML]{000000} \textbf{0.44}}} & \multicolumn{1}{c|}{{\color[HTML]{000000} \textbf{0.40}}} & \multicolumn{1}{c|}{{\color[HTML]{000000} 0.45}} & \multicolumn{1}{c|}{{\color[HTML]{000000} 0.78}} & \multicolumn{1}{c|}{{\color[HTML]{000000} 0.52}} & \multicolumn{1}{c|}{{\color[HTML]{000000} 0.40}} & \multicolumn{1}{c|}{{\color[HTML]{000000} 0.43}} & \multicolumn{1}{c|}{{\color[HTML]{000000} 0.41}} & \multicolumn{1}{c|}{{\color[HTML]{000000} 0.44}} & {\color[HTML]{000000} 0.47} \\ \hline
{\color[HTML]{000000} EWC+PRE} & \multicolumn{1}{c|}{{\color[HTML]{000000} 0.43}} & \multicolumn{1}{c|}{{\color[HTML]{000000} 0.39}} & \multicolumn{1}{c|}{{\color[HTML]{000000} 0.44}} & \multicolumn{1}{c|}{{\color[HTML]{000000} 0.81}} & \multicolumn{1}{c|}{{\color[HTML]{000000} \textbf{0.55}}} & \multicolumn{1}{c|}{{\color[HTML]{000000} 0.40}} & \multicolumn{1}{c|}{{\color[HTML]{000000} 0.48}} & \multicolumn{1}{c|}{{\color[HTML]{000000} 0.42}} & \multicolumn{1}{c|}{{\color[HTML]{000000} \textbf{0.50}}} & {\color[HTML]{000000} \textbf{0.51}} \\ \hline
{\color[HTML]{000000} EWC+Reg.} & \multicolumn{1}{c|}{{\color[HTML]{000000} \textbf{0.44}}} & \multicolumn{1}{c|}{{\color[HTML]{000000} 0.38}} & \multicolumn{1}{c|}{{\color[HTML]{000000} \textbf{0.49}}} & \multicolumn{1}{c|}{{\color[HTML]{000000} \textbf{0.94}}} & \multicolumn{1}{c|}{{\color[HTML]{000000} 0.52}} & \multicolumn{1}{c|}{{\color[HTML]{000000} \textbf{0.41}}} & \multicolumn{1}{c|}{{\color[HTML]{000000} \textbf{0.49}}} & \multicolumn{1}{c|}{{\color[HTML]{000000} \textbf{0.44}}} & \multicolumn{1}{c|}{{\color[HTML]{000000} 0.42}} & {\color[HTML]{000000} 0.42} \\ \hline
\multicolumn{1}{|l|}{{\color[HTML]{000000} }} & \multicolumn{10}{l|}{{\color[HTML]{000000} MAP}} \\ \cline{2-11} 
\multicolumn{1}{|l|}{\multirow{-2}{*}{{\color[HTML]{000000} }}} & \multicolumn{1}{c|}{{\color[HTML]{000000} ST}} & \multicolumn{1}{c|}{{\color[HTML]{000000} NS}} & \multicolumn{1}{c|}{{\color[HTML]{000000} ST}} & \multicolumn{1}{c|}{{\color[HTML]{000000} NS}} & \multicolumn{1}{c|}{{\color[HTML]{000000} ST}} & \multicolumn{1}{c|}{{\color[HTML]{000000} NS}} & \multicolumn{1}{c|}{{\color[HTML]{000000} ST}} & \multicolumn{1}{c|}{{\color[HTML]{000000} NS}} & \multicolumn{1}{c|}{{\color[HTML]{000000} ST}} & {\color[HTML]{000000} NS} \\ \hline
{\color[HTML]{000000} EWC+Churn} & \multicolumn{1}{c|}{{\color[HTML]{000000} \textbf{0.04}}} & \multicolumn{1}{c|}{{\color[HTML]{000000} \textbf{0.04}}} & \multicolumn{1}{c|}{{\color[HTML]{000000} 0.04}} & \multicolumn{1}{c|}{{\color[HTML]{000000} 0.08}} & \multicolumn{1}{c|}{{\color[HTML]{000000} \textbf{0.05}}} & \multicolumn{1}{c|}{{\color[HTML]{000000} \textbf{0.04}}} & \multicolumn{1}{c|}{{\color[HTML]{000000} 0.04}} & \multicolumn{1}{c|}{{\color[HTML]{000000} \textbf{0.04}}} & \multicolumn{1}{c|}{{\color[HTML]{000000} 0.04}} & {\color[HTML]{000000} \textbf{0.05}} \\ \hline
{\color[HTML]{000000} EWC+PRE} & \multicolumn{1}{c|}{{\color[HTML]{000000} \textbf{0.04}}} & \multicolumn{1}{c|}{{\color[HTML]{000000} \textbf{0.04}}} & \multicolumn{1}{c|}{{\color[HTML]{000000} 0.04}} & \multicolumn{1}{c|}{{\color[HTML]{000000} 0.08}} & \multicolumn{1}{c|}{{\color[HTML]{000000} \textbf{0.05}}} & \multicolumn{1}{c|}{{\color[HTML]{000000} \textbf{0.04}}} & \multicolumn{1}{c|}{{\color[HTML]{000000} \textbf{0.05}}} & \multicolumn{1}{c|}{{\color[HTML]{000000} \textbf{0.04}}} & \multicolumn{1}{c|}{{\color[HTML]{000000} \textbf{0.05}}} & {\color[HTML]{000000} \textbf{0.05}} \\ \hline
{\color[HTML]{000000} EWC+Reg.} & \multicolumn{1}{c|}{{\color[HTML]{000000} \textbf{0.04}}} & \multicolumn{1}{c|}{{\color[HTML]{000000} \textbf{0.04}}} & \multicolumn{1}{c|}{{\color[HTML]{000000} \textbf{0.05}}} & \multicolumn{1}{c|}{{\color[HTML]{000000} \textbf{0.10}}} & \multicolumn{1}{c|}{{\color[HTML]{000000} \textbf{0.05}}} & \multicolumn{1}{c|}{{\color[HTML]{000000} \textbf{0.04}}} & \multicolumn{1}{c|}{{\color[HTML]{000000} \textbf{0.05}}} & \multicolumn{1}{c|}{{\color[HTML]{000000} \textbf{0.04}}} & \multicolumn{1}{c|}{{\color[HTML]{000000} 0.04}} & {\color[HTML]{000000} \textbf{0.05}} \\ \hline
\multicolumn{1}{|l|}{{\color[HTML]{000000} }} & \multicolumn{10}{l|}{{\color[HTML]{000000} MRR}} \\ \cline{2-11} 
\multicolumn{1}{|l|}{\multirow{-2}{*}{{\color[HTML]{000000} }}} & \multicolumn{1}{c|}{{\color[HTML]{000000} ST}} & \multicolumn{1}{c|}{{\color[HTML]{000000} NS}} & \multicolumn{1}{c|}{{\color[HTML]{000000} ST}} & \multicolumn{1}{c|}{{\color[HTML]{000000} NS}} & \multicolumn{1}{c|}{{\color[HTML]{000000} ST}} & \multicolumn{1}{c|}{{\color[HTML]{000000} NS}} & \multicolumn{1}{c|}{{\color[HTML]{000000} ST}} & \multicolumn{1}{c|}{{\color[HTML]{000000} NS}} & \multicolumn{1}{c|}{{\color[HTML]{000000} ST}} & {\color[HTML]{000000} NS} \\ \hline
{\color[HTML]{000000} EWC+Churn} & \multicolumn{1}{c|}{{\color[HTML]{000000} \textbf{0.31}}} & \multicolumn{1}{c|}{{\color[HTML]{000000} 0.23}} & \multicolumn{1}{c|}{{\color[HTML]{000000} 0.33}} & \multicolumn{1}{c|}{{\color[HTML]{000000} 0.44}} & \multicolumn{1}{c|}{{\color[HTML]{000000} 0.23}} & \multicolumn{1}{c|}{{\color[HTML]{000000} \textbf{0.38}}} & \multicolumn{1}{c|}{{\color[HTML]{000000} 0.36}} & \multicolumn{1}{c|}{{\color[HTML]{000000} 0.17}} & \multicolumn{1}{c|}{{\color[HTML]{000000} 0.25}} & {\color[HTML]{000000} \textbf{0.32}} \\ \hline
{\color[HTML]{000000} EWC+PRE} & \multicolumn{1}{c|}{{\color[HTML]{000000} \textbf{0.31}}} & \multicolumn{1}{c|}{{\color[HTML]{000000} \textbf{0.28}}} & \multicolumn{1}{c|}{{\color[HTML]{000000} 0.34}} & \multicolumn{1}{c|}{{\color[HTML]{000000} 0.44}} & \multicolumn{1}{c|}{{\color[HTML]{000000} 0.38}} & \multicolumn{1}{c|}{{\color[HTML]{000000} 0.32}} & \multicolumn{1}{c|}{{\color[HTML]{000000} 0.30}} & \multicolumn{1}{c|}{{\color[HTML]{000000} 0.17}} & \multicolumn{1}{c|}{{\color[HTML]{000000} \textbf{0.29}}} & {\color[HTML]{000000} 0.21} \\ \hline
{\color[HTML]{000000} EWC+Reg.} & \multicolumn{1}{c|}{{\color[HTML]{000000} 0.26}} & \multicolumn{1}{c|}{{\color[HTML]{000000} \textbf{0.28}}} & \multicolumn{1}{c|}{{\color[HTML]{000000} \textbf{0.36}}} & \multicolumn{1}{c|}{{\color[HTML]{000000} \textbf{0.48}}} & \multicolumn{1}{c|}{{\color[HTML]{000000} \textbf{0.45}}} & \multicolumn{1}{c|}{{\color[HTML]{000000} 0.26}} & \multicolumn{1}{c|}{{\color[HTML]{000000} \textbf{0.43}}} & \multicolumn{1}{c|}{{\color[HTML]{000000} \textbf{0.42}}} & \multicolumn{1}{c|}{{\color[HTML]{000000} 0.18}} & {\color[HTML]{000000} 0.26} \\ \hline
\end{tabular}
}
%\begin{tablenotes}
 %    \item In bold are the best average performances per algorithm per metric for each project. 
 %\end{tablenotes}
 %\vspace{-1em}
\end{table*}
Moreover, since CL agents may suffer from catastrophic forgetting \cite{mccloskey1989catastrophic}, we adopt rehearsal and regularization based techniques (CLEAR and EWC respectively) that mitigate catastrophic forgetting through policies that adjust distribution shifts in the data to be learned.\\
%\Amin{we need to motivate this RQ with referring to concept drift like the intro} \Paulina{Please check the revised paragraph}The purpose of this RQ is to evaluate the performance of CL agents on both stationary and non-stationary data across all datasets. In \textbf{RQ1}, empirical evaluation shows that current DL based techniques have difficulties adapting to non-stationary data. One way to solve it is sequential training on both stationary and non-stationary data. Mccloskey et al. \cite{mccloskey1989catastrophic} demonstrated that DL models can suffer catastrophic forgetting \Amin{why again catastrophic forgetting? .} \Paulina{Please view the revised sentences} \Paulina{Here since it is a sequential training, like in CL to adress concept drift, I think we can say catastrophic forgetting ?}when the model is trained sequentially on multiple tasks (e.g., stationary data then non-stationary data) as the weights in the model that are important for one task change to meet the objectives of the other tasks. To address this issue, we use CL techniques that mitigate catastrophic forgetting through a multitask training setup and policies that correct distribution shifts in the data to be learned. 
\begin{table*}[t]
\caption{Ablation study 3: different bug report metrics added to CLEAR. In bold are the best average performances per algorithm per metric for each project. In bold are the best average performances per algorithm per metric for each project.}
\label{tab:Ablation study3: different bug report metrics added to CLEAR}
\resizebox{0.75\textwidth}{!}{
\centering
%put it here|c|ccccc|ccccc|ccccc|ccccc|ccccc|
\begin{tabular}{|c|cccccccccc|}
\hline
                  & \multicolumn{2}{c|}{Eclipse}                                            & \multicolumn{2}{c|}{SWT}                                                & \multicolumn{2}{c|}{Birt}                                               & \multicolumn{2}{c|}{AspectJ}                                            & \multicolumn{2}{c|}{Tomcat}                        \\ \hline
\multirow{2}{*}{} & \multicolumn{10}{l|}{top@1}                                                                                                                                                                                                                                                                                                                                \\ \cline{2-11} 
                  & \multicolumn{1}{c|}{ST}            & \multicolumn{1}{c|}{NS}            & \multicolumn{1}{c|}{ST}            & \multicolumn{1}{c|}{NS}            & \multicolumn{1}{c|}{ST}            & \multicolumn{1}{c|}{NS}            & \multicolumn{1}{c|}{ST}            & \multicolumn{1}{c|}{NS}            & \multicolumn{1}{c|}{ST}            & NS            \\ \hline
CLEAR+LOC         & \multicolumn{1}{c|}{\textbf{0.04}} & \multicolumn{1}{c|}{\textbf{0.04}} & \multicolumn{1}{c|}{0.04}          & \multicolumn{1}{c|}{0.12}          & \multicolumn{1}{c|}{0.05}          & \multicolumn{1}{c|}{0.03}          & \multicolumn{1}{c|}{\textbf{0.04}} & \multicolumn{1}{c|}{0.02}          & \multicolumn{1}{c|}{0.05}          & 0.04          \\ \hline
CLEAR+MLOC        & \multicolumn{1}{c|}{\textbf{0.04}} & \multicolumn{1}{c|}{\textbf{0.04}} & \multicolumn{1}{c|}{0.04}          & \multicolumn{1}{c|}{0.12}          & \multicolumn{1}{c|}{\textbf{0.07}} & \multicolumn{1}{c|}{0.03}          & \multicolumn{1}{c|}{\textbf{0.04}} & \multicolumn{1}{c|}{0.03}          & \multicolumn{1}{c|}{0.04}          & 0.08          \\ \hline
CLEAR+VG          & \multicolumn{1}{c|}{\textbf{0.04}} & \multicolumn{1}{c|}{\textbf{0.04}} & \multicolumn{1}{c|}{\textbf{0.05}} & \multicolumn{1}{c|}{0.12}          & \multicolumn{1}{c|}{0.06}          & \multicolumn{1}{c|}{0.04}          & \multicolumn{1}{c|}{\textbf{0.04}} & \multicolumn{1}{c|}{\textbf{0.04}} & \multicolumn{1}{c|}{0.05}          & \textbf{0.09} \\ \hline
CLEAR+ALL         & \multicolumn{1}{c|}{\textbf{0.04}} & \multicolumn{1}{c|}{\textbf{0.04}} & \multicolumn{1}{c|}{\textbf{0.05}} & \multicolumn{1}{c|}{0.10}          & \multicolumn{1}{c|}{0.05}          & \multicolumn{1}{c|}{0.04}          & \multicolumn{1}{c|}{\textbf{0.04}} & \multicolumn{1}{c|}{0.03}          & \multicolumn{1}{c|}{\textbf{0.06}} & 0.05          \\ \hline
CLEAR+Reg.        & \multicolumn{1}{c|}{\textbf{0.04}} & \multicolumn{1}{c|}{\textbf{0.04}} & \multicolumn{1}{c|}{\textbf{0.05}} & \multicolumn{1}{c|}{\textbf{0.13}} & \multicolumn{1}{c|}{0.05}          & \multicolumn{1}{c|}{\textbf{0.06}} & \multicolumn{1}{c|}{\textbf{0.04}} & \multicolumn{1}{c|}{\textbf{0.04}} & \multicolumn{1}{c|}{\textbf{0.04}} & 0.06          \\ \hline
\multirow{2}{*}{} & \multicolumn{10}{l|}{top@5}                                                                                                                                                                                                                                                                                                                                \\ \cline{2-11} 
                  & \multicolumn{1}{c|}{ST}            & \multicolumn{1}{c|}{NS}            & \multicolumn{1}{c|}{ST}            & \multicolumn{1}{c|}{NS}            & \multicolumn{1}{c|}{ST}            & \multicolumn{1}{c|}{NS}            & \multicolumn{1}{c|}{ST}            & \multicolumn{1}{c|}{NS}            & \multicolumn{1}{c|}{ST}            & NS            \\ \hline
CLEAR+LOC         & \multicolumn{1}{c|}{\textbf{0.23}} & \multicolumn{1}{c|}{0.23}          & \multicolumn{1}{c|}{0.23}          & \multicolumn{1}{c|}{0.36}          & \multicolumn{1}{c|}{\textbf{0.28}} & \multicolumn{1}{c|}{0.18}          & \multicolumn{1}{c|}{0.21}          & \multicolumn{1}{c|}{0.14}          & \multicolumn{1}{c|}{0.22}          & 0.21          \\ \hline
CLEAR+MLOC        & \multicolumn{1}{c|}{0.22}          & \multicolumn{1}{c|}{0.17}          & \multicolumn{1}{c|}{0.21}          & \multicolumn{1}{c|}{0.40}          & \multicolumn{1}{c|}{0.26}          & \multicolumn{1}{c|}{0.20}          & \multicolumn{1}{c|}{0.23}          & \multicolumn{1}{c|}{0.18}          & \multicolumn{1}{c|}{\textbf{0.24}} & \textbf{0.30} \\ \hline
CLEAR+VG          & \multicolumn{1}{c|}{0.20}          & \multicolumn{1}{c|}{0.19}          & \multicolumn{1}{c|}{0.23}          & \multicolumn{1}{c|}{\textbf{0.52}} & \multicolumn{1}{c|}{0.27}          & \multicolumn{1}{c|}{\textbf{0.29}} & \multicolumn{1}{c|}{0.22}          & \multicolumn{1}{c|}{\textbf{0.27}} & \multicolumn{1}{c|}{0.20}          & 0.29          \\ \hline
CLEAR+ALL         & \multicolumn{1}{c|}{\textbf{0.23}} & \multicolumn{1}{c|}{\textbf{0.20}} & \multicolumn{1}{c|}{\textbf{0.24}} & \multicolumn{1}{c|}{0.35}          & \multicolumn{1}{c|}{0.25}          & \multicolumn{1}{c|}{0.23}          & \multicolumn{1}{c|}{0.20}          & \multicolumn{1}{c|}{0.22}          & \multicolumn{1}{c|}{0.22}          & 0.26          \\ \hline
CLEAR+Reg.        & \multicolumn{1}{c|}{0.21}          & \multicolumn{1}{c|}{\textbf{0.20}} & \multicolumn{1}{c|}{\textbf{0.24}} & \multicolumn{1}{c|}{0.41}          & \multicolumn{1}{c|}{0.25}          & \multicolumn{1}{c|}{0.25}          & \multicolumn{1}{c|}{\textbf{0.24}} & \multicolumn{1}{c|}{0.17}          & \multicolumn{1}{c|}{0.21}          & 0.29          \\ \hline
\multirow{2}{*}{} & \multicolumn{10}{l|}{top@10}                                                                                                                                                                                                                                                                                                                               \\ \cline{2-11} 
                  & \multicolumn{1}{c|}{ST}            & \multicolumn{1}{c|}{NS}            & \multicolumn{1}{c|}{ST}            & \multicolumn{1}{c|}{NS}            & \multicolumn{1}{c|}{ST}            & \multicolumn{1}{c|}{NS}            & \multicolumn{1}{c|}{ST}            & \multicolumn{1}{c|}{NS}            & \multicolumn{1}{c|}{ST}            & NS            \\ \hline
CLEAR+LOC         & \multicolumn{1}{c|}{0.43}          & \multicolumn{1}{c|}{\textbf{0.44}} & \multicolumn{1}{c|}{0.45}          & \multicolumn{1}{c|}{0.79}          & \multicolumn{1}{c|}{0.53}          & \multicolumn{1}{c|}{0.35}          & \multicolumn{1}{c|}{0.47}          & \multicolumn{1}{c|}{0.33}          & \multicolumn{1}{c|}{\textbf{0.45}} & 0.40          \\ \hline
CLEAR+MLOC        & \multicolumn{1}{c|}{\textbf{0.47}} & \multicolumn{1}{c|}{0.39}          & \multicolumn{1}{c|}{0.43}          & \multicolumn{1}{c|}{0.75}          & \multicolumn{1}{c|}{\textbf{0.54}} & \multicolumn{1}{c|}{0.42}          & \multicolumn{1}{c|}{0.45}          & \multicolumn{1}{c|}{0.38}          & \multicolumn{1}{c|}{0.44}          & \textbf{0.51} \\ \hline
CLEAR+VG          & \multicolumn{1}{c|}{0.35}          & \multicolumn{1}{c|}{0.34}          & \multicolumn{1}{c|}{0.43}          & \multicolumn{1}{c|}{0.87}          & \multicolumn{1}{c|}{0.43}          & \multicolumn{1}{c|}{\textbf{0.54}} & \multicolumn{1}{c|}{0.40}          & \multicolumn{1}{c|}{0.45}          & \multicolumn{1}{c|}{0.28}          & 0.43          \\ \hline
CLEAR+ALL         & \multicolumn{1}{c|}{0.45}          & \multicolumn{1}{c|}{0.42}          & \multicolumn{1}{c|}{\textbf{0.46}} & \multicolumn{1}{c|}{0.73}          & \multicolumn{1}{c|}{0.50}          & \multicolumn{1}{c|}{0.42}          & \multicolumn{1}{c|}{0.44}          & \multicolumn{1}{c|}{\textbf{0.48}} & \multicolumn{1}{c|}{0.42}          & 0.47          \\ \hline
CLEAR+Reg.        & \multicolumn{1}{c|}{0.35}          & \multicolumn{1}{c|}{0.31}          & \multicolumn{1}{c|}{0.43}          & \multicolumn{1}{c|}{\textbf{0.91}} & \multicolumn{1}{c|}{0.49}          & \multicolumn{1}{c|}{0.47}          & \multicolumn{1}{c|}{\textbf{0.48}} & \multicolumn{1}{c|}{0.37}          & \multicolumn{1}{c|}{0.32}          & 0.43          \\ \hline
\multirow{2}{*}{} & \multicolumn{10}{l|}{MAP}                                                                                                                                                                                                                                                                                                                                  \\ \cline{2-11} 
                  & \multicolumn{1}{c|}{ST}            & \multicolumn{1}{c|}{NS}            & \multicolumn{1}{c|}{ST}            & \multicolumn{1}{c|}{NS}            & \multicolumn{1}{c|}{ST}            & \multicolumn{1}{c|}{NS}            & \multicolumn{1}{c|}{ST}            & \multicolumn{1}{c|}{NS}            & \multicolumn{1}{c|}{ST}            & NS            \\ \hline
CLEAR+LOC         & \multicolumn{1}{c|}{0.04}          & \multicolumn{1}{c|}{\textbf{0.05}} & \multicolumn{1}{c|}{\textbf{0.05}} & \multicolumn{1}{c|}{\textbf{0.09}} & \multicolumn{1}{c|}{\textbf{0.05}} & \multicolumn{1}{c|}{0.04}          & \multicolumn{1}{c|}{\textbf{0.05}} & \multicolumn{1}{c|}{0.03}          & \multicolumn{1}{c|}{0.04}          & 0.04          \\ \hline
CLEAR+MLOC        & \multicolumn{1}{c|}{0.04}          & \multicolumn{1}{c|}{0.04}          & \multicolumn{1}{c|}{0.04}          & \multicolumn{1}{c|}{0.07}          & \multicolumn{1}{c|}{\textbf{0.05}} & \multicolumn{1}{c|}{0.04}          & \multicolumn{1}{c|}{\textbf{0.05}} & \multicolumn{1}{c|}{0.04}          & \multicolumn{1}{c|}{\textbf{0.04}} & 0.05          \\ \hline
CLEAR+VG          & \multicolumn{1}{c|}{0.04}          & \multicolumn{1}{c|}{0.04}          & \multicolumn{1}{c|}{0.04}          & \multicolumn{1}{c|}{\textbf{0.09}} & \multicolumn{1}{c|}{\textbf{0.05}} & \multicolumn{1}{c|}{\textbf{0.05}} & \multicolumn{1}{c|}{\textbf{0.05}} & \multicolumn{1}{c|}{\textbf{0.10}} & \multicolumn{1}{c|}{\textbf{0.04}} & \textbf{0.08} \\ \hline
CLEAR+ALL         & \multicolumn{1}{c|}{\textbf{0.04}} & \multicolumn{1}{c|}{0.04}          & \multicolumn{1}{c|}{\textbf{0.05}} & \multicolumn{1}{c|}{\textbf{0.09}} & \multicolumn{1}{c|}{\textbf{0.05}} & \multicolumn{1}{c|}{0.04}          & \multicolumn{1}{c|}{0.04}          & \multicolumn{1}{c|}{0.04}          & \multicolumn{1}{c|}{\textbf{0.04}} & 0.05          \\ \hline
CLEAR+Reg.        & \multicolumn{1}{c|}{\textbf{0.04}} & \multicolumn{1}{c|}{\textbf{0.04}} & \multicolumn{1}{c|}{\textbf{0.05}} & \multicolumn{1}{c|}{\textbf{0.09}} & \multicolumn{1}{c|}{\textbf{0.05}} & \multicolumn{1}{c|}{\textbf{0.05}} & \multicolumn{1}{c|}{\textbf{0.05}} & \multicolumn{1}{c|}{0.04}          & \multicolumn{1}{c|}{\textbf{0.04}} & 0.06          \\ \hline
\multirow{2}{*}{} & \multicolumn{10}{l|}{MRR}                                                                                                                                                                                                                                                                                                                                  \\ \cline{2-11} 
                  & \multicolumn{1}{c|}{ST}            & \multicolumn{1}{c|}{NS}            & \multicolumn{1}{c|}{ST}            & \multicolumn{1}{c|}{NS}            & \multicolumn{1}{c|}{ST}            & \multicolumn{1}{c|}{NS}            & \multicolumn{1}{c|}{ST}            & \multicolumn{1}{c|}{NS}            & \multicolumn{1}{c|}{ST}            & NS            \\ \hline
CLEAR+LOC         & \multicolumn{1}{c|}{0.33}          & \multicolumn{1}{c|}{0.37}          & \multicolumn{1}{c|}{0.26}          & \multicolumn{1}{c|}{0.41}          & \multicolumn{1}{c|}{0.27}          & \multicolumn{1}{c|}{0.26}          & \multicolumn{1}{c|}{0.35}          & \multicolumn{1}{c|}{0.34}          & \multicolumn{1}{c|}{0.25}          & 0.24          \\ \hline
CLEAR+MLOC        & \multicolumn{1}{c|}{0.31}          & \multicolumn{1}{c|}{\textbf{0.40}} & \multicolumn{1}{c|}{0.20}          & \multicolumn{1}{c|}{\textbf{0.48}} & \multicolumn{1}{c|}{\textbf{0.51}} & \multicolumn{1}{c|}{0.22}          & \multicolumn{1}{c|}{\textbf{0.39}} & \multicolumn{1}{c|}{0.30}          & \multicolumn{1}{c|}{0.32}          & 0.24          \\ \hline
CLEAR+VG          & \multicolumn{1}{c|}{0.33}          & \multicolumn{1}{c|}{0.23}          & \multicolumn{1}{c|}{\textbf{0.42}} & \multicolumn{1}{c|}{0.36}          & \multicolumn{1}{c|}{0.32}          & \multicolumn{1}{c|}{\textbf{0.36}} & \multicolumn{1}{c|}{0.28}          & \multicolumn{1}{c|}{\textbf{0.39}} & \multicolumn{1}{c|}{0.26}          & \textbf{0.37} \\ \hline
CLEAR+ALL         & \multicolumn{1}{c|}{0.30}          & \multicolumn{1}{c|}{0.28}          & \multicolumn{1}{c|}{0.32}          & \multicolumn{1}{c|}{0.24}          & \multicolumn{1}{c|}{0.36}          & \multicolumn{1}{c|}{0.31}          & \multicolumn{1}{c|}{0.32}          & \multicolumn{1}{c|}{0.24}          & \multicolumn{1}{c|}{\textbf{0.33}} & 0.29          \\ \hline
CLEAR+Reg.        & \multicolumn{1}{c|}{\textbf{0.34}} & \multicolumn{1}{c|}{0.24}          & \multicolumn{1}{c|}{0.24}          & \multicolumn{1}{c|}{0.41}          & \multicolumn{1}{c|}{0.36}          & \multicolumn{1}{c|}{0.33}          & \multicolumn{1}{c|}{\textbf{0.39}} & \multicolumn{1}{c|}{0.26}          & \multicolumn{1}{c|}{0.31}          & \textbf{0.28} \\ \hline
\end{tabular}
}
%\begin{tablenotes}
%     \item In bold are the best average performances per algorithm per metric for each project. 
% \end{tablenotes}
 %\vspace{-1em}
\end{table*}

\textbf{Method:} We calculate our evaluation metrics for each dataset and compare the performance of the CL agents against all baselines and each other (CLEAR vs. EWC). Additionally, for each CL agent, we calculate the performance differences between stationary and non-stationary data. Our goal is to ensure that the agents effectively localize bugs in both types of environments.\\
\textbf{Results: }Table \ref{tab:Tomcat}, \ref{tab:Birt}, \ref{tab:Eclipse}, \ref{tab:AspectJ}, and \ref{tab:SWT} show the performance of the CL agents (i.e., CLEAR and EWC) on stationary and non-stationary data (i.e., changeset-files and hunks) across the studied datasets against FLIM and RLOCATOR. CLEAR and EWC show up to a 35\% performance improvement over RLOCATOR (with and without entropy) on the stationary data of the AspectJ project, measured by top@5, top@10, MRR, and MAP metrics. In contrast, for other projects (i.e., Tomcat, Birt, SWT, and Eclipse), CLEAR and EWC perform similarly to RLOCATOR (with and without entropy) in top@1, top@5, and MAP metrics. 
%\Amin{compared to RLOCATOR? any explanation? with this, how can we argue for good performance of CL agents? do I miss anything?}.
The CL agents show better or competitive performance compared to RLOCATOR (with and without entropy),  indicating their capability to retain their effectiveness on stationary data even when the software project undergoes significant changes (non-stationary data) and concept drift. This is due to the CL agents' capability of protecting knowledge from previous cycles during training. %\Amin{as I understand, we are implying forgetting things by the model not handling concept drift} \Paulina{I do not understand. My point here is while handling concept drift by being trained sequentially, the CL agents protect the knowledge learned previously}.
%CLEAR significantly outperforms RLOCATOR with MRR metric on Tomcat project, and RLOCATOR with Entropy with top@10 metric on Birt and AspectJ projects. 
%Further EWC significantly outperforms RLOCATOR with MRR metric on Tomcat project, and RLOCATOR with Entropy with top@10 metric on  AspectJ projects. 
%On the SWT project, the performance between  CLEAR, EWC, FLIM  and RLOCATOR(with and without entropy) with all metrics is not significant, due to the limited non-stationary data available, nevertheless, CLEAR and EWC show better performance on average.
\begin{table*}[t]
\caption{Ablation study 4: different bug report metrics added to EWC. In bold are the best average performances per algorithm per metric for each project.}
\label{tab:Ablation study4: different bug report metrics added to EWC}
\resizebox{0.75\textwidth}{!}{
\centering
%put it here|c|ccccc|ccccc|ccccc|ccccc|ccccc|
\begin{tabular}{|c|cccccccccc|}
\hline
                                        & \multicolumn{2}{c|}{Eclipse}                                            & \multicolumn{2}{c|}{SWT}                                                & \multicolumn{2}{c|}{Birt}                                               & \multicolumn{2}{c|}{AspectJ}                                            & \multicolumn{2}{c|}{Tomcat}                        \\ \hline
\multirow{2}{*}{}                       & \multicolumn{10}{l|}{top@1}                                                                                                                                                                                                                                                                                                                                \\ \cline{2-11} 
                                        & \multicolumn{1}{c|}{ST}            & \multicolumn{1}{c|}{NS}            & \multicolumn{1}{c|}{ST}            & \multicolumn{1}{c|}{NS}            & \multicolumn{1}{c|}{ST}            & \multicolumn{1}{c|}{NS}            & \multicolumn{1}{c|}{ST}            & \multicolumn{1}{c|}{NS}            & \multicolumn{1}{c|}{ST}            & NS            \\ \hline
EWC+LOC                                 & \multicolumn{1}{c|}{\textbf{0.04}} & \multicolumn{1}{c|}{\textbf{0.04}} & \multicolumn{1}{c|}{\textbf{0.05}} & \multicolumn{1}{c|}{\textbf{0.21}} & \multicolumn{1}{c|}{0.04}          & \multicolumn{1}{c|}{0.05}          & \multicolumn{1}{c|}{0.05}          & \multicolumn{1}{c|}{0.04}          & \multicolumn{1}{c|}{0.04}          & 0.04          \\ \hline
EWC+MLOC                                & \multicolumn{1}{c|}{\textbf{0.04}} & \multicolumn{1}{c|}{0.03}          & \multicolumn{1}{c|}{\textbf{0.05}} & \multicolumn{1}{c|}{0.17}          & \multicolumn{1}{c|}{0.05}          & \multicolumn{1}{c|}{\textbf{0.06}} & \multicolumn{1}{c|}{0.05}          & \multicolumn{1}{c|}{0.04}          & \multicolumn{1}{c|}{0.04}          & 0.04          \\ \hline
EWC+VG                                  & \multicolumn{1}{c|}{\textbf{0.04}} & \multicolumn{1}{c|}{0.03}          & \multicolumn{1}{c|}{0.04}          & \multicolumn{1}{c|}{0.17}          & \multicolumn{1}{c|}{0.05}          & \multicolumn{1}{c|}{0.05}          & \multicolumn{1}{c|}{0.04}          & \multicolumn{1}{c|}{0.04}          & \multicolumn{1}{c|}{\textbf{0.05}} & \textbf{0.05} \\ \hline
EWC+ALL                                 & \multicolumn{1}{c|}{\textbf{0.04}} & \multicolumn{1}{c|}{0.03}          & \multicolumn{1}{c|}{\textbf{0.05}} & \multicolumn{1}{c|}{0.13}          & \multicolumn{1}{c|}{0.04}          & \multicolumn{1}{c|}{0.03}          & \multicolumn{1}{c|}{\textbf{0.06}} & \multicolumn{1}{c|}{0.02}          & \multicolumn{1}{c|}{\textbf{0.05}} & 0.03          \\ \hline
EWC+Reg.                                & \multicolumn{1}{c|}{\textbf{0.04}} & \multicolumn{1}{c|}{\textbf{0.04}} & \multicolumn{1}{c|}{\textbf{0.05}} & \multicolumn{1}{c|}{0.13}          & \multicolumn{1}{c|}{\textbf{0.06}} & \multicolumn{1}{c|}{\textbf{0.06}} & \multicolumn{1}{c|}{0.04}          & \multicolumn{1}{c|}{\textbf{0.05}} & \multicolumn{1}{c|}{0.04}          & 0.04          \\ \hline
\multicolumn{1}{|l|}{\multirow{2}{*}{}} & \multicolumn{10}{l|}{top@5}                                                                                                                                                                                                                                                                                                                                \\ \cline{2-11} 
\multicolumn{1}{|l|}{}                  & \multicolumn{1}{c|}{ST}            & \multicolumn{1}{c|}{NS}            & \multicolumn{1}{c|}{ST}            & \multicolumn{1}{c|}{NS}            & \multicolumn{1}{c|}{ST}            & \multicolumn{1}{c|}{NS}            & \multicolumn{1}{c|}{ST}            & \multicolumn{1}{c|}{NS}            & \multicolumn{1}{c|}{ST}            & NS            \\ \hline
EWC+LOC                                 & \multicolumn{1}{c|}{\textbf{0.24}} & \multicolumn{1}{c|}{0.16}          & \multicolumn{1}{c|}{0.23}          & \multicolumn{1}{c|}{\textbf{0.57}} & \multicolumn{1}{c|}{\textbf{0.27}} & \multicolumn{1}{c|}{0.18}          & \multicolumn{1}{c|}{0.25}          & \multicolumn{1}{c|}{0.21}          & \multicolumn{1}{c|}{\textbf{0.24}} & 0.21          \\ \hline
EWC+MLOC                                & \multicolumn{1}{c|}{0.23}          & \multicolumn{1}{c|}{0.18}          & \multicolumn{1}{c|}{0.22}          & \multicolumn{1}{c|}{0.44}          & \multicolumn{1}{c|}{0.26}          & \multicolumn{1}{c|}{\textbf{0.20}} & \multicolumn{1}{c|}{0.25}          & \multicolumn{1}{c|}{0.21}          & \multicolumn{1}{c|}{0.22}          & 0.18          \\ \hline
EWC+VG                                  & \multicolumn{1}{c|}{0.19}          & \multicolumn{1}{c|}{0.18}          & \multicolumn{1}{c|}{0.22}          & \multicolumn{1}{c|}{0.50}          & \multicolumn{1}{c|}{0.25}          & \multicolumn{1}{c|}{0.19}          & \multicolumn{1}{c|}{0.23}          & \multicolumn{1}{c|}{0.21}          & \multicolumn{1}{c|}{\textbf{0.24}} & 0.21          \\ \hline
EWC+ALL                                 & \multicolumn{1}{c|}{0.22}          & \multicolumn{1}{c|}{\textbf{0.20}} & \multicolumn{1}{c|}{0.21}          & \multicolumn{1}{c|}{0.52}          & \multicolumn{1}{c|}{0.23}          & \multicolumn{1}{c|}{\textbf{0.20}} & \multicolumn{1}{c|}{0.24}          & \multicolumn{1}{c|}{0.22}          & \multicolumn{1}{c|}{0.21}          & 0.22          \\ \hline
EWC+Reg.                                & \multicolumn{1}{c|}{0.22}          & \multicolumn{1}{c|}{0.19}          & \multicolumn{1}{c|}{\textbf{0.25}} & \multicolumn{1}{c|}{0.47}          & \multicolumn{1}{c|}{\textbf{0.27}} & \multicolumn{1}{c|}{\textbf{0.20}} & \multicolumn{1}{c|}{\textbf{0.26}} & \multicolumn{1}{c|}{\textbf{0.23}} & \multicolumn{1}{c|}{0.20}          & \textbf{0.24} \\ \hline
\multicolumn{1}{|l|}{\multirow{2}{*}{}} & \multicolumn{10}{l|}{top@10}                                                                                                                                                                                                                                                                                                                               \\ \cline{2-11} 
\multicolumn{1}{|l|}{}                  & \multicolumn{1}{c|}{ST}            & \multicolumn{1}{c|}{NS}            & \multicolumn{1}{c|}{ST}            & \multicolumn{1}{c|}{NS}            & \multicolumn{1}{c|}{ST}            & \multicolumn{1}{c|}{NS}            & \multicolumn{1}{c|}{ST}            & \multicolumn{1}{c|}{NS}            & \multicolumn{1}{c|}{ST}            & NS            \\ \hline
EWC+LOC                                 & \multicolumn{1}{c|}{\textbf{0.45}} & \multicolumn{1}{c|}{0.38}          & \multicolumn{1}{c|}{0.46}          & \multicolumn{1}{c|}{0.97}          & \multicolumn{1}{c|}{0.51}          & \multicolumn{1}{c|}{0.39}          & \multicolumn{1}{c|}{0.47}          & \multicolumn{1}{c|}{0.38}          & \multicolumn{1}{c|}{\textbf{0.48}} & 0.44          \\ \hline
EWC+MLOC                                & \multicolumn{1}{c|}{\textbf{0.45}} & \multicolumn{1}{c|}{\textbf{0.39}} & \multicolumn{1}{c|}{0.44}          & \multicolumn{1}{c|}{0.84}          & \multicolumn{1}{c|}{\textbf{0.53}} & \multicolumn{1}{c|}{0.39}          & \multicolumn{1}{c|}{\textbf{0.50}} & \multicolumn{1}{c|}{0.39}          & \multicolumn{1}{c|}{0.42}          & 0.41          \\ \hline
EWC+VG                                  & \multicolumn{1}{c|}{0.43}          & \multicolumn{1}{c|}{0.36}          & \multicolumn{1}{c|}{0.42}          & \multicolumn{1}{c|}{\textbf{1.0}}  & \multicolumn{1}{c|}{0.52}          & \multicolumn{1}{c|}{0.38}          & \multicolumn{1}{c|}{0.49}          & \multicolumn{1}{c|}{0.40}          & \multicolumn{1}{c|}{0.45}          & \textbf{0.45} \\ \hline
EWC+ALL                                 & \multicolumn{1}{c|}{\textbf{0.45}} & \multicolumn{1}{c|}{\textbf{0.39}} & \multicolumn{1}{c|}{0.43}          & \multicolumn{1}{c|}{\textbf{1.0}}  & \multicolumn{1}{c|}{0.48}          & \multicolumn{1}{c|}{\textbf{0.45}} & \multicolumn{1}{c|}{0.470}         & \multicolumn{1}{c|}{\textbf{0.47}} & \multicolumn{1}{c|}{0.47}          & \textbf{0.45} \\ \hline
EWC+Reg.                                & \multicolumn{1}{c|}{0.44}          & \multicolumn{1}{c|}{0.38}          & \multicolumn{1}{c|}{\textbf{0.49}} & \multicolumn{1}{c|}{0.94}          & \multicolumn{1}{c|}{0.52}          & \multicolumn{1}{c|}{0.41}          & \multicolumn{1}{c|}{0.49}          & \multicolumn{1}{c|}{0.44}          & \multicolumn{1}{c|}{0.42}          & 0.42          \\ \hline
\multicolumn{1}{|l|}{\multirow{2}{*}{}} & \multicolumn{10}{l|}{MAP}                                                                                                                                                                                                                                                                                                                                  \\ \cline{2-11} 
\multicolumn{1}{|l|}{}                  & \multicolumn{1}{c|}{ST}            & \multicolumn{1}{c|}{NS}            & \multicolumn{1}{c|}{ST}            & \multicolumn{1}{c|}{NS}            & \multicolumn{1}{c|}{ST}            & \multicolumn{1}{c|}{NS}            & \multicolumn{1}{c|}{ST}            & \multicolumn{1}{c|}{NS}            & \multicolumn{1}{c|}{ST}            & NS            \\ \hline
EWC+LOC                                 & \multicolumn{1}{c|}{\textbf{0.04}} & \multicolumn{1}{c|}{0.03}          & \multicolumn{1}{c|}{0.04}          & \multicolumn{1}{c|}{\textbf{0.10}} & \multicolumn{1}{c|}{\textbf{0.05}} & \multicolumn{1}{c|}{\textbf{0.04}} & \multicolumn{1}{c|}{0.04}          & \multicolumn{1}{c|}{\textbf{0.04}} & \multicolumn{1}{c|}{\textbf{0.04}} & 0.04          \\ \hline
EWC+MLOC                                & \multicolumn{1}{c|}{\textbf{0.04}} & \multicolumn{1}{c|}{0.03}          & \multicolumn{1}{c|}{0.04}          & \multicolumn{1}{c|}{0.08}          & \multicolumn{1}{c|}{\textbf{0.05}} & \multicolumn{1}{c|}{\textbf{0.04}} & \multicolumn{1}{c|}{\textbf{0.05}} & \multicolumn{1}{c|}{\textbf{0.04}} & \multicolumn{1}{c|}{\textbf{0.04}} & 0.04          \\ \hline
EWC+VG                                  & \multicolumn{1}{c|}{\textbf{0.04}} & \multicolumn{1}{c|}{0.03}          & \multicolumn{1}{c|}{0.04}          & \multicolumn{1}{c|}{0.09}          & \multicolumn{1}{c|}{\textbf{0.05}} & \multicolumn{1}{c|}{0.03}          & \multicolumn{1}{c|}{0.04}          & \multicolumn{1}{c|}{\textbf{0.04}} & \multicolumn{1}{c|}{\textbf{0.04}} & 0.04          \\ \hline
EWC+ALL                                 & \multicolumn{1}{c|}{\textbf{0.04}} & \multicolumn{1}{c|}{0.03}          & \multicolumn{1}{c|}{0.04}          & \multicolumn{1}{c|}{\textbf{0.10}} & \multicolumn{1}{c|}{0.04}          & \multicolumn{1}{c|}{\textbf{0.04}} & \multicolumn{1}{c|}{0.04}          & \multicolumn{1}{c|}{\textbf{0.04}} & \multicolumn{1}{c|}{\textbf{0.04}} & 0.04          \\ \hline
EWC+Reg.                                & \multicolumn{1}{c|}{\textbf{0.04}} & \multicolumn{1}{c|}{\textbf{0.04}} & \multicolumn{1}{c|}{\textbf{0.05}} & \multicolumn{1}{c|}{\textbf{0.10}} & \multicolumn{1}{c|}{\textbf{0.05}} & \multicolumn{1}{c|}{\textbf{0.04}} & \multicolumn{1}{c|}{\textbf{0.05}} & \multicolumn{1}{c|}{\textbf{0.04}} & \multicolumn{1}{c|}{\textbf{0.04}} & \textbf{0.05} \\ \hline
\multicolumn{1}{|l|}{\multirow{2}{*}{}} & \multicolumn{10}{l|}{MRR}                                                                                                                                                                                                                                                                                                                                  \\ \cline{2-11} 
\multicolumn{1}{|l|}{}                  & \multicolumn{1}{c|}{ST}            & \multicolumn{1}{c|}{NS}            & \multicolumn{1}{c|}{ST}            & \multicolumn{1}{c|}{NS}            & \multicolumn{1}{c|}{ST}            & \multicolumn{1}{c|}{NS}            & \multicolumn{1}{c|}{ST}            & \multicolumn{1}{c|}{NS}            & \multicolumn{1}{c|}{ST}            & NS            \\ \hline
EWC+LOC                                 & \multicolumn{1}{c|}{0.26}          & \multicolumn{1}{c|}{0.24}          & \multicolumn{1}{c|}{0.40}          & \multicolumn{1}{c|}{0.48}          & \multicolumn{1}{c|}{0.27}          & \multicolumn{1}{c|}{0.20}          & \multicolumn{1}{c|}{0.28}          & \multicolumn{1}{c|}{0.19}          & \multicolumn{1}{c|}{0.27}          & 0.28          \\ \hline
EWC+MLOC                                & \multicolumn{1}{c|}{0.25}          & \multicolumn{1}{c|}{0.34}          & \multicolumn{1}{c|}{0.42}          & \multicolumn{1}{c|}{0.38}          & \multicolumn{1}{c|}{0.36}          & \multicolumn{1}{c|}{\textbf{0.38}} & \multicolumn{1}{c|}{0.34}          & \multicolumn{1}{c|}{\textbf{0.42}} & \multicolumn{1}{c|}{0.26}          & 0.34          \\ \hline
EWC+VG                                  & \multicolumn{1}{c|}{0.25}          & \multicolumn{1}{c|}{0.30}          & \multicolumn{1}{c|}{\textbf{0.48}} & \multicolumn{1}{c|}{\textbf{0.62}} & \multicolumn{1}{c|}{0.42}          & \multicolumn{1}{c|}{0.24}          & \multicolumn{1}{c|}{0.32}          & \multicolumn{1}{c|}{0.18}          & \multicolumn{1}{c|}{0.21}          & 0.40          \\ \hline
EWC+ALL                                 & \multicolumn{1}{c|}{\textbf{0.42}} & \multicolumn{1}{c|}{\textbf{0.44}} & \multicolumn{1}{c|}{0.34}          & \multicolumn{1}{c|}{0.59}          & \multicolumn{1}{c|}{\textbf{0.52}} & \multicolumn{1}{c|}{0.37}          & \multicolumn{1}{c|}{0.35}          & \multicolumn{1}{c|}{0.13}          & \multicolumn{1}{c|}{\textbf{0.33}} & \textbf{0.43} \\ \hline
EWC+Reg.                                & \multicolumn{1}{c|}{0.26}          & \multicolumn{1}{c|}{0.28}          & \multicolumn{1}{c|}{0.36}          & \multicolumn{1}{c|}{0.48}          & \multicolumn{1}{c|}{0.45}          & \multicolumn{1}{c|}{0.26}          & \multicolumn{1}{c|}{\textbf{0.43}} & \multicolumn{1}{c|}{\textbf{0.42}} & \multicolumn{1}{c|}{0.18}          & 0.26          \\ \hline
\end{tabular}
}
%\begin{tablenotes}
%     \item In bold are the best average performances per algorithm per metric for each project. 
% \end{tablenotes}
 \end{table*}
For non-stationary data in Eclipse (top@1, top@5, top@10, and MRR metrics) and Tomcat (top@1, top@5, top@10, and MRR metrics), as well as Birt (top@5, MRR metrics), SWT (top@1, top@5, top@10, and MRR metrics) and AspectJ (top@1, MRR metrics), CLEAR and EWC outperform FLIM by at least 14\%. Similarly, CLEAR and EWC significantly outperform RLOCATOR (with and without entropy) on non-stationary data in  AspectJ for top@5, top@10, and MRR metrics. 
\begin{table}[t]
\caption{Ablation study 5: different bug report metrics added to CLEAR. In bold are the best average performances per algorithm per metric for each project.}
\label{tab:Ablation study5: different bug report metrics added to CLEAR.}
\resizebox{0.4\textwidth}{!}{
\centering
%put it here|c|ccccc|ccccc|ccccc|ccccc|ccccc|
\begin{tabular}{|c|cccc|}
\hline
\multirow{3}{*}{} & \multicolumn{2}{c|}{PDE}                                                & \multicolumn{2}{c|}{Zxing}                         \\ \cline{2-5} 
                  & \multicolumn{4}{l|}{top@1}                                                                                                   \\ \cline{2-5} 
                  & \multicolumn{1}{c|}{ST}            & \multicolumn{1}{c|}{NS}            & \multicolumn{1}{c|}{ST}            & NS            \\ \hline
CLEAR+Churn       & \multicolumn{1}{c|}{\textbf{0.08}} & \multicolumn{1}{c|}{0.08}          & \multicolumn{1}{c|}{0.11}          & \textbf{0.12} \\ \hline
CLEAR+PRE         & \multicolumn{1}{c|}{0.06}          & \multicolumn{1}{c|}{\textbf{0.10}} & \multicolumn{1}{c|}{0.11}          & 0.11          \\ \hline
CLEAR+Reg.        & \multicolumn{1}{c|}{\textbf{0.08}} & \multicolumn{1}{c|}{0.08}          & \multicolumn{1}{c|}{\textbf{0.12}} & \textbf{0.12} \\ \hline
\multirow{2}{*}{} & \multicolumn{4}{l|}{top@5}                                                                                                   \\ \cline{2-5} 
                  & \multicolumn{1}{c|}{ST}            & \multicolumn{1}{c|}{NS}            & \multicolumn{1}{c|}{ST}            & NS            \\ \hline
CLEAR+Churn       & \multicolumn{1}{c|}{0.40}          & \multicolumn{1}{c|}{0.41}          & \multicolumn{1}{c|}{0.62}          & 0.56          \\ \hline
CLEAR+PRE         & \multicolumn{1}{c|}{0.37}          & \multicolumn{1}{c|}{\textbf{0.47}} & \multicolumn{1}{c|}{0.61}          & 0.55          \\ \hline
CLEAR+Reg.        & \multicolumn{1}{c|}{\textbf{0.42}} & \multicolumn{1}{c|}{0.39}          & \multicolumn{1}{c|}{\textbf{0.63}} & \textbf{0.61} \\ \hline
\multirow{2}{*}{} & \multicolumn{4}{l|}{top@10}                                                                                                  \\ \cline{2-5} 
                  & \multicolumn{1}{c|}{ST}            & \multicolumn{1}{c|}{NS}            & \multicolumn{1}{c|}{ST}            & NS            \\ \hline
CLEAR+Churn       & \multicolumn{1}{c|}{0.64}          & \multicolumn{1}{c|}{0.67}          & \multicolumn{1}{c|}{\textbf{1.00}} & \textbf{1.00} \\ \hline
CLEAR+PRE         & \multicolumn{1}{c|}{0.72}          & \multicolumn{1}{c|}{\textbf{0.90}} & \multicolumn{1}{c|}{\textbf{1.00}} & \textbf{1.00} \\ \hline
CLEAR+Reg.        & \multicolumn{1}{c|}{\textbf{0.77}} & \multicolumn{1}{c|}{0.77}          & \multicolumn{1}{c|}{\textbf{1.00}} & \textbf{1.00} \\ \hline
\multirow{2}{*}{} & \multicolumn{4}{l|}{MAP}                                                                                                     \\ \cline{2-5} 
                  & \multicolumn{1}{c|}{ST}            & \multicolumn{1}{c|}{NS}            & \multicolumn{1}{c|}{ST}            & NS            \\ \hline
CLEAR+Churn       & \multicolumn{1}{c|}{\textbf{0.08}} & \multicolumn{1}{c|}{0.09}          & \multicolumn{1}{c|}{\textbf{0.12}} & 0.11          \\ \hline
CLEAR+PRE         & \multicolumn{1}{c|}{0.07}          & \multicolumn{1}{c|}{\textbf{0.10}} & \multicolumn{1}{c|}{\textbf{0.12}} & 0.11          \\ \hline
CLEAR+Reg.        & \multicolumn{1}{c|}{\textbf{0.08}} & \multicolumn{1}{c|}{0.08}          & \multicolumn{1}{c|}{\textbf{0.12}} & \textbf{0.12} \\ \hline
\multirow{2}{*}{} & \multicolumn{4}{l|}{MRR}                                                                                                     \\ \cline{2-5} 
                  & \multicolumn{1}{c|}{ST}            & \multicolumn{1}{c|}{NS}            & \multicolumn{1}{c|}{ST}            & NS            \\ \hline
CLEAR+Churn       & \multicolumn{1}{c|}{\textbf{0.49}} & \multicolumn{1}{c|}{0.34}          & \multicolumn{1}{c|}{\textbf{0.62}} & 0.47          \\ \hline
CLEAR+PRE         & \multicolumn{1}{c|}{0.36}          & \multicolumn{1}{c|}{0.38}          & \multicolumn{1}{c|}{0.56}          & 0.48          \\ \hline
CLEAR+Reg.        & \multicolumn{1}{c|}{0.33}          & \multicolumn{1}{c|}{\textbf{0.41}} & \multicolumn{1}{c|}{0.42}          & \textbf{0.49} \\ \hline
\end{tabular}
}
%\begin{tablenotes}
 %    \item In bold are the best average performances per algorithm per metric for each project. 
% \end{tablenotes}
 %\vspace{-1em}
\end{table}
\begin{table}[t]
\caption{Ablation study 6: different bug report metrics added to EWC. In bold are the best average performances per algorithm per metric for each project.}
\label{tab:Ablation study6: different bug report metrics added to EWC.}
\resizebox{0.4\textwidth}{!}{
\centering
%put it here|c|ccccc|ccccc|ccccc|ccccc|ccccc|
\begin{tabular}{|c|llcl|}
\hline
\multicolumn{1}{|l|}{\multirow{3}{*}{}} & \multicolumn{2}{c|}{PDE}                                                & \multicolumn{2}{c|}{Zxing}                                              \\ \cline{2-5} 
\multicolumn{1}{|l|}{}                  & \multicolumn{4}{l|}{top@1}                                                                                                                        \\ \cline{2-5} 
\multicolumn{1}{|l|}{}                  & \multicolumn{1}{c|}{ST}            & \multicolumn{1}{c|}{NS}            & \multicolumn{1}{c|}{ST}            & \multicolumn{1}{c|}{NS}            \\ \hline
EWC+Churn                               & \multicolumn{1}{l|}{\textbf{0.10}} & \multicolumn{1}{l|}{0.08}          & \multicolumn{1}{l|}{\textbf{0.13}} & 0.09                               \\ \hline
EWC+PRE                                 & \multicolumn{1}{l|}{\textbf{0.10}} & \multicolumn{1}{l|}{0.07}          & \multicolumn{1}{l|}{0.12}          & 0.11                               \\ \hline
EWC+Reg.                                & \multicolumn{1}{c|}{\textbf{0.10}} & \multicolumn{1}{c|}{\textbf{0.09}} & \multicolumn{1}{c|}{\textbf{0.13}} & \multicolumn{1}{c|}{\textbf{0.12}} \\ \hline
\multicolumn{1}{|l|}{\multirow{2}{*}{}} & \multicolumn{4}{l|}{top@5}                                                                                                                        \\ \cline{2-5} 
\multicolumn{1}{|l|}{}                  & \multicolumn{1}{c|}{ST}            & \multicolumn{1}{c|}{NS}            & \multicolumn{1}{c|}{ST}            & \multicolumn{1}{c|}{NS}            \\ \hline
EWC+Churn                               & \multicolumn{1}{l|}{\textbf{0.46}} & \multicolumn{1}{l|}{0.42}          & \multicolumn{1}{l|}{0.60}          & 0.54                               \\ \hline
EWC+PRE                                 & \multicolumn{1}{l|}{0.43}          & \multicolumn{1}{l|}{0.44}          & \multicolumn{1}{l|}{\textbf{0.64}} & 0.55                               \\ \hline
EWC+Reg.                                & \multicolumn{1}{c|}{0.43}          & \multicolumn{1}{c|}{\textbf{0.44}} & \multicolumn{1}{c|}{0.63}          & \multicolumn{1}{c|}{\textbf{0.59}} \\ \hline
\multicolumn{1}{|l|}{\multirow{2}{*}{}} & \multicolumn{4}{l|}{top@10}                                                                                                                       \\ \cline{2-5} 
\multicolumn{1}{|l|}{}                  & \multicolumn{1}{c|}{ST}            & \multicolumn{1}{c|}{NS}            & \multicolumn{1}{c|}{ST}            & \multicolumn{1}{c|}{NS}            \\ \hline
EWC+Churn                               & \multicolumn{1}{l|}{\textbf{0.85}} & \multicolumn{1}{l|}{0.84}          & \multicolumn{1}{l|}{\textbf{1.00}} & \textbf{1.00}                      \\ \hline
EWC+PRE                                 & \multicolumn{1}{l|}{0.83}          & \multicolumn{1}{l|}{\textbf{0.89}} & \multicolumn{1}{l|}{\textbf{1.00}} & \textbf{1.00}                      \\ \hline
EWC+Reg.                                & \multicolumn{1}{c|}{\textbf{0.85}} & \multicolumn{1}{c|}{0.86}          & \multicolumn{1}{c|}{\textbf{1.00}} & \multicolumn{1}{c|}{\textbf{1.00}} \\ \hline
\multicolumn{1}{|l|}{\multirow{2}{*}{}} & \multicolumn{4}{l|}{MAP}                                                                                                                          \\ \cline{2-5} 
\multicolumn{1}{|l|}{}                  & \multicolumn{1}{c|}{ST}            & \multicolumn{1}{c|}{NS}            & \multicolumn{1}{c|}{ST}            & \multicolumn{1}{c|}{NS}            \\ \hline
EWC+Churn                               & \multicolumn{1}{l|}{\textbf{0.09}} & \multicolumn{1}{l|}{0.08}          & \multicolumn{1}{l|}{0.12}          & 0.11                               \\ \hline
EWC+PRE                                 & \multicolumn{1}{l|}{0.08}          & \multicolumn{1}{l|}{\textbf{0.09}} & \multicolumn{1}{l|}{\textbf{0.13}} & 0.11                               \\ \hline
EWC+Reg.                                & \multicolumn{1}{c|}{\textbf{0.09}} & \multicolumn{1}{c|}{\textbf{0.09}} & \multicolumn{1}{c|}{\textbf{0.13}} & \multicolumn{1}{c|}{\textbf{0.12}} \\ \hline
\multicolumn{1}{|l|}{\multirow{2}{*}{}} & \multicolumn{4}{l|}{MRR}                                                                                                                          \\ \cline{2-5} 
\multicolumn{1}{|l|}{}                  & \multicolumn{1}{c|}{ST}            & \multicolumn{1}{c|}{NS}            & \multicolumn{1}{c|}{ST}            & \multicolumn{1}{c|}{NS}            \\ \hline
EWC+Churn                               & \multicolumn{1}{l|}{\textbf{0.52}} & \multicolumn{1}{l|}{0.36}          & \multicolumn{1}{l|}{\textbf{0.71}} & \textbf{0.64}                      \\ \hline
EWC+PRE                                 & \multicolumn{1}{l|}{0.43}          & \multicolumn{1}{l|}{0.36}          & \multicolumn{1}{l|}{0.49}          & 0.45                               \\ \hline
EWC+Reg.                                & \multicolumn{1}{c|}{0.32}          & \multicolumn{1}{c|}{\textbf{0.39}} & \multicolumn{1}{c|}{0.58}          & \multicolumn{1}{c|}{\textbf{0.64}} \\ \hline
\end{tabular}
}
%\begin{tablenotes}
 %    \item In bold are the best average performances per algorithm per metric for each project. 
 %\end{tablenotes}
 %\vspace{-1em}
\end{table}
For the Birt project, CLEAR outperforms RLOCATOR with entropy by 14\% (top@5, top@10, and MRR metrics), while EWC outperforms RLOCATOR without entropy by 87\% (MRR metric). In the Eclipse project, CLEAR outperforms RLOCATOR (with and without entropy) by 14\% (top@5 and MRR metrics). Furthermore, EWC significantly outperforms RLOCATOR without entropy by 14\% (top@10 and MRR metrics) in the Tomcat project. Based on the obtained results, CL agents show that they can adapt to software project changes that lead to non-stationary data distributions, ensuring that the bug localization process remains effective over time.

Table \ref{tab:Algorithms evaluation across software projects with different  data granularities} and \ref{tab:Algorithms evaluation across PDE and Zxing projects with different  data granularities} report the performance of the CL agents and FBL-BERT for QARC, QARCL and QD at changeset-files and hunks level across SWT, AspectJ, Tomcat, PDE and Zxing software projects - the performance on stationary and non-stationary data are averaged for comparison with FBL-BERT. At the changeset-files level, the CL agents significantly outperform FBL-BERT for QARC, QARCL, and QD on AspectJ (by at least 80\% in top@5 and top@10), on Zxing (by at least 89\% in top@10) and on SWT (by at least 80\% in top@10). On Tomcat, the CL agents significantly outperform QARCL and QD by at least 22\% in both top@5 and top@10. On PDE, the CL agents significantly outperform QARCL and QD by at least 13\% in top@1, top@5 and top@10.
At the hunk level across the SWT, Tomcat, PDE, Zxing and AspectJ projects, the CL agents significantly outperform QARC, QARCL, and QD in top@10 by at least 67\%. However, FBL-BERT for QARC, QARCL, and QD shows better performance than CL agents in terms of  MRR and top@1 at both the changeset-files and hunk level across all studied projects. 
%Compared to FBL-BERT for QARC, QARCL, and QD the CL agents show better performance except with the top@1 and MRR metrics at the changeset-files and hunk level. 
FBL-BERT first accounts for all changeset-files and hunks (stationary and non-stationary) committed to a software project up to the present point in time then uses the InVerted File with Product Quantization (IVFPQ) \cite{johnson2019billion} algorithm to refine the search space by identifying the top-N most relevant ones. This allows FBL-BERT to focus on a smaller, more relevant set of changeset-files (or hunks) for bug localization, improving precision. The refined search helps FBL-BERT to quickly identify the most relevant changeset-files (or hunks), leading to better MRR and top@1 performance compared to the CL agents. Nevertheless, the consistent performance of the CL agents, with top@5 and top@10 metrics, indicates their effectiveness in capturing a high proportion of the most relevant changesets within the top-ranked selections.

The performance difference on stationary vs. non-stationary at changeset-files level for CLEAR is up to 14\% on Eclipse, 40\% on the Zxing, 38\% on Birt, 59\% on  PDE, 57\% on Tomcat, 90\% AspectJ, and 109\% on SWT projects with all ranking metrics. As for EWC, it is up to 12\% on Tomcat, 15\% on AspectJ, 19\% on  Zxing, 33\% on  PDE, 45\% on Birt, 115\% on SWT, and 63\% on the Eclipse projects at changeset-files level. At the hunks level, the performance difference on stationary vs. non-stationary for CLEAR is up to 15\% on Birt, 19\% on PDE, 9\% on Zxing, 17\% on Eclipse, 30\% on Tomcat, 22\% on AspectJ, and 72\% on SWT projects. As for EWC, the performance difference is up to 28\% on Tomcat, 45\% on PDE, 10\% on Zxing, 44\% on AspectJ, 71\% on Birt, 100\% on SWT, and 41\% on the Eclipse projects at the hunks level. CLEAR seems to keep old and new policies close together better than EWC for all projects and all data granularities except Tomcat (at the changeset-files and hunks levels), AspectJ (at the changeset-files level), PDE (at the changeset-files level) and Zxing (at the changeset-files level). A possible explanation lies in the complexity of the projects. For complex projects such as Tomcat (given its large number of changed files as reported in Table \ref{tab:benchmark-dataset}), it can become challenging for the behavioral cloning used by CLEAR to keep track of all the new and old experiences while preventing the DNNs from drifting.  %The CL agents significantly outperform FBL-BERT  with the top@5 and top@10 metrics across AspectJ, SWT, and Tomcat, while FBL-BERT performs better with the MAP and top@1 metrics across all projects. due to the simple architecture of IMPALA \cite{espeholt2018impala}

Regarding the computational efforts of the CL agents, as shown in Table \ref{tab:Computational effort of studied algorithms across datasets} and \ref{tab:Computational effort  in hours of studied algorithms across datasets with different input data granularities}, they require up to 5x less time to get trained at different data granularities compared to the baseline studies across all software projects except the Zxing project (with the FBL-BERT baseline study). The Zxing project has the fewest number of bug reports (see Table \ref{tab:benchmark-dataset}) compared to other software projects, which explains its fastest fine-tuning time on a BERT-based approach like FBL-BERT.
\begin{tcolorbox}
%\Foutse{what about non-stationary data?}
    \textbf{Finding 2:} The CL agents demonstrate substantial performance enhancements over RLOCATOR (with and without entropy), on stationary data for the AspectJ project, as measured by top@5, top@10, MRR, and MAP metrics, while achieving comparable performance on other projects. On non-stationary data, CLEAR and EWC outperform FLIM and  RLOCATOR (with and without entropy) on AspectJ and SWT projects as measured by top@5, top@10, and MRR metrics. Finally, they significantly outperform FBL-BERT with the top@5 and top@10 metrics across AspectJ, SWT, PDE, Zxing, and Tomcat at the hunks level, while FBL-BERT shows better performance as measured by MRR and top@1 metrics. 
\end{tcolorbox}
%Additionally, they significantly outperform the baseline studies on non-stationary at the changeset-files level for (3 out of 5) projects, across (3 out of 5) metrics while requiring much less computational effort.

\subsection{\textbf{RQ3: Can prior knowledge about bug-inducing factors improve the performance of CL techniques?}} \label{sec:rq3}
\textbf{Motivation:} The purpose of this RQ is to evaluate the performance of the CL agents on both stationary and non-stationary data (i.e., changeset-files and hunks) by integrating prior knowledge of bug-inducing factors into their training phase. Bug-inducing factors are good indicators of whether changeset-files and hunks are likely to be buggy or not \cite{taba2013predicting,wen2016locus}. Thus, incorporating them in any bug localization technique can improve its performance.\\
\textbf{Method:} We calculate our evaluation metrics for each dataset and compare the performance of the CL agents against the baselines, both before and after including the bug-inducing factors. Similarly to \textbf{RQ2}, for each CL agent (with regression), we calculate the performance differences in percentage between stationary and non-stationary data.\\
\textbf{Results:}
Table \ref{tab:Tomcat}, \ref{tab:Birt}, \ref{tab:Eclipse}, \ref{tab:AspectJ}, \ref{tab:SWT}, and \ref{tab:PDEANDZXING} show the performance of the CL agents with regression across the studied software projects at the changeset-files and hunks levels. The performance difference on stationary vs. non-stationary data at the changeset-files level for CLEAR with regression is up to 18\% on the Birt project (see Table \ref{tab:Birt}), 34\% on the Eclipse project (see Table \ref{tab:Eclipse}), 29\% on the Tomcat project (see Table \ref{tab:Tomcat}), 40\% AspectJ project (see Table \ref{tab:AspectJ}), 15\% on the Zxing project (see Table \ref{tab:PDEANDZXING}), 21\% PDE project (see Table \ref{tab:PDEANDZXING}), and 85\% on SWT project (see Table \ref{tab:SWT}) as measured by all ranking metrics. As for EWC with regression, the performance difference is up to 36\% on the Tomcat project, 10\% on the AspectJ project, 53\% on the Birt project, 62\% on SWT,  6\% on the Zxing project (see Table \ref{tab:PDEANDZXING}), 19\% PDE project (see Table \ref{tab:PDEANDZXING}) and 14\% on the Eclipse project as measured by all metrics at the changeset-files level. The performance difference on stationary vs. non-stationary at the hunks level for CLEAR with regression is up to 27\% on the Birt project, 30\% on the Eclipse project, 21\% on the Tomcat project, 18\% on the AspectJ project, 9\% on the Zxing project (see Table \ref{tab:PDEANDZXING}), 5\% on the PDE project (see Table \ref{tab:PDEANDZXING}) and 74\% on SWT as measured by all metrics. As for EWC with regression, the performance difference is up to 43\% on the Tomcat project, 33\% on the AspectJ project, 98\% on the Birt project, 114\% on the SWT project, 8\% on the Zxing project (see Table \ref{tab:PDEANDZXING}), 3\% on the PDE project (see Table \ref{tab:PDEANDZXING}), and 22\% on the Eclipse project as measured by all metrics at the hunks level. 
\begin{table}[t]
\caption{Computational effort (mean ± std) in hours of studied algorithms across software projects with changeset-files data granularity.}
\label{tab:Computational effort of studied algorithms across datasets}
\resizebox{0.8\textwidth}{!}{
\centering
\begin{tabular}{|c|c|c|c|c|c|}
\hline
{\color[HTML]{000000} } & {\color[HTML]{000000} AspectJ} & {\color[HTML]{000000} SWT} & {\color[HTML]{000000} Birt} & {\color[HTML]{000000} Eclipse} & {\color[HTML]{000000} Tomcat} \\ \hline
{\color[HTML]{000000} FLIM} & {\color[HTML]{000000} 16.1 ± 2.4} & {\color[HTML]{000000} 11.2 ± 0.1} & {\color[HTML]{000000} 11.4 ± 0.4} & {\color[HTML]{000000} 12.0 ± 0.1} & {\color[HTML]{000000} 10.5 ± 0.2} \\ \hline
{\color[HTML]{000000} RLO.} & {\color[HTML]{000000} 8.5 ± 0.07} & {\color[HTML]{000000} 13.6 ± 1.6} & {\color[HTML]{000000} 10.7 ± 0.8} & {\color[HTML]{000000} 12.4 ± 0.2} & {\color[HTML]{000000} 9.1 ± 0.04} \\ \hline
{\color[HTML]{000000} RLO. + Reg.} & {\color[HTML]{000000} 4.4 ± 1.0} & {\color[HTML]{000000} 14.0 ± 1.5} & {\color[HTML]{000000} 10.1 ± 0.7} & {\color[HTML]{000000} 13.1 ± 1.5} & {\color[HTML]{000000} 9.0 ± 0.2} \\ \hline
{\color[HTML]{000000} RLO. + Ent.} & {\color[HTML]{000000} 12.6 ± 1.2} & {\color[HTML]{000000} 16.8 ± 2.2} & \cellcolor[HTML]{FFFFFF}{\color[HTML]{000000} 15.6 ± 2.3} & {\color[HTML]{000000} 12.3 ± 0.1} & {\color[HTML]{000000} 9.8 ± 0.9} \\ \hline
{\color[HTML]{000000} RLO. + Ent. + Reg.} & {\color[HTML]{000000} 11.4 ± 1.2} & {\color[HTML]{000000} 16.7 ± 0.4} & {\color[HTML]{000000} 15.8 ± 0.4} & {\color[HTML]{000000} 10.6 ± 0.05} & {\color[HTML]{000000} 11.0 ± 1.5} \\ \hline
{\color[HTML]{000000} CLEAR} & \cellcolor[HTML]{656565}{\color[HTML]{000000} 0.32±0.01} & \cellcolor[HTML]{656565}{\color[HTML]{000000} 0.70±0.05} & \cellcolor[HTML]{656565}{\color[HTML]{000000} 0.43 ± 0.00} & \cellcolor[HTML]{656565}{\color[HTML]{000000} 0.52 ± 0.00} & \cellcolor[HTML]{656565}{\color[HTML]{000000} 0.71±0.05} \\ \hline
{\color[HTML]{000000} CLEAR + Reg.} & {\color[HTML]{000000} 0.68 ± 0.04} & {\color[HTML]{000000} 1.38 ± 0.30} & {\color[HTML]{000000} 0.80 ± 0.02} & {\color[HTML]{000000} 3.15 ± 0.31} & {\color[HTML]{000000} 1.46 ± 0.09} \\ \hline
{\color[HTML]{000000} EWC} & \cellcolor[HTML]{C0C0C0}{\color[HTML]{000000} 0.36±0.05} & \cellcolor[HTML]{C0C0C0}{\color[HTML]{000000} 0.77±0.02} & \cellcolor[HTML]{C0C0C0}{\color[HTML]{000000} 0.47 ± 0.01} & \cellcolor[HTML]{C0C0C0}{\color[HTML]{000000} 0.59 ± 0.02} & \cellcolor[HTML]{C0C0C0}{\color[HTML]{000000} 0.79±0.02} \\ \hline
{\color[HTML]{000000} EWC + Reg.} & {\color[HTML]{000000} 0.60 ± 0.09} & {\color[HTML]{000000} 1.89 ± 0.50} & {\color[HTML]{000000} 0.88 ± 0.06} & {\color[HTML]{000000} 2.09 ± 1.04} & {\color[HTML]{000000} 1.48 ± 0.06} \\ \hline
\end{tabular}
}
\begin{tablenotes}
     \item "RLO." refers to the baseline RLOCATOR. "+ Reg" indicates the inclusion of logistic regression, while "+ Ent" denotes the addition of entropy. The best and second best average performances are highlighted in dark grey and light grey, respectively. 
 \end{tablenotes}
 %\vspace{-1em}
\end{table}
%Tables \ref{tab:Algorithm evaluation on Tomcat project}, \ref{tab:Algorithm evaluation on Birt project}, \ref{tab:Algorithm evaluation on AspectJ project}, \ref{tab:Algorithm evaluation on SWT project}, \ref{tab:Algorithm evaluation on Eclipse project} show the performance of the CL agents (i.e. CLEAR and EWC) on stationary and non-stationary data across the five studied datasets. 

Our results show an improvement in performance difference on stationary and non-stationary data for up to 38\% across AspectJ, Birt, PDE, Zxing, and Tomcat projects when incorporating bug-inducing factors as prior knowledge of CLEAR at the changeset-files and hunks levels. With EWC, our results show a performance improvement of up to 42\% on AspectJ, PDE, Zxing, and Eclipse projects at the changeset-files and hunks levels. Suggesting that prior knowledge about bug-inducing factors can provide a critical appropriate context that helps the CL agents better understand and adapt to changes in the data of software projects.
\begin{table}[t]
\caption{Computational effort (mean ± std) in hours of studied algorithms across software projects with different  data granularities. The best and second best average performances are highlighted in dark grey and light grey, respectively.}
\label{tab:Computational effort  in hours of studied algorithms across datasets with different input data granularities}
\resizebox{0.9\textwidth}{!}{
\centering
\begin{tabular}{|c|c|c|c|c|c|c|}
\hline
                             & Granularity     & SWT                               & Tomcat                            & AspectJ                           & PDE                              & Zxing                            \\ \hline
                             & hunks           & \cellcolor[HTML]{656565}0.54±0.01 & \cellcolor[HTML]{656565}0.49±0.03 & \cellcolor[HTML]{656565}0.32±0.08 & \cellcolor[HTML]{656565}0.41±0.0 & 0.26±0.0                         \\ \cline{2-7} 
\multirow{-2}{*}{CLEAR}      & changeset-files & 0.77±0.02                         & 0.79±0.02                         & 0.36±0.05                         & 0.42±0.0                         & 0.28±0.0                         \\ \hline
                             & hunks           & \cellcolor[HTML]{C0C0C0}0.55±0.03 & \cellcolor[HTML]{C0C0C0}0.56±0.02 & \cellcolor[HTML]{C0C0C0}0.33±0.04 & \cellcolor[HTML]{C0C0C0}0.44±0.0 & 0.29±0.0                         \\ \cline{2-7} 
\multirow{-2}{*}{EWC}        & changeset-files & 1.25±0.0                          & 1.68±0.1                          & 3.18±0.07                         & 0.46±0.0                         & 0.31±0.0                         \\ \hline
                             & hunks           & 0.66±0.01                         & 0.76±0.03                         & 0.49±0.0                          & 0.44±0.02                        & 0.26±0.0                         \\ \cline{2-7} 
\multirow{-2}{*}{CLEAR+Reg.} & changeset-files & 1.38±0.3                          & 1.46±0.09                         & 0.68±0.04                         & 0.49±0.01                        & 0.27±0.0                         \\ \hline
                             & hunks           & 0.71±0.01                         & 0.80±0.01                         & 0.54±0.02                         & 0.45±0.0                         & 0.28±0.0                         \\ \cline{2-7} 
\multirow{-2}{*}{EWC+Reg.}   & changeset-files & 1.89±0.5                          & 1.48±0.06                         & 0.60±0.09                         & 0.47±0.0                         & 0.30±0.0                         \\ \hline
                             & hunks           & 2.34±0.1                          & 1.97±0.01                         & 3.17±0.1                          & 1.56±0.0                         & 0.08±0.0                         \\ \cline{2-7} 
\multirow{-2}{*}{QARC}       & changeset-files & 1.20±0.02                         & 1.60±0.0                          & 3.28±0.07                         & 0.97±0.0                         & \cellcolor[HTML]{656565}0.05±0.0 \\ \hline
                             & hunks           & 2.31±0.12                         & 1.88±0.02                         & 3.16±0.1                          & 1.43±0.02                        & 0.08±0.0                         \\ \cline{2-7} 
\multirow{-2}{*}{QARCL}      & changeset-files & 1.20±0.05                         & 1.59±0.01                         & 3.23±0.1                          & 0.96±0.0                         & \cellcolor[HTML]{656565}0.05±0.0 \\ \hline
                             & hunks           & 2.13±0.0                          & 1.82±0.03                         & 3.29±0.1                          & 1.41±0.0                         & 0.08±0.0                         \\ \cline{2-7} 
\multirow{-2}{*}{QD}         & changeset-files & 1.19±0.04                         & 1.59±0.01                         & 3.22±0.12                         & 0.91±0.0                         & \cellcolor[HTML]{C0C0C0}0.06±0.0 \\ \hline
\end{tabular}
}
%\begin{tablenotes}
%     \item The best and second best average performances are highlighted in dark grey and light grey, respectively. 
% \end{tablenotes}
\end{table}

The CL agents with regression significantly outperform RLOCATOR (with and without entropy) on stationary data of Birt, and AspectJ projects at the changeset-files level, as measured by the MRR metric by up to 50\%. On stationary data of the other projects (at the changeset-files level), they achieve comparable performance against RLOCATOR (with and without entropy) as measured by top@1 and top@5 metrics. Given that we adopted the same environment as RLOCATOR, we enhanced it as well with prior knowledge of bug-inducing factors. Table \ref{tab:Tomcat}, \ref{tab:Birt}, \ref{tab:AspectJ}, \ref{tab:SWT}, and \ref{tab:Eclipse} show the performance of RLOCATOR (with and without entropy) across Tomcat, Birt, AspectJ, SWT and Eclipse projects when incorporating prior knowledge of bug inducing factors through regression at the changeset-files level. 
Our results show that the CL agents with regression significantly outperform RLOCATOR (with and without entropy) with regression on stationary data of Birt, AspectJ projects with the MRR metric by up to 29\%, suggesting CLEAR and EWC (with regression) utilize bug-inducing factors better than RLOCATOR. This is likely due to our CL mechanisms that are capable of retaining and applying learned knowledge.
\begin{table}[t]
\caption{Ablation study 7: different bug report metrics added to CLEAR. In bold are the best average performances per algorithm per metric for each project.}
\label{tab:Ablation study7: different bug report metrics added to CLEAR.}
\resizebox{0.4\textwidth}{!}{
\centering
%put it here|c|ccccc|ccccc|ccccc|ccccc|ccccc|
\begin{tabular}{|c|cccc|}
\hline
\multirow{3}{*}{} & \multicolumn{2}{c|}{PDE}                                                & \multicolumn{2}{c|}{Zxing}                         \\ \cline{2-5} 
                  & \multicolumn{4}{l|}{top@1}                                                                                                   \\ \cline{2-5} 
                  & \multicolumn{1}{c|}{ST}            & \multicolumn{1}{c|}{NS}            & \multicolumn{1}{c|}{ST}            & NS            \\ \hline
CLEAR+LOC         & \multicolumn{1}{c|}{0.09}          & \multicolumn{1}{c|}{\textbf{0.10}} & \multicolumn{1}{c|}{0.11}          & 0.12          \\ \hline
CLEAR+MLOC        & \multicolumn{1}{c|}{0.07}          & \multicolumn{1}{c|}{\textbf{0.10}} & \multicolumn{1}{c|}{0.08}          & 0.11          \\ \hline
CLEAR+VG          & \multicolumn{1}{c|}{0.09}          & \multicolumn{1}{c|}{\textbf{0.10}} & \multicolumn{1}{c|}{\textbf{0.12}} & 0.12          \\ \hline
CLEAR+ALL         & \multicolumn{1}{c|}{\textbf{0.10}} & \multicolumn{1}{c|}{\textbf{0.10}} & \multicolumn{1}{c|}{0.10}          & \textbf{0.13} \\ \hline
CLEAR+Reg.        & \multicolumn{1}{c|}{0.08}          & \multicolumn{1}{c|}{0.08}          & \multicolumn{1}{c|}{\textbf{0.12}} & 0.12          \\ \hline
\multirow{2}{*}{} & \multicolumn{4}{l|}{top@5}                                                                                                   \\ \cline{2-5} 
                  & \multicolumn{1}{c|}{ST}            & \multicolumn{1}{c|}{NS}            & \multicolumn{1}{c|}{ST}            & NS            \\ \hline
CLEAR+LOC         & \multicolumn{1}{c|}{0.41}          & \multicolumn{1}{c|}{\textbf{0.48}} & \multicolumn{1}{c|}{0.60}          & 0.55          \\ \hline
CLEAR+MLOC        & \multicolumn{1}{c|}{0.41}          & \multicolumn{1}{c|}{\textbf{0.48}} & \multicolumn{1}{c|}{0.53}          & 0.58          \\ \hline
CLEAR+VG          & \multicolumn{1}{c|}{0.42}          & \multicolumn{1}{c|}{0.45}          & \multicolumn{1}{c|}{0.48}          & 0.47          \\ \hline
CLEAR+ALL         & \multicolumn{1}{c|}{\textbf{0.45}} & \multicolumn{1}{c|}{0.42}          & \multicolumn{1}{c|}{0.54}          & 0.60          \\ \hline
CLEAR+Reg.        & \multicolumn{1}{c|}{0.42}          & \multicolumn{1}{c|}{0.39}          & \multicolumn{1}{c|}{\textbf{0.63}} & \textbf{0.61} \\ \hline
\multirow{2}{*}{} & \multicolumn{4}{l|}{top@10}                                                                                                  \\ \cline{2-5} 
                  & \multicolumn{1}{c|}{ST}            & \multicolumn{1}{c|}{NS}            & \multicolumn{1}{c|}{ST}            & NS            \\ \hline
CLEAR+LOC         & \multicolumn{1}{c|}{0.81}          & \multicolumn{1}{c|}{0.85}          & \multicolumn{1}{c|}{\textbf{1.00}} & \textbf{1.00} \\ \hline
CLEAR+MLOC        & \multicolumn{1}{c|}{0.80}          & \multicolumn{1}{c|}{\textbf{0.91}} & \multicolumn{1}{c|}{\textbf{1.00}} & \textbf{1.00} \\ \hline
CLEAR+VG          & \multicolumn{1}{c|}{0.66}          & \multicolumn{1}{c|}{0.72}          & \multicolumn{1}{c|}{\textbf{1.00}} & \textbf{1.00} \\ \hline
CLEAR+ALL         & \multicolumn{1}{c|}{\textbf{0.86}} & \multicolumn{1}{c|}{0.82}          & \multicolumn{1}{c|}{\textbf{1.00}} & \textbf{1.00} \\ \hline
CLEAR+Reg.        & \multicolumn{1}{c|}{0.77}          & \multicolumn{1}{c|}{0.77}          & \multicolumn{1}{c|}{\textbf{1.00}} & \textbf{1.00} \\ \hline
\multirow{2}{*}{} & \multicolumn{4}{l|}{MAP}                                                                                                     \\ \cline{2-5} 
                  & \multicolumn{1}{c|}{ST}            & \multicolumn{1}{c|}{NS}            & \multicolumn{1}{c|}{ST}            & NS            \\ \hline
CLEAR+LOC         & \multicolumn{1}{c|}{\textbf{0.09}} & \multicolumn{1}{c|}{\textbf{0.10}} & \multicolumn{1}{c|}{\textbf{0.12}} & 0.11          \\ \hline
CLEAR+MLOC        & \multicolumn{1}{c|}{0.08}          & \multicolumn{1}{c|}{0.09}          & \multicolumn{1}{c|}{0.11}          & 0.11          \\ \hline
CLEAR+VG          & \multicolumn{1}{c|}{\textbf{0.09}} & \multicolumn{1}{c|}{0.09}          & \multicolumn{1}{c|}{\textbf{0.12}} & 0.11          \\ \hline
CLEAR+ALL         & \multicolumn{1}{c|}{\textbf{0.09}} & \multicolumn{1}{c|}{0.09}          & \multicolumn{1}{c|}{0.11}          & \textbf{0.12} \\ \hline
CLEAR+Reg.        & \multicolumn{1}{c|}{0.08}          & \multicolumn{1}{c|}{0.08}          & \multicolumn{1}{c|}{\textbf{0.12}} & \textbf{0.12} \\ \hline
\multirow{2}{*}{} & \multicolumn{4}{l|}{MRR}                                                                                                     \\ \cline{2-5} 
                  & \multicolumn{1}{c|}{ST}            & \multicolumn{1}{c|}{NS}            & \multicolumn{1}{c|}{ST}            & NS            \\ \hline
CLEAR+LOC         & \multicolumn{1}{c|}{\textbf{0.60}} & \multicolumn{1}{c|}{0.46}          & \multicolumn{1}{c|}{0.43}          & \textbf{0.58} \\ \hline
CLEAR+MLOC        & \multicolumn{1}{c|}{0.37}          & \multicolumn{1}{c|}{0.30}          & \multicolumn{1}{c|}{\textbf{0.56}} & 0.50          \\ \hline
CLEAR+VG          & \multicolumn{1}{c|}{0.29}          & \multicolumn{1}{c|}{0.33}          & \multicolumn{1}{c|}{0.50}          & 0.47          \\ \hline
CLEAR+ALL         & \multicolumn{1}{c|}{0.48}          & \multicolumn{1}{c|}{\textbf{0.48}} & \multicolumn{1}{c|}{0.48}          & 0.44          \\ \hline
CLEAR+Reg.        & \multicolumn{1}{c|}{0.33}          & \multicolumn{1}{c|}{0.41}          & \multicolumn{1}{c|}{0.42}          & 0.49          \\ \hline
\end{tabular}
}
%\begin{tablenotes}
 %    \item In bold are the best average performances per algorithm per metric for each project. 
 %\end{tablenotes}
 %\vspace{-1em}
\end{table}
\begin{table}[t]
\caption{Ablation study 8: different bug report metrics added to EWC. In bold are the best average performances per algorithm per metric for each project.}
\label{tab:Ablation study8: different bug report metrics added to EWC.}
\resizebox{0.4\textwidth}{!}{
\centering
%put it here|c|ccccc|ccccc|ccccc|ccccc|ccccc|
\begin{tabular}{|c|cccc|}
\hline
\multicolumn{1}{|l|}{\multirow{3}{*}{}} & \multicolumn{2}{c|}{PDE}                                                & \multicolumn{2}{c|}{Zxing}                         \\ \cline{2-5} 
\multicolumn{1}{|l|}{}                  & \multicolumn{4}{l|}{top@1}                                                                                                   \\ \cline{2-5} 
\multicolumn{1}{|l|}{}                  & \multicolumn{1}{c|}{ST}            & \multicolumn{1}{c|}{NS}            & \multicolumn{1}{c|}{ST}            & NS            \\ \hline
EWC+LOC                                 & \multicolumn{1}{c|}{0.08}          & \multicolumn{1}{c|}{0.08}          & \multicolumn{1}{c|}{0.13}          & 0.11          \\ \hline
EWC+MLOC                                & \multicolumn{1}{c|}{0.09}          & \multicolumn{1}{c|}{0.08}          & \multicolumn{1}{c|}{0.12}          & \textbf{0.12} \\ \hline
\multicolumn{1}{|l|}{EWC+VG}            & \multicolumn{1}{c|}{0.08}          & \multicolumn{1}{c|}{\textbf{0.10}} & \multicolumn{1}{c|}{\textbf{0.14}} & 0.10          \\ \hline
\multicolumn{1}{|l|}{EWC+ALL}           & \multicolumn{1}{c|}{0.09}          & \multicolumn{1}{c|}{0.07}          & \multicolumn{1}{c|}{0.11}          & 0.11          \\ \hline
EWC+Reg.                                & \multicolumn{1}{c|}{\textbf{0.10}} & \multicolumn{1}{c|}{0.09}          & \multicolumn{1}{c|}{0.13}          & \textbf{0.12} \\ \hline
\multicolumn{1}{|l|}{\multirow{2}{*}{}} & \multicolumn{4}{l|}{top@5}                                                                                                   \\ \cline{2-5} 
\multicolumn{1}{|l|}{}                  & \multicolumn{1}{c|}{ST}            & \multicolumn{1}{c|}{NS}            & \multicolumn{1}{c|}{ST}            & NS            \\ \hline
EWC+LOC                                 & \multicolumn{1}{c|}{\textbf{0.43}} & \multicolumn{1}{c|}{0.42}          & \multicolumn{1}{c|}{0.60}          & 0.57          \\ \hline
EWC+MLOC                                & \multicolumn{1}{c|}{\textbf{0.43}} & \multicolumn{1}{c|}{0.44}          & \multicolumn{1}{c|}{0.57}          & 0.58          \\ \hline
\multicolumn{1}{|l|}{EWC+VG}            & \multicolumn{1}{c|}{0.39}          & \multicolumn{1}{c|}{\textbf{0.45}} & \multicolumn{1}{c|}{\textbf{0.64}} & \textbf{0.59} \\ \hline
\multicolumn{1}{|l|}{EWC+ALL}           & \multicolumn{1}{c|}{\textbf{0.43}} & \multicolumn{1}{c|}{0.44}          & \multicolumn{1}{c|}{0.57}          & \textbf{0.59} \\ \hline
EWC+Reg.                                & \multicolumn{1}{c|}{\textbf{0.43}} & \multicolumn{1}{c|}{0.44}          & \multicolumn{1}{c|}{0.63}          & \textbf{0.59} \\ \hline
\multicolumn{1}{|l|}{\multirow{2}{*}{}} & \multicolumn{4}{l|}{top@10}                                                                                                  \\ \cline{2-5} 
\multicolumn{1}{|l|}{}                  & \multicolumn{1}{c|}{ST}            & \multicolumn{1}{c|}{NS}            & \multicolumn{1}{c|}{ST}            & NS            \\ \hline
EWC+LOC                                 & \multicolumn{1}{c|}{0.81}          & \multicolumn{1}{c|}{0.82}          & \multicolumn{1}{c|}{\textbf{1.00}} & \textbf{1.00} \\ \hline
EWC+MLOC                                & \multicolumn{1}{c|}{0.87}          & \multicolumn{1}{c|}{0.88}          & \multicolumn{1}{c|}{\textbf{1.00}} & \textbf{1.00} \\ \hline
EWC+VG                                  & \multicolumn{1}{c|}{0.79}          & \multicolumn{1}{c|}{\textbf{0.91}} & \multicolumn{1}{c|}{\textbf{1.00}} & \textbf{1.00} \\ \hline
EWC+ALL                                 & \multicolumn{1}{c|}{\textbf{0.86}} & \multicolumn{1}{c|}{0.88}          & \multicolumn{1}{c|}{\textbf{1.00}} & \textbf{1.00} \\ \hline
EWC+Reg.                                & \multicolumn{1}{c|}{0.85}          & \multicolumn{1}{c|}{0.86}          & \multicolumn{1}{c|}{\textbf{1.00}} & \textbf{1.00} \\ \hline
\multicolumn{1}{|l|}{\multirow{2}{*}{}} & \multicolumn{4}{l|}{MAP}                                                                                                     \\ \cline{2-5} 
\multicolumn{1}{|l|}{}                  & \multicolumn{1}{c|}{ST}            & \multicolumn{1}{c|}{NS}            & \multicolumn{1}{c|}{ST}            & NS            \\ \hline
EWC+LOC                                 & \multicolumn{1}{c|}{0.08}          & \multicolumn{1}{c|}{0.08}          & \multicolumn{1}{c|}{0.12}          & \textbf{0.12} \\ \hline
EWC+MLOC                                & \multicolumn{1}{c|}{\textbf{0.09}} & \multicolumn{1}{c|}{\textbf{0.09}} & \multicolumn{1}{c|}{0.12}          & 0.11          \\ \hline
EWC+VG                                  & \multicolumn{1}{c|}{0.08}          & \multicolumn{1}{c|}{\textbf{0.09}} & \multicolumn{1}{c|}{0.12}          & \textbf{0.12} \\ \hline
EWC+ALL                                 & \multicolumn{1}{c|}{\textbf{0.09}} & \multicolumn{1}{c|}{\textbf{0.09}} & \multicolumn{1}{c|}{0.12}          & \textbf{0.12} \\ \hline
EWC+Reg.                                & \multicolumn{1}{c|}{\textbf{0.09}} & \multicolumn{1}{c|}{\textbf{0.09}} & \multicolumn{1}{c|}{\textbf{0.13}} & \textbf{0.12} \\ \hline
\multicolumn{1}{|l|}{\multirow{2}{*}{}} & \multicolumn{4}{l|}{MRR}                                                                                                     \\ \cline{2-5} 
\multicolumn{1}{|l|}{}                  & \multicolumn{1}{c|}{ST}            & \multicolumn{1}{c|}{NS}            & \multicolumn{1}{c|}{ST}            & NS            \\ \hline
EWC+LOC                                 & \multicolumn{1}{c|}{\textbf{0.53}} & \multicolumn{1}{c|}{\textbf{0.45}} & \multicolumn{1}{c|}{0.57}          & 0.36          \\ \hline
EWC+MLOC                                & \multicolumn{1}{c|}{0.52}          & \multicolumn{1}{c|}{0.37}          & \multicolumn{1}{c|}{\textbf{0.59}} & 0.55          \\ \hline
EWC+VG                                  & \multicolumn{1}{c|}{0.27}          & \multicolumn{1}{c|}{0.31}          & \multicolumn{1}{c|}{0.57}          & 0.63          \\ \hline
EWC+ALL                                 & \multicolumn{1}{c|}{0.41}          & \multicolumn{1}{c|}{0.30}          & \multicolumn{1}{c|}{0.48}          & 0.48          \\ \hline
EWC+Reg.                                & \multicolumn{1}{c|}{0.32}          & \multicolumn{1}{c|}{0.39}          & \multicolumn{1}{c|}{0.58}          & \textbf{0.64} \\ \hline
\end{tabular}
}
%\begin{tablenotes}
 %    \item In bold are the best average performances per algorithm per metric for each project. 
 %\end{tablenotes}
 %\vspace{-1em}
\end{table}
%Rows RLOCATOR + Reg. and RLOCATOR + Ent. + Reg. refers to the enhanced

On non-stationary data, CLEAR with regression significantly outperforms RLOCATOR (with and without regression as well as with entropy) on Tomcat (with top@1 and top@5 metrics), Birt (top@5 and MRR metrics), AspectJ (top@1 metric), and Eclipse (top@5 metric) projects at the changeset-files level. %When adding entropy to RLOCATOR as well as regression CLEAR performs significantly better on Tomcat (top@1 metric), and AspectJ (MRR metric) projects.
Similarly, EWC CLEAR with regression significantly outperforms RLOCATOR (with and without regression as well as with entropy) on Birt (top@1 metric), AspectJ (top@1, top@5, top@10 and MRR metrics), and Eclipse (top@1 metric) projects at the changeset-files level. As of the comparison with FLIM, both CLEAR and EWC with regression perform better on non-stationary data of all projects except AspectJ (top@5 and top@10 metrics), and Birt (top@1, top@10, and MAP metrics). Only a small subset of source code files in the Tomcat, SWT, and AspectJ projects have Pre-released Bugs greater than 0 (see Section \ref{sec:rq1}). This limitation prevents FLIM from capturing sufficient information to accurately localize bugs in non-stationary settings. In contrast, our CL agents demonstrate better adaptation to new knowledge in these non-stationary settings. Similarly to their versions without regression, CLEAR and EWC require much less computational effort (up to 5x less effort). 
%Table \ref{tab:Algorithms evaluation across software projects with different  data granularities} reports the performance of the CL agents and FBL-BERT for QARC, QARCL and QD at changeset-files and hunks level across SWT, AspectJ and Tomcat software projects - the performance on stationary and non-stationary data are averaged for comparison with FBL-BERT. Compared to FBL-BERT for QARC, QARCL, and QD the CL agents with regression show similar performance as their performance without regression at both changeset-files and hunks level. They perform significantly better than the FBL-BERT metrics across the SWT project (with top@10), Tomcat project (with top@10 metric), and AspectJ (with top@5 and top@10 metrics). 

Table \ref{tab:Algorithms evaluation across software projects with different data granularity when incorporating bug-inducing factors on CL agents} and \ref{tab:Algorithms evaluation across PDE and Zxing projects with different data granularity when incorporating bug-inducing factors on CL agents} report the performance of the CL agents with regression and FBL-BERT for QARC, QARCL and QD at changeset-files and hunks levels across SWT, AspectJ, Tomcat, PDE, and Zxing software projects - the performance on stationary and non-stationary data are averaged for comparison with FBL-BERT. Compared to FBL-BERT for QARC, QARCL, and QD, the CL agents with regression show similar performance as their performance without regression at both changeset-files and hunks levels. They perform significantly better than the FBL-BERT metrics across the SWT project (with top@10), Tomcat project (with top@10 metric), AspectJ project (with top@5 and top@10 metrics), PDE project (with top@5 and top@10 metrics) and Zxing project (with top@5 and top@10 metrics). 

Given the diverse nature of software projects, developers must select the CL agent that best fits their project. Our results indicate significant performance differences between the CL agents, suggesting that the characteristics of the software project might be influencing this. To better assist developers in choosing the appropriate CL agent, we evaluate the overall forgetting performance of EWC and CLEAR with and without regression across the studied software projects at the changeset-files and hunks levels. Table \ref{tab:Forgetting statistics  of the CL agents across datasets.} shows the performance of the CL agents in terms of forgetting across the software projects at the changeset-files and hunks levels.
The range of forgetting values obtained when incorporating bug-inducing factors into the CL agents is similar to the results reported by Powers et al. \cite{powers2022cora}. This indicates that EWC+Reg. and CLEAR+Reg. at the changeset-files and hunks levels are more effective compared to EWC and CLEAR at mitigating forgetting while addressing the concept drift associated with data across software projects, which confirms the findings of our study. Moreover, the forgetting values of CL agents suggest that for complex projects such as Tomcat — given its high number of file changes in commits between the time a bug is fixed and is reported (reported in Table \ref{tab:benchmark-dataset}), any of the CL agents with regression applied might be more suited.
\begin{tcolorbox}
    \textbf{Finding 3:} Incorporating prior knowledge about bug-inducing factors through regression improves the performance of both CLEAR and EWC, outperforming their non-regression counterparts on four out of seven projects at the changeset-files level and three out of seven projects at the hunks level across all ranking metrics.
\end{tcolorbox}
\subsection{\textbf{Ablation study on using bug-inducing factors}} \label{sec:Ablation study on using bug-inducing factors}
To validate our regression approach for enhancing CL agents, we analyze the impact of different bug-inducing factor configurations, as shown in Table \ref{tab:measured-metrics}, when added to the reward function. For each CL agent (i.e., CLEAR and EWC), we consider as an additional component of their reward functions 1) each bug-inducing factor value (i.e., LOC, MLOC, VG, PRE, Churn), 2) the sum of all bug-inducing factor values (i.e., LOC+MLOC+VG+PRE+Churn), and 3) the output of the logistic regression model with PRE and Churn as independent, statistically significant and non-collinear variables (the configuration used in this paper to answer our RQs).
\begin{table}[t]
\caption{Forgetting ($\mathcal{F}$) statistics (mean ± std) of the CL agents across datasets.}
\label{tab:Forgetting statistics  of the CL agents across datasets.}
\resizebox{\textwidth}{!}{
\centering
\begin{tabular}{|c|c|c|c|c|c|c|c|c|c|}
\hline
-files  & CLEAR     & CLEAR + Reg.                      & EWC       & EWC + Reg.                        & hunks   & CLEAR     & CLEAR + Reg.                    & EWC       & EWC + Reg.                        \\ \hline
AspectJ & 3.9±0.1   & 1.3 ± 0.3                         & 3.6±0.1   & \cellcolor[HTML]{C0C0C0}1.1 ± 0.0 & AspectJ & 3.7±0.2   & \cellcolor[HTML]{656565}1.2±0.2 & 3.6±0.0   & 1.1±0.1                           \\ \hline
SWT     & 3.1 ± 1.4 & \cellcolor[HTML]{656565}0.9 ± 0.2 & 3.6 ± 0.3 & \cellcolor[HTML]{C0C0C0}1.0 ± 0.1 & SWT     & 3.6±0.1   & \cellcolor[HTML]{C0C0C0}1.1±0.1 & 3.2±1.5   & \cellcolor[HTML]{C0C0C0}1.0±0.1   \\ \hline
Birt    & 3.7±0.3   & \cellcolor[HTML]{656565}0.7±0.0   & 3.6±0.3   & \cellcolor[HTML]{C0C0C0}0.7 ± 0.0 & Birt    & 3.9±0.5   & \cellcolor[HTML]{656565}0.7±0.0 & 3.7±0.2   & 0.8±0.4                           \\ \hline
Eclipse & 3.6 ± 0.2 & 1.1±0.2                           & 3.5 ± 0.2 & \cellcolor[HTML]{C0C0C0}0.5 ± 0.2 & Eclipse & 3.7±0.2   & 0.8 ± 0.4                       & 3.6±0.2   & \cellcolor[HTML]{656565}0.2±0.1   \\ \hline
Tomcat  & 3.1 ± 1.1 & 1.1 ± 0.1                         & 3.7±0.4   & 0.7 ± 0.5                         & Tomcat  & 3.9±0.1   & \cellcolor[HTML]{C0C0C0}0.6±0.5 & 3.8±0.1   & \cellcolor[HTML]{656565}0.5±0.5   \\ \hline
PDE     & 5.6 ± 0.9 & 0.7 ± 0.1                         & 5.3 ± 2.6 & \cellcolor[HTML]{C0C0C0}0.5 ± 0.3 & PDE     & 5.6 ± 2.1 & 0.8 ± 0.1                       & 6.0 ± 1.5 & \cellcolor[HTML]{656565}0.3 ± 0.3 \\ \hline
Zxing   & 4.8 ± 0.7 & 0.5 ± 0.2                         & 3.5 ± 1.4 & \cellcolor[HTML]{C0C0C0}0.3 ± 0.3 & Zxing   & 5.3 ± 1.2 & 0.5 ± 0.3                       & 4.4 ± 0.4 & \cellcolor[HTML]{656565}0.1 ± 0.1 \\ \hline
\end{tabular}
}
\begin{tablenotes}
     \item The best results are highlighted in dark grey.  "-files"=changeset-files. 
 \end{tablenotes}
\end{table}

Table \ref{tab:Ablation study1: different bug report metrics added to CLEAR.}, \ref{tab:Ablation study2: different bug report metrics added to EWC}, \ref{tab:Ablation study3: different bug report metrics added to CLEAR}, \ref{tab:Ablation study4: different bug report metrics added to EWC}, \ref{tab:Ablation study5: different bug report metrics added to CLEAR.}, \ref{tab:Ablation study6: different bug report metrics added to EWC.}, \ref{tab:Ablation study7: different bug report metrics added to CLEAR.} and \ref{tab:Ablation study8: different bug report metrics added to EWC.} report the performance of the CL agents, enhanced with different configurations of the bug-inducing factors, across the studied software projects on stationary and non-stationary changeset-files. Compared to other configurations, CLEAR with regression shows improvements of up to 26\% in top@1 (across SWT, Birt, and Tomcat), 27\% in top@5 (across Eclipse, SWT, Zxing and AspectJ), 21\% in top@10 (across SWT, Zxing and AspectJ), 25\% in MAP (across all projects), and 26\% in MRR (across Eclipse and Tomcat). Similarly, compared to other configurations, EWC with regression shows improvements of up to 73\% in top@1 (across  SWT, and Birt and AspectJ), 18\% in top@5 (across Eclipse, SWT, and AspectJ), 15\% in top@10 (across SWT and Zxing ), 22\% in MAP (across all projects), and 105\% in MRR (across AspectJ). Therefore, boosting the CL agents with bug-inducing factors through regression is suitable for bug localization.

\section{Discussions}
\label{sec:Discussions}
DL-based bug localization techniques are becoming more prevalent to assist developers in maintaining software projects. Nevertheless, they often do not consider the dynamic nature of software projects, typically depending on stationary data. Our research shows that CL agents can perform at the same level or better than traditional DL-based techniques while significantly lowering the computational effort needed for training by at least half. Further, without additional cost to the training, we incorporate bug-inducing factors into the reward function of CL agents. Our results show an improvement in the performance of the CL agents.

Beyond the bug localization task, other software engineering tasks (e.g., software defect prediction \cite{gangwar2023concept}, malware detection \cite{singh2012tracking}, and test case prioritization \cite{bagherzadeh2021reinforcement}) might be affected by concept drift associated with data that can cause DL-based techniques to drift and degrade in performance \cite{Olewicki2024Costs,gangwar2023concept,ekanayake2009tracking,singh2012tracking,bagherzadeh2021reinforcement}. For example, Gangwar et al. \cite{gangwar2023concept} show that software defect prediction models become outdated due to the concept drift associated with software metrics data. We recommend utilizing CL agents as a tool for software defect prediction, training them on temporally diverse data from the given project to predict defects in future releases. Singh et al. \cite{singh2012tracking} investigate how malware evolves over time to defeat detection, which can lead to non-stationary malware data. These non-stationary malware data can nullify the performance of existing DL-based techniques for malware detection which are trained on static malware data \cite{singh2012tracking}. In this context, CL agents trained on malware data, collected over time, can learn useful malware patterns to detect future malware. Bagherzadeh et al. \cite{bagherzadeh2021reinforcement} employ DRL to learn a test case prioritization strategy. The test case prioritization approach trains the model continuously on continuous integration cycles to address the concept drift associated with the evolution of test cases. Their results show up to 177.5 hours of training for some datasets, which can be more for larger datasets. The simple architecture of the CL agents can be an option in the context of test case prioritization to reduce the computational effort for training while remaining effective at prioritizing test cases. Our proposed solution can be applied to other SE tasks, as well as our findings that CL agents can adapt to changes in data and retain appropriate context when evaluated on either stationary or non-stationary data.
%Software defect prediction models are trained on static software metrics data, which can unexpectedly change over time \cite{ekanayake2009tracking}.

\section{Threats to validity}
\label{threats}
Our study has some limitations that we discuss in detail in this section.

\emph{Conclusion validity.} In general, the conclusion’s limitations concern the degree of accuracy of the statistical findings on the different steps in our study. Regarding the logistic regression model, we remove the insignificant variable based on a threshold \emph{p-value} of 0.05, as commonly used by other studies. In the same way, to handle the multicollinearity of the model's variables, we compute the VIF considering a maximum value of 2.5, as suggested in the literature. We use Welch's ANOVA post-hoc tests as statistical tests. The significance level is set to 0.05 which is standard across the literature as shown by Welch et al. \cite{welch1947generalization} and Games et al. \cite{games1976pairwise}. The non-deterministic nature of DL-based algorithms can also threaten the conclusions made in this work. We address this by collecting results from 5 independent runs for all our experiments and reporting the average. 

\emph{Internal validity.} In this study, we excluded the JDT project from our evaluation because we could not collect relevant non-stationary data for it. Specifically, when collecting non-stationary data from the JDT project (i.e., all changeset-files modified by commits made between the dates of the bug report and the corresponding bug-fixing commit), we found that none of the changeset-files matched our oracle. Here, the oracle refers to the ground truth list of changeset-files that are directly related to resolving a specific bug report.

%We could not use the whole dataset, as non-stationary data was collected for the project JDT. As we collect the changes applied on the oracle \Foutse{what do you mean by 'oracle' of source files? please clarify this sentence!} of source code files reported in the bug report, although we collected the commits between \Foutse{what are these dates exactly? what is the bug report? do you mean reported date of the bug? what is a commit bug date? are you referring to bug inducing commit? bug fixing commit? it is unclear, please be precise!} the \emph{bug report} and \emph{commit bug} dates, we did not observe changes in these commits applied in the target source code files.

\emph{Construct validity.} Regarding evaluating our model, the measures adopted here might not reflect real-world situations. 
Such a threat is mitigated by considering evaluation measures commonly and largely used by related studies~\cite{chakraborty2023rlocator, ye2015mapping,liang2022modeling}. Furthermore, these measures represent the best available ones for measuring and comparing the performance of our model and related studies on information retrieval-based bug localization.

\emph{External validity.} 
Our results are based on a dataset of bugs from open-source projects written in Java, which may limit their generalizability to other projects. To further validate our findings, additional evaluations are needed on projects written in different programming languages and developed in varied settings with diverse characteristics. However, previous studies have also conducted their investigations based on the same dataset evaluated here and the assessed methods are language-agnostic.

\emph{Reliability validity.} 
To allow other researchers to reproduce or build on our research, we provide a replication package \cite{replication-package}. 

\section{Related work} 
\label{Related work}

In this section, we discuss related work to our current study. First, we discuss different techniques for bug localization that have been applied over time. Second, we discuss the challenges of dealing with concept drift in machine learning models.

\subsection{Bug Localization Techniques}
%\Foutse{please carefully proofread and revise this section, there are many description of related work here that are hard to understand!}
Previous studies have investigated different approaches to deal with bug localization.
Unlike our approach, some previous studies investigate the effort spent by developers to manually reproduce bugs, covering aspects related to bug monitoring, replaying, and reporting.
For example, Moran et al.~\cite{moran2016automatically} propose a tool to augment crash detection reports by exploring input generation for Android apps.
Yu et al.~\cite{yu2017descry} present a tool to support the reproduction of bugs based on log messages collected from system executions.
Roehm et al.~\cite{roehm2013monitoring} explores the interaction between users and their applications, focusing on predicting failures based on a taxonomy of previously monitored user interactions.

Over time, many studies have investigated different information retrieval techniques, exploring fine-grained information at different levels while associating them with a bug~\cite{wang2013improving, rath2018analyzing, mills2018bug}.
%The results show that the proposed hybrid multi-objective approach achieves success for 78\% of the bug reports analyzed. %, when a given current project does not have enough information.
Ranking changeset-files associated with a bug was one of the primary techniques used for bug localization.
Chakraborty et al. \cite{chakraborty2023rlocator} and Liang et al. \cite{liang2022modeling}, our baseline studies, are DL-based techniques that rank a set of changeset-files based on their relevance to a bug report.
Ye et al.~\cite{ye2015mapping} evaluate their approach on the same software projects we consider for our evaluation in this study. Specifically, they leverage project-specific characteristic such as functional decomposition, API descriptions, bug-fixing history, code change history, and file dependency graphs to rank changeset-files relevant to a bug report using a learning-to-rank technique trained on previously resolved bug reports.
%To rank the files, they leverage the knowledge from a given project by exploring the functional decomposition of source code, API descriptions of libraries, bug-fixing, and history of code changes.
Wang et al. \cite{wang2020multi} present a CNN model based on their initial exploration of features from different dimensions between a bug report and a changeset-file. 
Huo et al.~\cite{huo2019deep} take a different route by proposing TRANP-CNN, a DL approach for extracting semantic features from source code projects and using them for cross-project bug localization.
Similarly, Miryeganeh et al.~\cite{miryeganeh2021globug} present Globug, a framework that aims to improve the accuracy of previously pre-trained models for fault localization. 
The authors claim to leverage external knowledge to a given project, considering such an initial training dataset may not be rich enough (historical bug reports). 
This way, they advocate exploring external data (other projects) to calculate better the
textual similarities of a new bug report to other historical bug reports or source code elements.
These external data contain several open-source projects, later used to evaluate the potential of adopting a word embedding technique based on Doc2Vec. 
%\Foutse{based on models pre-trained on global data from bug reports? what is global data?.} based on pre-trained models on global data from bug reports. These data\Foutse{which data?} tackle \Foutse{contain?} several open-source projects used to evaluate the potential of adopting a word embedding technique based on Doc2Vec. 
Further exploring fine-grained information from source code, previous studies have investigated ranking methods associated with bugs rather than changeset-files, as recommending methods can significantly reduce the developer's inspection effort \cite{almhana2021method}. 
Almhana et al.~\cite{almhana2021method} propose to rank the methods associated with a bug based on (i) the history of changes and bug-fixing, and (ii) the lexical similarity between the bug report description and the API documentation. 

%Moving forward \Foutse{what do you mean by 'moving forward'? from where?}, 
Previous studies also explore bug localization techniques based on data that considers both source code changes and associated metadata.
For example, changesets and hunks are information extracted from version control systems that provide metadata about ongoing changes.
Ciborowska and Damevski \cite{ciborowska2022fast} investigate how to leverage %\Foutse{how to leverage BERT?} 
BERT to match the semantics reported in the bug reports with the inducing changesets (e.g., changeset-files and hunks). 
Similarly, Wu et al. \cite{wu2018changelocator} propose ChangeLocator, a model for locating crash-inducing changes (changesets) based on crash reports.
Du and Yu \cite{du2023pre} propose a Semantic Flow Graph, responsible for capturing the source code semantics.
Wen et al. \cite{wen2016locus} investigate how hunks, which are fine-grained segments of information consisting of changed and context lines within changeset files, could be used in a Vector Space Model (VSM) to retrieve bug-related information.
Although the previous studies present interesting and promising results, they do not cover the problem of concept drift, as we explore here.

\subsection{Addressing Concept Drift}
Similar to our study, Olewicki et al.~\cite{Olewicki2024Costs} investigated how to overcome concept drift caused by the changing nature of development activities. They explore lifelong learning, which is a paradigm that dynamically evolves a given model based on a stream of information~\cite{parisi2019continual}.
Olewicki et al. conducted a case study with two different tools, responsible for detecting brown builds~\cite{olewicki2022towards} and just-in-time risk prediction~\cite{nayrolles2018clever}.
They report that the lifelong learning framework helps speed up the training process; resulting in updates taking 3.3 - 13.7 times less data than the state-of-the-art framework.

Regarding the catastrophic forgetting problem,  Kirkpatrick et al.~\cite{kirkpatrick2017overcoming} proposed an elastic weight consolidation algorithm, which remembers the knowledge of old tasks by selectively decreasing the plasticity of the weights important for the tasks under evaluation.
Specifically in the context of bug localization problems, Lee et al.~\cite{lee2022light} investigate how to assign appropriate developers to fix specific bugs, presenting the framework LBT-P.
For that, the authors developed a fine-tuning method to preserve language knowledge.
Similar to previous studies, in this work, we aim to handle the catastrophic forgetting problem by exploring continual learning and showing relevant results compared to related work.

%\lm{New Related Work \cite{du2023pre, ciborowska2022fast, wang2020multi, ye2015mapping, wu2018changelocator}}

\section{Conclusion} \label{Conclusion}
In this paper, we propose a CL framework for bug localization. We compare the effectiveness of two configurations of  CL agents, with and without regression, against three baseline studies on stationary and non-stationary data (i.e., changeset-files and hunks). The CL agents employ rehearsal and regularization approaches to mitigate catastrophic forgetting while being trained cyclically in sequence on stationary and non-stationary data from seven software projects. Our results show that CL agents can achieve comparable or better performance than the baselines on both stationary and non-stationary data across five software projects while requiring less computational effort during training (on six out of the seven projects). Additionally, incorporating bug-inducing factors as prior knowledge for the CL agents significantly improves their performance. 

While our work advances the state of the art in the use of CL agents for bug localization, some issues remain open that should be tackled by future work. In the following, we present the most important ones:
\begin{itemize}
    \item We leverage CL techniques with their default hyperparameters. Therefore,  optimization and tuning to get the best CL agents for the bug localization task is a natural next step in this work. Hyperparameters tuning frameworks like Optuna \cite{akiba2019optuna} can be used.
    \item During our evaluation, we used datasets from existing software projects made available by previous studies. We observe that the performance of CL agents varies depending on the size of available data. The availability of more diverse datasets could be useful in this research area to evaluate existing techniques.
    \item Concept drift is also prevalent in other software engineering tasks, hence there is a need for effective techniques that can adapt to data that evolves over time. Another next step for this work is to use CL agents for other software engineering tasks.
    \item In this paper, we evaluate CL agents at two data granularity levels: changeset-files and hunks. While the commits level (i.e., a set of hunks), which spans code changes across multiple files, has also been studied \cite{ciborowska2022fast}, we focus on selecting and ranking individual files or hunks. Future work could explore environments where agents select and rank entire commits level data.
\end{itemize}

\bibliographystyle{ACM-Reference-Format}
\bibliography{sample-base}


\begin{thebibliography}{83}

%%% the Web address.

\ifx \showCODEN    \undefined \def \showCODEN     #1{\unskip}     \fi
\ifx \showDOI      \undefined \def \showDOI       #1{#1}\fi
\ifx \showISBNx    \undefined \def \showISBNx     #1{\unskip}     \fi
\ifx \showISBNxiii \undefined \def \showISBNxiii  #1{\unskip}     \fi
\ifx \showISSN     \undefined \def \showISSN      #1{\unskip}     \fi
\ifx \showLCCN     \undefined \def \showLCCN      #1{\unskip}     \fi
\ifx \shownote     \undefined \def \shownote      #1{#1}          \fi
\ifx \showarticletitle \undefined \def \showarticletitle #1{#1}   \fi
\ifx \showURL      \undefined \def \showURL       {\relax}        \fi
% invisible to TeX
\providecommand\bibfield[2]{#2}
\providecommand\bibinfo[2]{#2}
\providecommand\natexlab[1]{#1}
\providecommand\showeprint[2][]{arXiv:#2}

\bibitem[Lin(2012)]%
        {Link-for-Bug-384108}
 \bibinfo{year}{2012}\natexlab{}.
\newblock \bibinfo{title}{Link for Bug - 384108}.
\newblock \bibinfo{howpublished}{\url{https://bugs.eclipse.org/bugs/show\_bug.cgi?id=384108}}.
\newblock


\bibitem[Lin(2013)]%
        {Link-for-Bug-420210}
 \bibinfo{year}{2013}\natexlab{}.
\newblock \bibinfo{title}{Link for Bug - 420210}.
\newblock \bibinfo{howpublished}{\url{https://bugs.eclipse.org/bugs/show_bug.cgi?id=420210}}.
\newblock


\bibitem[All(2017)]%
        {Alliance}
 \bibinfo{year}{2017}\natexlab{}.
\newblock \bibinfo{title}{Alliance}.
\newblock \bibinfo{howpublished}{\url{https://docs.alliancecan.ca/wiki/Cedar}}.
\newblock


\bibitem[scc(2018)]%
        {scc-tool}
 \bibinfo{year}{2018}\natexlab{}.
\newblock \bibinfo{title}{scc tool}.
\newblock \bibinfo{howpublished}{\url{https://github.com/boyter/scc}}.
\newblock


\bibitem[Ecl(2022)]%
        {Eclipse-Apache-project}
 \bibinfo{year}{2022}\natexlab{}.
\newblock \bibinfo{title}{Eclipse Apache project}.
\newblock \bibinfo{howpublished}{\url{https://github.com/eclipse-platform/eclipse.platform.ui.git}}.
\newblock


\bibitem[rep(2024)]%
        {replication-package}
 \bibinfo{year}{2024}\natexlab{}.
\newblock \bibinfo{title}{Replication Package}.
\newblock \bibinfo{howpublished}{\url{https://zenodo.org/records/14271134}}.
\newblock


\bibitem[Abel et~al\mbox{.}(2024)]%
        {abel2024definition}
\bibfield{author}{\bibinfo{person}{David Abel}, \bibinfo{person}{Andr{\'e} Barreto}, \bibinfo{person}{Benjamin Van~Roy}, \bibinfo{person}{Doina Precup}, \bibinfo{person}{Hado~P van Hasselt}, {and} \bibinfo{person}{Satinder Singh}.} \bibinfo{year}{2024}\natexlab{}.
\newblock \showarticletitle{A definition of continual reinforcement learning}.
\newblock \bibinfo{journal}{\emph{Advances in Neural Information Processing Systems}}  \bibinfo{volume}{36} (\bibinfo{year}{2024}).
\newblock


\bibitem[Akiba et~al\mbox{.}(2019)]%
        {akiba2019optuna}
\bibfield{author}{\bibinfo{person}{Takuya Akiba}, \bibinfo{person}{Shotaro Sano}, \bibinfo{person}{Toshihiko Yanase}, \bibinfo{person}{Takeru Ohta}, {and} \bibinfo{person}{Masanori Koyama}.} \bibinfo{year}{2019}\natexlab{}.
\newblock \showarticletitle{Optuna: A next-generation hyperparameter optimization framework}. In \bibinfo{booktitle}{\emph{Proceedings of the 25th ACM SIGKDD international conference on knowledge discovery \& data mining}}. \bibinfo{pages}{2623--2631}.
\newblock


\bibitem[Almhana et~al\mbox{.}(2021)]%
        {almhana2021method}
\bibfield{author}{\bibinfo{person}{Rafi Almhana}, \bibinfo{person}{Marouane Kessentini}, {and} \bibinfo{person}{Wiem Mkaouer}.} \bibinfo{year}{2021}\natexlab{}.
\newblock \showarticletitle{Method-level bug localization using hybrid multi-objective search}.
\newblock \bibinfo{journal}{\emph{Information and Software Technology}}  \bibinfo{volume}{131} (\bibinfo{year}{2021}), \bibinfo{pages}{106474}.
\newblock


\bibitem[Antoniol and Gu{\'e}h{\'e}neuc(2005)]%
        {antoniol2005feature}
\bibfield{author}{\bibinfo{person}{Giuliano Antoniol} {and} \bibinfo{person}{Y-G Gu{\'e}h{\'e}neuc}.} \bibinfo{year}{2005}\natexlab{}.
\newblock \showarticletitle{Feature identification: a novel approach and a case study}. In \bibinfo{booktitle}{\emph{21st IEEE International Conference on Software Maintenance (ICSM'05)}}. IEEE, \bibinfo{pages}{357--366}.
\newblock


\bibitem[Bagherzadeh et~al\mbox{.}(2021)]%
        {bagherzadeh2021reinforcement}
\bibfield{author}{\bibinfo{person}{Mojtaba Bagherzadeh}, \bibinfo{person}{Nafiseh Kahani}, {and} \bibinfo{person}{Lionel Briand}.} \bibinfo{year}{2021}\natexlab{}.
\newblock \showarticletitle{Reinforcement learning for test case prioritization}.
\newblock \bibinfo{journal}{\emph{IEEE Transactions on Software Engineering}} \bibinfo{volume}{48}, \bibinfo{number}{8} (\bibinfo{year}{2021}), \bibinfo{pages}{2836--2856}.
\newblock


\bibitem[Bellemare et~al\mbox{.}(2013)]%
        {bellemare2013arcade}
\bibfield{author}{\bibinfo{person}{Marc~G Bellemare}, \bibinfo{person}{Yavar Naddaf}, \bibinfo{person}{Joel Veness}, {and} \bibinfo{person}{Michael Bowling}.} \bibinfo{year}{2013}\natexlab{}.
\newblock \showarticletitle{The arcade learning environment: An evaluation platform for general agents}.
\newblock \bibinfo{journal}{\emph{Journal of Artificial Intelligence Research}}  \bibinfo{volume}{47} (\bibinfo{year}{2013}), \bibinfo{pages}{253--279}.
\newblock


\bibitem[Chakraborty et~al\mbox{.}(2023)]%
        {chakraborty2023rlocator}
\bibfield{author}{\bibinfo{person}{Partha Chakraborty}, \bibinfo{person}{Mahmoud Alfadel}, {and} \bibinfo{person}{Meiyappan Nagappan}.} \bibinfo{year}{2023}\natexlab{}.
\newblock \showarticletitle{RLocator: Reinforcement Learning for Bug Localization}.
\newblock \bibinfo{journal}{\emph{arXiv preprint arXiv:2305.05586}} (\bibinfo{year}{2023}).
\newblock


\bibitem[Chen et~al\mbox{.}(2022)]%
        {chen2022adaptive}
\bibfield{author}{\bibinfo{person}{Xiaoyu Chen}, \bibinfo{person}{Xiangming Zhu}, \bibinfo{person}{Yufeng Zheng}, \bibinfo{person}{Pushi Zhang}, \bibinfo{person}{Li Zhao}, \bibinfo{person}{Wenxue Cheng}, \bibinfo{person}{Peng Cheng}, \bibinfo{person}{Yongqiang Xiong}, \bibinfo{person}{Tao Qin}, \bibinfo{person}{Jianyu Chen}, {et~al\mbox{.}}} \bibinfo{year}{2022}\natexlab{}.
\newblock \showarticletitle{An adaptive deep rl method for non-stationary environments with piecewise stable context}.
\newblock \bibinfo{journal}{\emph{Advances in Neural Information Processing Systems}}  \bibinfo{volume}{35} (\bibinfo{year}{2022}), \bibinfo{pages}{35449--35461}.
\newblock


\bibitem[Chen and Lin(2014)]%
        {chen2014big}
\bibfield{author}{\bibinfo{person}{Xue-Wen Chen} {and} \bibinfo{person}{Xiaotong Lin}.} \bibinfo{year}{2014}\natexlab{}.
\newblock \showarticletitle{Big data deep learning: challenges and perspectives}.
\newblock \bibinfo{journal}{\emph{IEEE access}}  \bibinfo{volume}{2} (\bibinfo{year}{2014}), \bibinfo{pages}{514--525}.
\newblock


\bibitem[Chidamber and Kemerer(1994)]%
        {chidamber1994metrics}
\bibfield{author}{\bibinfo{person}{Shyam~R Chidamber} {and} \bibinfo{person}{Chris~F Kemerer}.} \bibinfo{year}{1994}\natexlab{}.
\newblock \showarticletitle{A metrics suite for object oriented design}.
\newblock \bibinfo{journal}{\emph{IEEE Transactions on software engineering}} \bibinfo{volume}{20}, \bibinfo{number}{6} (\bibinfo{year}{1994}), \bibinfo{pages}{476--493}.
\newblock


\bibitem[Ciborowska and Damevski(2022)]%
        {ciborowska2022fast}
\bibfield{author}{\bibinfo{person}{Agnieszka Ciborowska} {and} \bibinfo{person}{Kostadin Damevski}.} \bibinfo{year}{2022}\natexlab{}.
\newblock \showarticletitle{Fast changeset-based bug localization with BERT}. In \bibinfo{booktitle}{\emph{Proceedings of the 44th International Conference on Software Engineering}}. \bibinfo{pages}{946--957}.
\newblock


\bibitem[Corley et~al\mbox{.}(2018)]%
        {corley2018changeset}
\bibfield{author}{\bibinfo{person}{Christopher~S Corley}, \bibinfo{person}{Kostadin Damevski}, {and} \bibinfo{person}{Nicholas~A Kraft}.} \bibinfo{year}{2018}\natexlab{}.
\newblock \showarticletitle{Changeset-based topic modeling of software repositories}.
\newblock \bibinfo{journal}{\emph{IEEE Transactions on Software Engineering}} \bibinfo{volume}{46}, \bibinfo{number}{10} (\bibinfo{year}{2018}), \bibinfo{pages}{1068--1080}.
\newblock


\bibitem[Cotroneo et~al\mbox{.}(2016)]%
        {cotroneo2016bugs}
\bibfield{author}{\bibinfo{person}{Domenico Cotroneo}, \bibinfo{person}{Roberto Pietrantuono}, \bibinfo{person}{Stefano Russo}, {and} \bibinfo{person}{Kishor Trivedi}.} \bibinfo{year}{2016}\natexlab{}.
\newblock \showarticletitle{How do bugs surface? A comprehensive study on the characteristics of software bugs manifestation}.
\newblock \bibinfo{journal}{\emph{Journal of Systems and Software}}  \bibinfo{volume}{113} (\bibinfo{year}{2016}), \bibinfo{pages}{27--43}.
\newblock


\bibitem[Da~Silva et~al\mbox{.}(2024)]%
        {da2024detecting}
\bibfield{author}{\bibinfo{person}{L{\'e}uson Da~Silva}, \bibinfo{person}{Paulo Borba}, \bibinfo{person}{Toni Maciel}, \bibinfo{person}{Wardah Mahmood}, \bibinfo{person}{Thorsten Berger}, \bibinfo{person}{Jo{\~a}o Moisakis}, \bibinfo{person}{Aldiberg Gomes}, {and} \bibinfo{person}{Vin{\'\i}cius Leite}.} \bibinfo{year}{2024}\natexlab{}.
\newblock \showarticletitle{Detecting semantic conflicts with unit tests}.
\newblock \bibinfo{journal}{\emph{Journal of Systems and Software}}  \bibinfo{volume}{214} (\bibinfo{year}{2024}), \bibinfo{pages}{112070}.
\newblock


\bibitem[Da~Silva et~al\mbox{.}(2022)]%
        {da2022build}
\bibfield{author}{\bibinfo{person}{L{\'e}uson Da~Silva}, \bibinfo{person}{Paulo Borba}, {and} \bibinfo{person}{Arthur Pires}.} \bibinfo{year}{2022}\natexlab{}.
\newblock \showarticletitle{Build conflicts in the wild}.
\newblock \bibinfo{journal}{\emph{Journal of Software: Evolution and Process}} \bibinfo{volume}{34}, \bibinfo{number}{4} (\bibinfo{year}{2022}), \bibinfo{pages}{e2441}.
\newblock


\bibitem[Du and Yu(2023)]%
        {du2023pre}
\bibfield{author}{\bibinfo{person}{Yali Du} {and} \bibinfo{person}{Zhongxing Yu}.} \bibinfo{year}{2023}\natexlab{}.
\newblock \showarticletitle{Pre-training code representation with semantic flow graph for effective bug localization}. In \bibinfo{booktitle}{\emph{Proceedings of the 31st ACM Joint European Software Engineering Conference and Symposium on the Foundations of Software Engineering}}. \bibinfo{pages}{579--591}.
\newblock


\bibitem[Ekanayake et~al\mbox{.}(2009)]%
        {ekanayake2009tracking}
\bibfield{author}{\bibinfo{person}{Jayalath Ekanayake}, \bibinfo{person}{Jonas Tappolet}, \bibinfo{person}{Harald~C Gall}, {and} \bibinfo{person}{Abraham Bernstein}.} \bibinfo{year}{2009}\natexlab{}.
\newblock \showarticletitle{Tracking concept drift of software projects using defect prediction quality}. In \bibinfo{booktitle}{\emph{2009 6th IEEE International Working Conference on Mining Software Repositories}}. IEEE, \bibinfo{pages}{51--60}.
\newblock


\bibitem[Espeholt et~al\mbox{.}(2018)]%
        {espeholt2018impala}
\bibfield{author}{\bibinfo{person}{Lasse Espeholt}, \bibinfo{person}{Hubert Soyer}, \bibinfo{person}{Remi Munos}, \bibinfo{person}{Karen Simonyan}, \bibinfo{person}{Vlad Mnih}, \bibinfo{person}{Tom Ward}, \bibinfo{person}{Yotam Doron}, \bibinfo{person}{Vlad Firoiu}, \bibinfo{person}{Tim Harley}, \bibinfo{person}{Iain Dunning}, {et~al\mbox{.}}} \bibinfo{year}{2018}\natexlab{}.
\newblock \showarticletitle{Impala: Scalable distributed deep-rl with importance weighted actor-learner architectures}. In \bibinfo{booktitle}{\emph{International conference on machine learning}}. PMLR, \bibinfo{pages}{1407--1416}.
\newblock


\bibitem[Feng et~al\mbox{.}(2020)]%
        {feng2020codebert}
\bibfield{author}{\bibinfo{person}{Zhangyin Feng}, \bibinfo{person}{Daya Guo}, \bibinfo{person}{Duyu Tang}, \bibinfo{person}{Nan Duan}, \bibinfo{person}{Xiaocheng Feng}, \bibinfo{person}{Ming Gong}, \bibinfo{person}{Linjun Shou}, \bibinfo{person}{Bing Qin}, \bibinfo{person}{Ting Liu}, \bibinfo{person}{Daxin Jiang}, {et~al\mbox{.}}} \bibinfo{year}{2020}\natexlab{}.
\newblock \showarticletitle{Codebert: A pre-trained model for programming and natural languages}.
\newblock \bibinfo{journal}{\emph{arXiv preprint arXiv:2002.08155}} (\bibinfo{year}{2020}).
\newblock


\bibitem[Games and Howell(1976)]%
        {games1976pairwise}
\bibfield{author}{\bibinfo{person}{Paul~A Games} {and} \bibinfo{person}{John~F Howell}.} \bibinfo{year}{1976}\natexlab{}.
\newblock \showarticletitle{Pairwise multiple comparison procedures with unequal n’s and/or variances: a Monte Carlo study}.
\newblock \bibinfo{journal}{\emph{Journal of Educational Statistics}} \bibinfo{volume}{1}, \bibinfo{number}{2} (\bibinfo{year}{1976}), \bibinfo{pages}{113--125}.
\newblock


\bibitem[Gangwar and Kumar(2023)]%
        {gangwar2023concept}
\bibfield{author}{\bibinfo{person}{Arvind~Kumar Gangwar} {and} \bibinfo{person}{Sandeep Kumar}.} \bibinfo{year}{2023}\natexlab{}.
\newblock \showarticletitle{Concept Drift in Software Defect Prediction: A Method for Detecting and Handling the Drift}.
\newblock \bibinfo{journal}{\emph{ACM Transactions on Internet Technology}} \bibinfo{volume}{23}, \bibinfo{number}{2} (\bibinfo{year}{2023}), \bibinfo{pages}{1--28}.
\newblock


\bibitem[Gormley and Tong(2015)]%
        {gormley2015elasticsearch}
\bibfield{author}{\bibinfo{person}{Clinton Gormley} {and} \bibinfo{person}{Zachary Tong}.} \bibinfo{year}{2015}\natexlab{}.
\newblock \bibinfo{booktitle}{\emph{Elasticsearch: the definitive guide: a distributed real-time search and analytics engine}}.
\newblock \bibinfo{publisher}{" O'Reilly Media, Inc."}.
\newblock


\bibitem[Huo et~al\mbox{.}(2019)]%
        {huo2019deep}
\bibfield{author}{\bibinfo{person}{Xuan Huo}, \bibinfo{person}{Ferdian Thung}, \bibinfo{person}{Ming Li}, \bibinfo{person}{David Lo}, {and} \bibinfo{person}{Shu-Ting Shi}.} \bibinfo{year}{2019}\natexlab{}.
\newblock \showarticletitle{Deep transfer bug localization}.
\newblock \bibinfo{journal}{\emph{IEEE Transactions on software engineering}} \bibinfo{volume}{47}, \bibinfo{number}{7} (\bibinfo{year}{2019}), \bibinfo{pages}{1368--1380}.
\newblock


\bibitem[Johnson et~al\mbox{.}(2019)]%
        {johnson2019billion}
\bibfield{author}{\bibinfo{person}{Jeff Johnson}, \bibinfo{person}{Matthijs Douze}, {and} \bibinfo{person}{Herv{\'e} J{\'e}gou}.} \bibinfo{year}{2019}\natexlab{}.
\newblock \showarticletitle{Billion-scale similarity search with GPUs}.
\newblock \bibinfo{journal}{\emph{IEEE Transactions on Big Data}} \bibinfo{volume}{7}, \bibinfo{number}{3} (\bibinfo{year}{2019}), \bibinfo{pages}{535--547}.
\newblock


\bibitem[Kim et~al\mbox{.}(2013)]%
        {kim2013should}
\bibfield{author}{\bibinfo{person}{Dongsun Kim}, \bibinfo{person}{Yida Tao}, \bibinfo{person}{Sunghun Kim}, {and} \bibinfo{person}{Andreas Zeller}.} \bibinfo{year}{2013}\natexlab{}.
\newblock \showarticletitle{Where should we fix this bug? a two-phase recommendation model}.
\newblock \bibinfo{journal}{\emph{IEEE transactions on software Engineering}} \bibinfo{volume}{39}, \bibinfo{number}{11} (\bibinfo{year}{2013}), \bibinfo{pages}{1597--1610}.
\newblock


\bibitem[Kirkpatrick et~al\mbox{.}(2017)]%
        {kirkpatrick2017overcoming}
\bibfield{author}{\bibinfo{person}{James Kirkpatrick}, \bibinfo{person}{Razvan Pascanu}, \bibinfo{person}{Neil Rabinowitz}, \bibinfo{person}{Joel Veness}, \bibinfo{person}{Guillaume Desjardins}, \bibinfo{person}{Andrei~A Rusu}, \bibinfo{person}{Kieran Milan}, \bibinfo{person}{John Quan}, \bibinfo{person}{Tiago Ramalho}, \bibinfo{person}{Agnieszka Grabska-Barwinska}, {et~al\mbox{.}}} \bibinfo{year}{2017}\natexlab{}.
\newblock \showarticletitle{Overcoming catastrophic forgetting in neural networks}.
\newblock \bibinfo{journal}{\emph{Proceedings of the national academy of sciences}} \bibinfo{volume}{114}, \bibinfo{number}{13} (\bibinfo{year}{2017}), \bibinfo{pages}{3521--3526}.
\newblock


\bibitem[Kurle et~al\mbox{.}(2019)]%
        {kurle2019continual}
\bibfield{author}{\bibinfo{person}{Richard Kurle}, \bibinfo{person}{Botond Cseke}, \bibinfo{person}{Alexej Klushyn}, \bibinfo{person}{Patrick Van Der~Smagt}, {and} \bibinfo{person}{Stephan G{\"u}nnemann}.} \bibinfo{year}{2019}\natexlab{}.
\newblock \showarticletitle{Continual learning with bayesian neural networks for non-stationary data}. In \bibinfo{booktitle}{\emph{International Conference on Learning Representations}}.
\newblock


\bibitem[Lam et~al\mbox{.}(2017)]%
        {lam2017bug}
\bibfield{author}{\bibinfo{person}{An~Ngoc Lam}, \bibinfo{person}{Anh~Tuan Nguyen}, \bibinfo{person}{Hoan~Anh Nguyen}, {and} \bibinfo{person}{Tien~N Nguyen}.} \bibinfo{year}{2017}\natexlab{}.
\newblock \showarticletitle{Bug localization with combination of deep learning and information retrieval}. In \bibinfo{booktitle}{\emph{2017 IEEE/ACM 25th International Conference on Program Comprehension (ICPC)}}. IEEE, \bibinfo{pages}{218--229}.
\newblock


\bibitem[Lee et~al\mbox{.}(2022)]%
        {lee2022light}
\bibfield{author}{\bibinfo{person}{Jaehyung Lee}, \bibinfo{person}{Kisun Han}, {and} \bibinfo{person}{Hwanjo Yu}.} \bibinfo{year}{2022}\natexlab{}.
\newblock \showarticletitle{A light bug triage framework for applying large pre-trained language model}. In \bibinfo{booktitle}{\emph{Proceedings of the 37th IEEE/ACM International Conference on Automated Software Engineering}}. \bibinfo{pages}{1--11}.
\newblock


\bibitem[Li(2017)]%
        {li2017deep}
\bibfield{author}{\bibinfo{person}{Yuxi Li}.} \bibinfo{year}{2017}\natexlab{}.
\newblock \showarticletitle{Deep reinforcement learning: An overview}.
\newblock \bibinfo{journal}{\emph{arXiv preprint arXiv:1701.07274}} (\bibinfo{year}{2017}).
\newblock


\bibitem[Liang et~al\mbox{.}(2022)]%
        {liang2022modeling}
\bibfield{author}{\bibinfo{person}{Hongliang Liang}, \bibinfo{person}{Dengji Hang}, {and} \bibinfo{person}{Xiangyu Li}.} \bibinfo{year}{2022}\natexlab{}.
\newblock \showarticletitle{Modeling function-level interactions for file-level bug localization}.
\newblock \bibinfo{journal}{\emph{Empirical Software Engineering}} \bibinfo{volume}{27}, \bibinfo{number}{7} (\bibinfo{year}{2022}), \bibinfo{pages}{186}.
\newblock


\bibitem[Lu et~al\mbox{.}(2018)]%
        {lu2018learning}
\bibfield{author}{\bibinfo{person}{Jie Lu}, \bibinfo{person}{Anjin Liu}, \bibinfo{person}{Fan Dong}, \bibinfo{person}{Feng Gu}, \bibinfo{person}{Joao Gama}, {and} \bibinfo{person}{Guangquan Zhang}.} \bibinfo{year}{2018}\natexlab{}.
\newblock \showarticletitle{Learning under concept drift: A review}.
\newblock \bibinfo{journal}{\emph{IEEE transactions on knowledge and data engineering}} \bibinfo{volume}{31}, \bibinfo{number}{12} (\bibinfo{year}{2018}), \bibinfo{pages}{2346--2363}.
\newblock


\bibitem[Mccabe(1996)]%
        {mccabe1996cyclomatic}
\bibfield{author}{\bibinfo{person}{Thomas Mccabe}.} \bibinfo{year}{1996}\natexlab{}.
\newblock \showarticletitle{Cyclomatic complexity and the year 2000}.
\newblock \bibinfo{journal}{\emph{IEEE Software}} \bibinfo{volume}{13}, \bibinfo{number}{3} (\bibinfo{year}{1996}), \bibinfo{pages}{115--117}.
\newblock


\bibitem[McCloskey and Cohen(1989)]%
        {mccloskey1989catastrophic}
\bibfield{author}{\bibinfo{person}{Michael McCloskey} {and} \bibinfo{person}{Neal~J Cohen}.} \bibinfo{year}{1989}\natexlab{}.
\newblock \showarticletitle{Catastrophic interference in connectionist networks: The sequential learning problem}.
\newblock In \bibinfo{booktitle}{\emph{Psychology of learning and motivation}}. Vol.~\bibinfo{volume}{24}. \bibinfo{publisher}{Elsevier}, \bibinfo{pages}{109--165}.
\newblock


\bibitem[Mills et~al\mbox{.}(2018)]%
        {mills2018bug}
\bibfield{author}{\bibinfo{person}{Chris Mills}, \bibinfo{person}{Jevgenija Pantiuchina}, \bibinfo{person}{Esteban Parra}, \bibinfo{person}{Gabriele Bavota}, {and} \bibinfo{person}{Sonia Haiduc}.} \bibinfo{year}{2018}\natexlab{}.
\newblock \showarticletitle{Are bug reports enough for text retrieval-based bug localization?}. In \bibinfo{booktitle}{\emph{2018 IEEE International Conference on Software Maintenance and Evolution (ICSME)}}. IEEE, \bibinfo{pages}{381--392}.
\newblock


\bibitem[Mindom et~al\mbox{.}(2022)]%
        {mindom2022comparison}
\bibfield{author}{\bibinfo{person}{Paulina Stevia~Nouwou Mindom}, \bibinfo{person}{Amin Nikanjam}, {and} \bibinfo{person}{Foutse Khomh}.} \bibinfo{year}{2022}\natexlab{}.
\newblock \showarticletitle{A comparison of reinforcement learning frameworks for software testing tasks}.
\newblock \bibinfo{journal}{\emph{arXiv preprint arXiv:2208.12136}} (\bibinfo{year}{2022}).
\newblock


\bibitem[Miryeganeh et~al\mbox{.}(2021)]%
        {miryeganeh2021globug}
\bibfield{author}{\bibinfo{person}{Nima Miryeganeh}, \bibinfo{person}{Sepehr Hashtroudi}, {and} \bibinfo{person}{Hadi Hemmati}.} \bibinfo{year}{2021}\natexlab{}.
\newblock \showarticletitle{GloBug: Using global data in fault localization}.
\newblock \bibinfo{journal}{\emph{Journal of Systems and Software}}  \bibinfo{volume}{177} (\bibinfo{year}{2021}), \bibinfo{pages}{110961}.
\newblock


\bibitem[Moran et~al\mbox{.}(2016)]%
        {moran2016automatically}
\bibfield{author}{\bibinfo{person}{Kevin Moran}, \bibinfo{person}{Mario Linares-V{\'a}squez}, \bibinfo{person}{Carlos Bernal-C{\'a}rdenas}, \bibinfo{person}{Christopher Vendome}, {and} \bibinfo{person}{Denys Poshyvanyk}.} \bibinfo{year}{2016}\natexlab{}.
\newblock \showarticletitle{Automatically discovering, reporting and reproducing android application crashes}. In \bibinfo{booktitle}{\emph{2016 IEEE international conference on software testing, verification and validation (icst)}}. IEEE, \bibinfo{pages}{33--44}.
\newblock


\bibitem[Moser et~al\mbox{.}(2008)]%
        {moser2008comparative}
\bibfield{author}{\bibinfo{person}{Raimund Moser}, \bibinfo{person}{Witold Pedrycz}, {and} \bibinfo{person}{Giancarlo Succi}.} \bibinfo{year}{2008}\natexlab{}.
\newblock \showarticletitle{A comparative analysis of the efficiency of change metrics and static code attributes for defect prediction}. In \bibinfo{booktitle}{\emph{Proceedings of the 30th international conference on Software engineering}}. \bibinfo{pages}{181--190}.
\newblock


\bibitem[Munappy et~al\mbox{.}(2019)]%
        {munappy2019data}
\bibfield{author}{\bibinfo{person}{Aiswarya Munappy}, \bibinfo{person}{Jan Bosch}, \bibinfo{person}{Helena~Holmstr{\"o}m Olsson}, \bibinfo{person}{Anders Arpteg}, {and} \bibinfo{person}{Bj{\"o}rn Brinne}.} \bibinfo{year}{2019}\natexlab{}.
\newblock \showarticletitle{Data management challenges for deep learning}. In \bibinfo{booktitle}{\emph{2019 45th Euromicro Conference on Software Engineering and Advanced Applications (SEAA)}}. IEEE, \bibinfo{pages}{140--147}.
\newblock


\bibitem[Murali et~al\mbox{.}(2021)]%
        {murali2021industry}
\bibfield{author}{\bibinfo{person}{Vijayaraghavan Murali}, \bibinfo{person}{Lee Gross}, \bibinfo{person}{Rebecca Qian}, {and} \bibinfo{person}{Satish Chandra}.} \bibinfo{year}{2021}\natexlab{}.
\newblock \showarticletitle{Industry-scale ir-based bug localization: A perspective from facebook}. In \bibinfo{booktitle}{\emph{2021 IEEE/ACM 43rd International Conference on Software Engineering: Software Engineering in Practice (ICSE-SEIP)}}. IEEE, \bibinfo{pages}{188--197}.
\newblock


\bibitem[Nayrolles and Hamou-Lhadj(2018)]%
        {nayrolles2018clever}
\bibfield{author}{\bibinfo{person}{Mathieu Nayrolles} {and} \bibinfo{person}{Abdelwahab Hamou-Lhadj}.} \bibinfo{year}{2018}\natexlab{}.
\newblock \showarticletitle{Clever: Combining code metrics with clone detection for just-in-time fault prevention and resolution in large industrial projects}. In \bibinfo{booktitle}{\emph{Proceedings of the 15th international conference on mining software repositories}}. \bibinfo{pages}{153--164}.
\newblock


\bibitem[Nguyen et~al\mbox{.}(2011)]%
        {nguyen2011topic}
\bibfield{author}{\bibinfo{person}{Anh~Tuan Nguyen}, \bibinfo{person}{Tung~Thanh Nguyen}, \bibinfo{person}{Jafar Al-Kofahi}, \bibinfo{person}{Hung~Viet Nguyen}, {and} \bibinfo{person}{Tien~N Nguyen}.} \bibinfo{year}{2011}\natexlab{}.
\newblock \showarticletitle{A topic-based approach for narrowing the search space of buggy files from a bug report}. In \bibinfo{booktitle}{\emph{2011 26th IEEE/ACM International Conference on Automated Software Engineering (ASE 2011)}}. IEEE, \bibinfo{pages}{263--272}.
\newblock


\bibitem[Olewicki et~al\mbox{.}(2023)]%
        {olewicki2023towards}
\bibfield{author}{\bibinfo{person}{Doriane Olewicki}, \bibinfo{person}{Sarra Habchi}, \bibinfo{person}{Mathieu Nayrolles}, \bibinfo{person}{Mojtaba Faramarzi}, \bibinfo{person}{Sarath Chandar}, {and} \bibinfo{person}{Bram Adams}.} \bibinfo{year}{2023}\natexlab{}.
\newblock \showarticletitle{Towards Lifelong Learning for Software Analytics Models: Empirical Study on Brown Build and Risk Prediction}.
\newblock \bibinfo{journal}{\emph{arXiv preprint arXiv:2305.09824}} (\bibinfo{year}{2023}).
\newblock


\bibitem[Olewicki et~al\mbox{.}(2024)]%
        {Olewicki2024Costs}
\bibfield{author}{\bibinfo{person}{Doriane Olewicki}, \bibinfo{person}{Sarra Habchi}, \bibinfo{person}{Mathieu Nayrolles}, \bibinfo{person}{Mojtaba Faramarzi}, \bibinfo{person}{Sarath Chandar}, {and} \bibinfo{person}{Bram Adams}.} \bibinfo{year}{2024}\natexlab{}.
\newblock \showarticletitle{On the Costs and Benefits of Adopting Lifelong Learning for Software Analytics - Empirical Study on Brown Build and Risk Prediction}. In \bibinfo{booktitle}{\emph{Proceedings of the 46th International Conference on Software Engineering: Software Engineering in Practice}} \emph{(\bibinfo{series}{ICSE-SEIP '24})}.
\newblock
\urldef\tempurl%
\url{https://doi.org/10.1145/3639477.3639717}
\showDOI{\tempurl}


\bibitem[Olewicki et~al\mbox{.}(2022)]%
        {olewicki2022towards}
\bibfield{author}{\bibinfo{person}{Doriane Olewicki}, \bibinfo{person}{Mathieu Nayrolles}, {and} \bibinfo{person}{Bram Adams}.} \bibinfo{year}{2022}\natexlab{}.
\newblock \showarticletitle{Towards language-independent brown build detection}. In \bibinfo{booktitle}{\emph{Proceedings of the 44th International Conference on Software Engineering}}. \bibinfo{pages}{2177--2188}.
\newblock


\bibitem[Parisi et~al\mbox{.}(2019)]%
        {parisi2019continual}
\bibfield{author}{\bibinfo{person}{German~I Parisi}, \bibinfo{person}{Ronald Kemker}, \bibinfo{person}{Jose~L Part}, \bibinfo{person}{Christopher Kanan}, {and} \bibinfo{person}{Stefan Wermter}.} \bibinfo{year}{2019}\natexlab{}.
\newblock \showarticletitle{Continual lifelong learning with neural networks: A review}.
\newblock \bibinfo{journal}{\emph{Neural networks}}  \bibinfo{volume}{113} (\bibinfo{year}{2019}), \bibinfo{pages}{54--71}.
\newblock


\bibitem[Petrillo et~al\mbox{.}(2016)]%
        {petrillo2016towards}
\bibfield{author}{\bibinfo{person}{Fabio Petrillo}, \bibinfo{person}{Z{\'e}phyrin Soh}, \bibinfo{person}{Foutse Khomh}, \bibinfo{person}{Marcelo Pimenta}, \bibinfo{person}{Carla Freitas}, {and} \bibinfo{person}{Yann-Ga{\"e}l Gu{\'e}h{\'e}neuc}.} \bibinfo{year}{2016}\natexlab{}.
\newblock \showarticletitle{Towards understanding interactive debugging}. In \bibinfo{booktitle}{\emph{2016 IEEE International Conference on Software Quality, Reliability and Security (QRS)}}. IEEE, \bibinfo{pages}{152--163}.
\newblock


\bibitem[Powers et~al\mbox{.}(2022)]%
        {powers2022cora}
\bibfield{author}{\bibinfo{person}{Sam Powers}, \bibinfo{person}{Eliot Xing}, \bibinfo{person}{Eric Kolve}, \bibinfo{person}{Roozbeh Mottaghi}, {and} \bibinfo{person}{Abhinav Gupta}.} \bibinfo{year}{2022}\natexlab{}.
\newblock \showarticletitle{Cora: Benchmarks, baselines, and metrics as a platform for continual reinforcement learning agents}. In \bibinfo{booktitle}{\emph{Conference on Lifelong Learning Agents}}. PMLR, \bibinfo{pages}{705--743}.
\newblock


\bibitem[Rath et~al\mbox{.}(2018)]%
        {rath2018analyzing}
\bibfield{author}{\bibinfo{person}{Michael Rath}, \bibinfo{person}{David Lo}, {and} \bibinfo{person}{Patrick M{\"a}der}.} \bibinfo{year}{2018}\natexlab{}.
\newblock \showarticletitle{Analyzing requirements and traceability information to improve bug localization}. In \bibinfo{booktitle}{\emph{Proceedings of the 15th International Conference on Mining Software Repositories}}. \bibinfo{pages}{442--453}.
\newblock


\bibitem[Roehm et~al\mbox{.}(2013)]%
        {roehm2013monitoring}
\bibfield{author}{\bibinfo{person}{Tobias Roehm}, \bibinfo{person}{Nigar Gurbanova}, \bibinfo{person}{Bernd Bruegge}, \bibinfo{person}{Christophe Joubert}, {and} \bibinfo{person}{Walid Maalej}.} \bibinfo{year}{2013}\natexlab{}.
\newblock \showarticletitle{Monitoring user interactions for supporting failure reproduction}. In \bibinfo{booktitle}{\emph{2013 21st International Conference on Program Comprehension (ICPC)}}. IEEE, \bibinfo{pages}{73--82}.
\newblock


\bibitem[Rolnick et~al\mbox{.}(2019)]%
        {rolnick2019experience}
\bibfield{author}{\bibinfo{person}{David Rolnick}, \bibinfo{person}{Arun Ahuja}, \bibinfo{person}{Jonathan Schwarz}, \bibinfo{person}{Timothy Lillicrap}, {and} \bibinfo{person}{Gregory Wayne}.} \bibinfo{year}{2019}\natexlab{}.
\newblock \showarticletitle{Experience replay for continual learning}.
\newblock \bibinfo{journal}{\emph{Advances in neural information processing systems}}  \bibinfo{volume}{32} (\bibinfo{year}{2019}).
\newblock


\bibitem[Rosen et~al\mbox{.}(2015)]%
        {rosen2015commit}
\bibfield{author}{\bibinfo{person}{Christoffer Rosen}, \bibinfo{person}{Ben Grawi}, {and} \bibinfo{person}{Emad Shihab}.} \bibinfo{year}{2015}\natexlab{}.
\newblock \showarticletitle{Commit guru: analytics and risk prediction of software commits}. In \bibinfo{booktitle}{\emph{Proceedings of the 2015 10th joint meeting on foundations of software engineering}}. \bibinfo{pages}{966--969}.
\newblock


\bibitem[Sauder and DeMars(2019)]%
        {sauder2019updated}
\bibfield{author}{\bibinfo{person}{Derek~C Sauder} {and} \bibinfo{person}{Christine~E DeMars}.} \bibinfo{year}{2019}\natexlab{}.
\newblock \showarticletitle{An updated recommendation for multiple comparisons}.
\newblock \bibinfo{journal}{\emph{Advances in Methods and Practices in Psychological Science}} \bibinfo{volume}{2}, \bibinfo{number}{1} (\bibinfo{year}{2019}), \bibinfo{pages}{26--44}.
\newblock


\bibitem[Savor et~al\mbox{.}(2016)]%
        {savor2016continuous}
\bibfield{author}{\bibinfo{person}{Tony Savor}, \bibinfo{person}{Mitchell Douglas}, \bibinfo{person}{Michael Gentili}, \bibinfo{person}{Laurie Williams}, \bibinfo{person}{Kent Beck}, {and} \bibinfo{person}{Michael Stumm}.} \bibinfo{year}{2016}\natexlab{}.
\newblock \showarticletitle{Continuous deployment at Facebook and OANDA}. In \bibinfo{booktitle}{\emph{Proceedings of the 38th International Conference on software engineering companion}}. \bibinfo{pages}{21--30}.
\newblock


\bibitem[Silver et~al\mbox{.}(2013)]%
        {silver2013lifelong}
\bibfield{author}{\bibinfo{person}{Daniel~L Silver}, \bibinfo{person}{Qiang Yang}, {and} \bibinfo{person}{Lianghao Li}.} \bibinfo{year}{2013}\natexlab{}.
\newblock \showarticletitle{Lifelong machine learning systems: Beyond learning algorithms}. In \bibinfo{booktitle}{\emph{2013 AAAI spring symposium series}}.
\newblock


\bibitem[Singh et~al\mbox{.}(2012)]%
        {singh2012tracking}
\bibfield{author}{\bibinfo{person}{Anshuman Singh}, \bibinfo{person}{Andrew Walenstein}, {and} \bibinfo{person}{Arun Lakhotia}.} \bibinfo{year}{2012}\natexlab{}.
\newblock \showarticletitle{Tracking concept drift in malware families}. In \bibinfo{booktitle}{\emph{Proceedings of the 5th ACM workshop on Security and artificial intelligence}}. \bibinfo{pages}{81--92}.
\newblock


\bibitem[Sutton and Barto(1998)]%
        {sutton1998reinforcement}
\bibfield{author}{\bibinfo{person}{Richard~S Sutton} {and} \bibinfo{person}{Andrew~G Barto}.} \bibinfo{year}{1998}\natexlab{}.
\newblock \showarticletitle{Reinforcement learning: an introduction MIT Press}.
\newblock \bibinfo{journal}{\emph{Cambridge, MA}}  \bibinfo{volume}{22447} (\bibinfo{year}{1998}).
\newblock


\bibitem[Taba et~al\mbox{.}(2013)]%
        {taba2013predicting}
\bibfield{author}{\bibinfo{person}{Seyyed Ehsan~Salamati Taba}, \bibinfo{person}{Foutse Khomh}, \bibinfo{person}{Ying Zou}, \bibinfo{person}{Ahmed~E Hassan}, {and} \bibinfo{person}{Meiyappan Nagappan}.} \bibinfo{year}{2013}\natexlab{}.
\newblock \showarticletitle{Predicting bugs using antipatterns}. In \bibinfo{booktitle}{\emph{2013 IEEE International Conference on Software Maintenance}}. IEEE, \bibinfo{pages}{270--279}.
\newblock


\bibitem[Wang et~al\mbox{.}(2020)]%
        {wang2020multi}
\bibfield{author}{\bibinfo{person}{Bei Wang}, \bibinfo{person}{Ling Xu}, \bibinfo{person}{Meng Yan}, \bibinfo{person}{Chao Liu}, {and} \bibinfo{person}{Ling Liu}.} \bibinfo{year}{2020}\natexlab{}.
\newblock \showarticletitle{Multi-dimension convolutional neural network for bug localization}.
\newblock \bibinfo{journal}{\emph{IEEE Transactions on Services Computing}} \bibinfo{volume}{15}, \bibinfo{number}{3} (\bibinfo{year}{2020}), \bibinfo{pages}{1649--1663}.
\newblock


\bibitem[Wang et~al\mbox{.}(2023)]%
        {wang2023systematic}
\bibfield{author}{\bibinfo{person}{Di Wang}, \bibinfo{person}{Matthias Galster}, {and} \bibinfo{person}{Miguel Morales-Trujillo}.} \bibinfo{year}{2023}\natexlab{}.
\newblock \showarticletitle{A systematic mapping study of bug reproduction and localization}.
\newblock \bibinfo{journal}{\emph{Information and Software Technology}} (\bibinfo{year}{2023}), \bibinfo{pages}{107338}.
\newblock


\bibitem[Wang et~al\mbox{.}(2013)]%
        {wang2013improving}
\bibfield{author}{\bibinfo{person}{Shaohua Wang}, \bibinfo{person}{Foutse Khomh}, {and} \bibinfo{person}{Ying Zou}.} \bibinfo{year}{2013}\natexlab{}.
\newblock \showarticletitle{Improving bug localization using correlations in crash reports}. In \bibinfo{booktitle}{\emph{2013 10th Working Conference on Mining Software Repositories (MSR)}}. IEEE, \bibinfo{pages}{247--256}.
\newblock


\bibitem[Welch(1947)]%
        {welch1947generalization}
\bibfield{author}{\bibinfo{person}{Bernard~L Welch}.} \bibinfo{year}{1947}\natexlab{}.
\newblock \showarticletitle{The generalization of ‘STUDENT'S’problem when several different population varlances are involved}.
\newblock \bibinfo{journal}{\emph{Biometrika}} \bibinfo{volume}{34}, \bibinfo{number}{1-2} (\bibinfo{year}{1947}), \bibinfo{pages}{28--35}.
\newblock


\bibitem[Wen et~al\mbox{.}(2016)]%
        {wen2016locus}
\bibfield{author}{\bibinfo{person}{Ming Wen}, \bibinfo{person}{Rongxin Wu}, {and} \bibinfo{person}{Shing-Chi Cheung}.} \bibinfo{year}{2016}\natexlab{}.
\newblock \showarticletitle{Locus: Locating bugs from software changes}. In \bibinfo{booktitle}{\emph{Proceedings of the 31st IEEE/ACM International Conference on Automated Software Engineering}}. \bibinfo{pages}{262--273}.
\newblock


\bibitem[Wong et~al\mbox{.}(2010)]%
        {wong2010family}
\bibfield{author}{\bibinfo{person}{W~Eric Wong}, \bibinfo{person}{Vidroha Debroy}, {and} \bibinfo{person}{Byoungju Choi}.} \bibinfo{year}{2010}\natexlab{}.
\newblock \showarticletitle{A family of code coverage-based heuristics for effective fault localization}.
\newblock \bibinfo{journal}{\emph{Journal of Systems and Software}} \bibinfo{volume}{83}, \bibinfo{number}{2} (\bibinfo{year}{2010}), \bibinfo{pages}{188--208}.
\newblock


\bibitem[Wong et~al\mbox{.}(2016)]%
        {wong2016survey}
\bibfield{author}{\bibinfo{person}{W~Eric Wong}, \bibinfo{person}{Ruizhi Gao}, \bibinfo{person}{Yihao Li}, \bibinfo{person}{Rui Abreu}, {and} \bibinfo{person}{Franz Wotawa}.} \bibinfo{year}{2016}\natexlab{}.
\newblock \showarticletitle{A survey on software fault localization}.
\newblock \bibinfo{journal}{\emph{IEEE Transactions on Software Engineering}} \bibinfo{volume}{42}, \bibinfo{number}{8} (\bibinfo{year}{2016}), \bibinfo{pages}{707--740}.
\newblock


\bibitem[Wong et~al\mbox{.}(1997)]%
        {wong1997study}
\bibfield{author}{\bibinfo{person}{W~Eric Wong}, \bibinfo{person}{Joseph~R Horgan}, \bibinfo{person}{Saul London}, {and} \bibinfo{person}{Hiralal Agrawal}.} \bibinfo{year}{1997}\natexlab{}.
\newblock \showarticletitle{A study of effective regression testing in practice}. In \bibinfo{booktitle}{\emph{PROCEEDINGS The Eighth International Symposium On Software Reliability Engineering}}. IEEE, \bibinfo{pages}{264--274}.
\newblock


\bibitem[Wu et~al\mbox{.}(2018)]%
        {wu2018changelocator}
\bibfield{author}{\bibinfo{person}{Rongxin Wu}, \bibinfo{person}{Ming Wen}, \bibinfo{person}{Shing-Chi Cheung}, {and} \bibinfo{person}{Hongyu Zhang}.} \bibinfo{year}{2018}\natexlab{}.
\newblock \showarticletitle{Changelocator: locate crash-inducing changes based on crash reports}.
\newblock \bibinfo{journal}{\emph{Empirical Software Engineering}}  \bibinfo{volume}{23} (\bibinfo{year}{2018}), \bibinfo{pages}{2866--2900}.
\newblock


\bibitem[Xiao et~al\mbox{.}(2019)]%
        {xiao2019improving}
\bibfield{author}{\bibinfo{person}{Yan Xiao}, \bibinfo{person}{Jacky Keung}, \bibinfo{person}{Kwabena~E Bennin}, {and} \bibinfo{person}{Qing Mi}.} \bibinfo{year}{2019}\natexlab{}.
\newblock \showarticletitle{Improving bug localization with word embedding and enhanced convolutional neural networks}.
\newblock \bibinfo{journal}{\emph{Information and Software Technology}}  \bibinfo{volume}{105} (\bibinfo{year}{2019}), \bibinfo{pages}{17--29}.
\newblock


\bibitem[Ye et~al\mbox{.}(2014)]%
        {ye2014learning}
\bibfield{author}{\bibinfo{person}{Xin Ye}, \bibinfo{person}{Razvan Bunescu}, {and} \bibinfo{person}{Chang Liu}.} \bibinfo{year}{2014}\natexlab{}.
\newblock \showarticletitle{Learning to rank relevant files for bug reports using domain knowledge}. In \bibinfo{booktitle}{\emph{Proceedings of the 22nd ACM SIGSOFT international symposium on foundations of software engineering}}. \bibinfo{pages}{689--699}.
\newblock


\bibitem[Ye et~al\mbox{.}(2015)]%
        {ye2015mapping}
\bibfield{author}{\bibinfo{person}{Xin Ye}, \bibinfo{person}{Razvan Bunescu}, {and} \bibinfo{person}{Chang Liu}.} \bibinfo{year}{2015}\natexlab{}.
\newblock \showarticletitle{Mapping bug reports to relevant files: A ranking model, a fine-grained benchmark, and feature evaluation}.
\newblock \bibinfo{journal}{\emph{IEEE Transactions on Software Engineering}} \bibinfo{volume}{42}, \bibinfo{number}{4} (\bibinfo{year}{2015}), \bibinfo{pages}{379--402}.
\newblock


\bibitem[Yu et~al\mbox{.}(2017)]%
        {yu2017descry}
\bibfield{author}{\bibinfo{person}{Tingting Yu}, \bibinfo{person}{Tarannum~S Zaman}, {and} \bibinfo{person}{Chao Wang}.} \bibinfo{year}{2017}\natexlab{}.
\newblock \showarticletitle{DESCRY: reproducing system-level concurrency failures}. In \bibinfo{booktitle}{\emph{Proceedings of the 2017 11th Joint Meeting on Foundations of Software Engineering}}. \bibinfo{pages}{694--704}.
\newblock


\bibitem[Zhang et~al\mbox{.}(2015)]%
        {zhang2015survey}
\bibfield{author}{\bibinfo{person}{Jie Zhang}, \bibinfo{person}{Xiaoyin Wang}, \bibinfo{person}{Dan Hao}, \bibinfo{person}{Bing Xie}, \bibinfo{person}{Lu Zhang}, {and} \bibinfo{person}{Hong Mei}.} \bibinfo{year}{2015}\natexlab{}.
\newblock \showarticletitle{A survey on bug-report analysis.}
\newblock \bibinfo{journal}{\emph{Sci. China Inf. Sci.}} \bibinfo{volume}{58}, \bibinfo{number}{2} (\bibinfo{year}{2015}), \bibinfo{pages}{1--24}.
\newblock


\bibitem[Zhang et~al\mbox{.}(2019)]%
        {zhang2019cnn}
\bibfield{author}{\bibinfo{person}{Zhuo Zhang}, \bibinfo{person}{Yan Lei}, \bibinfo{person}{Xiaoguang Mao}, {and} \bibinfo{person}{Panpan Li}.} \bibinfo{year}{2019}\natexlab{}.
\newblock \showarticletitle{CNN-FL: An effective approach for localizing faults using convolutional neural networks}. In \bibinfo{booktitle}{\emph{2019 IEEE 26th International Conference on Software Analysis, Evolution and Reengineering (SANER)}}. IEEE, \bibinfo{pages}{445--455}.
\newblock


\bibitem[Zhou et~al\mbox{.}(2012)]%
        {zhou2012should}
\bibfield{author}{\bibinfo{person}{Jian Zhou}, \bibinfo{person}{Hongyu Zhang}, {and} \bibinfo{person}{David Lo}.} \bibinfo{year}{2012}\natexlab{}.
\newblock \showarticletitle{Where should the bugs be fixed? more accurate information retrieval-based bug localization based on bug reports}. In \bibinfo{booktitle}{\emph{2012 34th International conference on software engineering (ICSE)}}. IEEE, \bibinfo{pages}{14--24}.
\newblock


\bibitem[Zimmermann et~al\mbox{.}(2007)]%
        {zimmermann2007predicting}
\bibfield{author}{\bibinfo{person}{Thomas Zimmermann}, \bibinfo{person}{Rahul Premraj}, {and} \bibinfo{person}{Andreas Zeller}.} \bibinfo{year}{2007}\natexlab{}.
\newblock \showarticletitle{Predicting defects for eclipse}. In \bibinfo{booktitle}{\emph{Third International Workshop on Predictor Models in Software Engineering (PROMISE'07: ICSE Workshops 2007)}}. IEEE, \bibinfo{pages}{9--9}.
\newblock


\bibitem[Zou et~al\mbox{.}(2018)]%
        {zou2018practitioners}
\bibfield{author}{\bibinfo{person}{Weiqin Zou}, \bibinfo{person}{David Lo}, \bibinfo{person}{Zhenyu Chen}, \bibinfo{person}{Xin Xia}, \bibinfo{person}{Yang Feng}, {and} \bibinfo{person}{Baowen Xu}.} \bibinfo{year}{2018}\natexlab{}.
\newblock \showarticletitle{How practitioners perceive automated bug report management techniques}.
\newblock \bibinfo{journal}{\emph{IEEE Transactions on Software Engineering}} \bibinfo{volume}{46}, \bibinfo{number}{8} (\bibinfo{year}{2018}), \bibinfo{pages}{836--862}.
\newblock


\end{thebibliography}

\end{document}